\documentclass[11 pt,a4paper]{article}
\usepackage[utf8]{inputenc}
\usepackage[T1]{fontenc}
\usepackage[top=1in, bottom=1.5in, left=1in, right=1in]{geometry}
\pdfoutput=1 

\usepackage[skins,theorems]{tcolorbox}
\tcbset{highlight math style={enhanced,
  colframe=brown,colback=lightgray,arc=5pt,boxrule=1pt}}
\definecolor{lightgray}{gray}{0.92}
\usepackage[title,titletoc,toc,page]{appendix}
\usepackage{authblk}
\usepackage{hyperref}     
\usepackage{color}
\usepackage{graphicx}
\usepackage{amssymb}
\usepackage{amsmath}
\usepackage{amsfonts}
\usepackage{epstopdf}
\usepackage{paralist}
\usepackage{setspace}
\usepackage{tikz}
\usetikzlibrary{angles,quotes} 
\usepackage{caption}
\usepackage{empheq}
\def\lnf{\ln f(p,z)} 
\def\ha{\frac{1}{2}}
\def\s{\mathcal{S}}
\def\ppz{\phi(p,z)}

\newcommand{\p}{\partial}
\newcommand{\DDx}[2]{\frac{\dd #1}{\dd #2}}
\newcommand{\ddx}[2]{\frac{d #1}{d #2}}
\newcommand{\ppx}[2]{\frac{\p #1}{\p #2}}

\newcommand{\ppt}{\frac{\partial }{\partial t}}
\newcommand{\lo}{\Lambda_0}
\newcommand{\lm}{\Lambda}
\newcommand{\hf}{{1\over 2}}
\newcommand{\be}{\begin{equation}}
\newcommand{\br}{\begin{eqnarray}}
\newcommand{\er}{\end{eqnarray}}
\newcommand{\ee}{\end{equation}}
\newcommand{\bt}{\begin{tabular}}
\newcommand{\et}{\end{tabular}}
\newcommand{\bc}{\begin{tcolorbox}}
\newcommand{\ec}{\end{tcolorbox}}

\newcommand{\dd}{\delta}

\newcommand{\dsp}{\frac{\delta S}{\delta \phi(p)}}
\newcommand{\dsmp}{\frac{\delta S}{\delta \phi(-p)}}
\newcommand{\ddp}{\frac{\delta }{\delta \phi(p)}}
\newcommand{\ddmp}{\frac{\delta }{\delta \phi(-p)}}
\newcommand{\CD}{{\cal D}}
\newcommand{\Dp}{\frac{d^Dp}{(2\pi)^D}}

\newcommand{\eps}{\epsilon}

\newcommand{\ep}{\epsilon}

\newcommand{\ddt}{\frac{\mathrm d}{\mathrm d t}}
\numberwithin{equation}{section}
\makeatletter
\g@addto@macro\bfseries{\boldmath}
\makeatother

\begin{document}

\title{$SO(1,d+1)$ symmetry of the Exact RG equation.}
\author[1]{Semanti Dutta \thanks{\href{mailto:semanti.dutta2010@gmail.com}{semanti.dutta2010@gmail.com}}}
\author[2,3]{B. Sathiapalan \thanks{\href{mailto:bala@imsc.res.in}{bala@imsc.res.in}}}
\affil[1]{S.N. Bose National Center for Basic Sciences, JD Block, Sector III, Salt Lake, Kolkata 700106, India}
\affil[2]{Institute of Mathematical Sciences, CIT Campus, Tharamani, Chennai 600113, India}
\affil[3]{Homi Bhabha National Institute\\Training School Complex, Anushakti Nagar, Mumbai 400085, India}

\makeatletter
\g@addto@macro\bfseries{\boldmath}
\makeatother

{\let\newpage\relax\maketitle}
\begin{abstract}
There is a method for constructing from first principles, a holographic bulk dual action in Euclidean $AdS_{d+1}$ space for a $d$-dimensional Euclidean CFT on the boundary, starting from the Polchinski's Exact Renormalization Group~(ERG) equation that describes the RG evolution of the interaction part of the boundary Wilson action. The bulk action in $AdS_{d+1}$ has an $SO(1,d+1)$ symmetry and is obtained from the evolution operator of the Polchinski's ERG equation  by a map that involves a field redefinition and requires a {\em special} form of the UV cutoff function in the ERG equation. 
In this paper, we show that for {\em any form} of the cutoff function, the ERG evolution operator has an $SO(1,d+1)$  symmetry.  The generators of the special conformal transformation depend on the cutoff function. For the special cutoff function that maps to $AdS$ space, the transformations have the standard form of $AdS$ isometry. We also show that the ERG evolution operator for the {\em full} Wilson action can  be put in the same form as the Polchinski's ERG equation by a field redefinition and consequently also has an $SO(1,d+1)$ symmetry for any cutoff function. 

\end{abstract}

\newpage 
\tableofcontents 

\section{Introduction}

The $AdS/CFT$ correspondence \cite{Maldacena,Polyakov,Witten1,Witten2} relates a gravitational theory in $AdS$ space to a CFT living on the boundary of $AdS$ space and is a concrete example of holography\cite{tHooft:1993,Susskind:1994}.  Holographic renormalization group provides some insight into this correspondence
[\cite{Akhmedov}-\cite{SSLee:2012}]. A suggestion for deriving holographic renormalization group from exact renormalization group~(ERG) [\cite{Wilson}-\cite{Wilson2}] was made in \cite{Sathiapalan:2017,Sathiapalan:2019} where it was shown that the  action for a free scalar field in $AdS$ space is related by a field redefinition to the {\em evolution operator} of Polchinski's exact renormalization group equation \cite{Polchinski}, when written as a functional integral. This idea has been succesfully used to obtain bulk AdS actions involving scalars, Yang-Mills vector fields and spin 2 graviton in [\cite{Sathiapalan:2020}-\cite{Dharanipragada:2023}].

Schematically the flow of boundary low energy action $S_{t}$ along renormalization group scale $t$  is given as
\[
e^{-S_{t_f}} = \int_{t_i}^{t_f} dt \int \mathcal{D} \phi\, e^{\overbrace{-\s[\phi]}^{\text{\textit{bulk} action}}}\,e^{-S_{t_i}}.
\]
The scalar field \textit{evolution} action $\s[\phi]$ that goes into this  functional integral does not appear to have any connection to an action in $AdS$ space, but remarkably, after the field redefinition it acquires precisely such a form.

One of the explanations for why $AdS$ space is a natural candidate for a bulk dual is the fact that the isometry group of $AdS_{d+1}$ is $SO(2,d)$ (for Euclidean $AdS_{d+1}$ it is $SO(1,d+1)$), which also happens to be the conformal group of a $d$ dimensional CFT. Now the action used in the ERG evolution operator has a cutoff function, which can be chosen freely as long as it regulates the UV end of the theory. It was shown in \cite{Sathiapalan:2017,Sathiapalan:2019} that mapping to $AdS$ by means of a field redefinition requires, not surpisingly, a very special form of the cutoff function. And for this form of the cutoff function one expects the evolution operator to have the same $SO(1,d+1)$ symmetry. 

On the other hand for {\em any} form of the cutoff function, the ERG evolution operator has the property that it leaves a fixed point theory invariant. Indeed the form of the fixed point Wilson action is determined by solving the fixed point ERG equation - which imposes scale invariance. The fixed point theory is scale invariant and, with few exceptions, such theories are also conformally invariant. So we must conclude that the ERG evolution operator should preserve this conformal symmetry for {\em any} form of the cutoff function. We show this to be true in this paper. The evolution operator is invariant under a non standard $SO(1,d+1)$ ``conformal'' transformation that, not unexpectedly, depends on the cutoff function. This also makes it non polynomial in derivatives and hence quasi-local\footnote{By ``quasi local'' it is meant that there is a Taylor expansion in derivatives.}. For the special choice of the cutoff function that has to be made in order to get $AdS$ space, these $SO(1,d+1)$ transformations take on the standard form of the isometry of the $AdS$ metric. 

Let us make a small digression to discuss the issue of scale invariance of the boundary theory {\em in the presence of a cutoff parameter $\lm$}, where $\lm=\lm_0 e^{-t}$ and $\lm_0$ is the UV cutoff. One of the transformations contained in $SO(1,d+1)$ is dilatations. The fixed point Wilson action has a finite cutoff parameter $\lm$ which naively breaks this symmetry. But we know that physics of this theory is scale invariant inspite of that. We can resolve this issue as follows. In ERG formalism this symmetry is expressed in terms of composite operators \cite{Igarashi,Sonoda:2015}: the combination $\phi \DDx{}{\phi}$ that implement scale transformation of the field variable is replaced by the composite operator $[\phi \DDx{}{\phi}]$ for a finite cutoff. For instance consider the free Wilson action $S= \hf \int \Dp \phi(p) \frac{p^2}{K(p/\lm)}\phi(-p)$. This is not naively scale invariant because of the factor $K$. However one can show that, for the free theory, $[\phi \DDx{}{\phi}]= K(p/\lm) \phi \DDx{}{\phi}$. This removes the factor $K$ in the Wilson action and restores the naive scale invariance~(See Appendix \ref{FP} for details). We can show that the net effect of this modification involving composite operators in the implementation of scale transformations is to also scale the value  of $\lm$ along with $p$ so that the combination $p/\lm$ is invariant. Thus effectively there is a term $\lm \ddx{}{\lm}$ in the generator of dilatations in the {\em boundary} theory. Now, $\lm$ is a parameter (like mass), and is neither a field variable in the boundary theory nor a coordinate on which the fields depend, so one cannot legitimately add $\lm \ddx{}{\lm}$ to a scaling transformation law. So as explained in  Appendix \ref{FP}, this scaling is implemented as an RG evolution. On the other hand in the bulk theory $\lm$ is a coordinate and $\lm \ddx{}{\lm}$ can legitimately be included as a scaling transformation. Similar arguments hold for conformal symmetry - which is also naively broken by the presence of $\lm$. Thus the net effect is that the $SO(1,d+1)$ generators of the \textit{bulk} action have extra terms involving $\ppt$ (here $\lm = \lo e^{-t}$). We note a similar formulation of scale dependent conformal transformation as a systematic method to derive a renormalized operator in the scheme of \textit{bulk reconstruction}~\cite{Wang:2015}.

In this paper we demonstrate this symmetry using the functional formalism involving an action. We also study it as a Hamiltonian evolution in the Heisenberg picture. In the Heisenberg picture RG scale $t$ is a variable along with momentum $p$ (as also in the functional formalism) unlike the Schrodinger formalism where all the $t$ dependence is in the wave function. Therefore this symmetry is easier to see in the Heisenberg picture. The existence of this symmetry is the main result of this paper. 

The ERG equation that has been employed in \cite{Sathiapalan:2017,Sathiapalan:2019} (and in subsequent papers) is Polchinski equation which is for the RG evolution of the interaction term in the Wilson action \cite{Polchinski}. One of the original reasons for this was that this version of the ERG equation has a convenient form of a functional Schroedinger equation in Euclidean time, or equivalently a diffusion equation- for which the evolution operator has a simple ``path integral for a free particle''  form. The ERG equation for the evolution of the full Wilson action contains additional terms. In this paper we  show that with a field redefinition, this \textit{full ERG equation} can also be written as a diffusion equation. Hence all the techniques that we have used in relating ERG to Holographic RG so far can be adapted to the ERG equation for the full action. As a consequence we have also shown that the $SO(1,d+1)$ symmetry is present in this case with the appropriate changes in the transformation laws.\\

This paper is organized as follows. In section \ref{sec: review} we briefly review how holographic RG can be interpreted using exact RG in the boundary. We also show how composite operators can be used to define the symmetry transformations on the Wilson action with a finite cutoff scale. We find the symmetry transformations of bulk actions in section~\ref{sec: symmetry phi} and \ref{sec: symmetry y} respectively.  Further these symmetry transformations have been obtained using Hamiltonian formalism in section~\ref{sec: conformal Hamiltonian} and its group structure established in section~\ref{sec: conf algebra}. We map  the symmetry transformations of the bulk action to $AdS$ isometry in section~\ref{sec: map AdS}.  In Section~\ref{sec: full ERG} the equivalence of the full ERG equation with the Polchinski's version is shown. Section \ref{concl} contains a summary and conclusions.

\hspace{0.4 in} For convenience the details of the calculations have all been relegated to many appendices. Appendices \ref{FP} and \ref{Full} contain some background material about ERG. Appendix \ref{app: AdS isometry} summarizes the isometry group of AdS. Appendix \ref{app: conformal transformation phi details}-\ref{app: consistency} contains the derivations of the symmetry transformations of the evolution actions and their consistency checks. Appendix~\ref{app: Hamiltonian details} and \ref{app: conf algebra details} respectively contains the details of obtaining the symmetry transformations in Hamiltonian formalism and its group structure. In appendix \ref{app: full ERG} we have computed the symmetry transformation for the full ERG equation. Finally appendix \ref{app: Identities} contains some identities which are useful in consistency checks of symmetry transformations.

\section{Review}
\label{sec: review}
\subsection{ Exact RG and holographic RG}

We consider a low energy effective action in momentum space at cutoff parameter or RG scale $t$ in $d$ dimensional Euclidean space around a Gaussian fixed point.
\begin{eqnarray}\label{eq: full Wilson action}
S_t[\phi_l] = \ha \int \frac{d^d p}{(2 \pi)^d}\,\frac{\phi_l(p) \phi_l(-p)}{G(p,t)}+S_{I,t}[\phi_l].
\end{eqnarray}
This action is obtained by inegrating out the high momentum modes $p> \lm$ ($\lm=\lm_0 e^{-t}$, $\lm_0$ being the ultra-violet~(UV) cutoff). $\phi_l$ denotes the low momentum field modes, for notational convenience we will omit the subscript $l$. $G(p,t)$ serves as a cutoff function of the theory. We assume the functional dependence for  $G(p,t)$ as
\begin{equation}\label{eq: G functional form}
G(p,t)=\frac{K(p,t)}{p^{2 \nu}}.
\end{equation}
where~$\nu=1$ for free theory. $K(p,t)$ is chsoen to be analytic function such that its momentum dependence is only through $|p|$ and has the following properties.
\begin{enumerate}
\item
 $K(p,t)= 1$ for $\frac{p^2 e^{2t}}{\lm_0^2} <1$,
 \item 
 $K(p,t)$ decays raipdly as  $\frac{p^2 e^{2t}}{\lm_0^2}\rightarrow \infty$.
\end{enumerate}
We assume $K$ is function of dimensionless quantity $\frac{p\,e^t}{\lm_0}$, hence
\begin{align}\label{eq: K property}
\left( p^\mu \frac{\p}{\p p_\mu} -\frac{\p}{\p t}\right)K(p,t)=0.
\end{align}
We can derive the following properties of $G(p,t)$.
\begin{align}\label{eq: property p-sym G}
& |p|-\textit{dependence}:p_\mu \frac{\partial}{\partial p^\rho} \frac{1}{\dot{G}(p,t)}= p_\rho \frac{\partial}{\partial p^\mu} \frac{1}{\dot{G}(p,t)},\\
& \label{eq: property scale G}\textit{scaling relation}:\frac{\partial}{\partial t}\frac{1}{\dot{G}(p,t)}-p^\rho\frac{\partial}{\partial p^\rho}\frac{1}{\dot{G}(p,t)}= -\frac{2\nu}{\dot{G}(p,t)}.
\end{align}
This will be useful from time to time in this paper.

The flow of $S_t[\phi]$ along RG scale $t$ is given by the Exact RG equation. In terms of a quantity $\Psi = e^{-S_t[\phi]}$ this becomes
\be \label{eq: Polchinski full ERG}
\p_t \Psi[\phi(p),t] ={\cal G}_{RG}\Psi\equiv -\hf \int_p\dot G(p,t) (\frac{\dd }{\dd \phi(p)}(\frac{\dd}{ \dd \phi(-p)}+ \frac{2}{G(p,t)}\phi(p)) \Psi[\phi(p),t].
\ee
The RHS consists of the number operator or classical scaling dimension of the theory and (in perturbative picture) the disconnected and connected contribution in the low energy effective action. The overall multiplicative factor $\dot{ G}(p,t)$ acts as the low energy propagator. 
In this paper we will mostly use the Polchinski's version of ERG equation which describes the flow of the interaction term $S_{I,t}[\phi]$ in the Wilson action \eqref{eq: full Wilson action}. This equation rewritten for the quantity~$\psi[\phi]= e^{-S_{I,t}[\phi]}$ turns out to be a diffusion equation \cite{Sathiapalan:2017}.
\begin{eqnarray}\label{eq: Polchinski diffusion}
\frac{\p \psi[\phi] }{\p t}\, = \ha \int \frac{d^d p}{(2 \pi)^d}\, \dot{G}(p,\Lambda)\,\frac{\delta^2 \psi[\phi]}{\delta \phi(p)\delta \phi(-p)}.
\end{eqnarray}
The evolution operator for this equation can be written using  the \textit{functional} or \textit{path integral} formalism. The action functional corresponding to the  Hamiltonian in \eqref{eq: Polchinski diffusion} can be obtained as
\begin{align}\label{eq: Polchinski evolution int}
\mathcal{S}[\phi]= -\frac{1}{2} \int_{t_i}^{t_f} dt \int \frac{d^d p}{(2 \pi)^d}\, \frac{\dot{\phi}(p,t)\dot{\phi}(-p,t)}{\dot{G}(p,t)}.
\end{align}
We refer $\mathcal{S}[\phi]$ as the \textit{evolution action} in this paper.  This acts on the initial state defined by $\psi(\phi_i)$ to admit the final state $\psi(\phi_f)$. 
\begin{align}\label{eq: functional integral}
\psi(\phi_f) = \int dt \int_{\phi(t_i)=\phi_i,\phi(t_f)=\phi_f} \mathcal{D} \phi\, e^{\frac{1}{2} \,\int_{t_i}^{t_f} dt\, \int \frac{d^d p}{(2 \pi)^d}\, \frac{\dot{\phi}(p,t)\dot{\phi}(-p,t)}{\dot{G}(p,t)}}\,\psi(\phi_i).
\end{align}
where $\phi_i$ and $\phi_f$ denotes the fields $\phi(p,t)$ at initial and final scale $t_i$ and $t_f$ respectively. This can be termed as the holographic renormalization group form of the Wilson action. $\s[\phi]$ thus lives in $d+1$ dimensions and we can map this evolution action  in \eqref{eq: Polchinski evolution int} to the bulk $AdS$ action through the following redefinition of fields and choice of cutoff function \cite{Sathiapalan:2017}.
\begin{align}\label{eq: map to AdS}
&\textit{Scale to Poincare coordinate}: z=e^t,\\
& \notag \textit{Field transformation}:\phi(p,z)= f(p,z)\, y(p,z),~\text{with}~f^2(p,z)= -z^{-d} \,\dot{G}(p,z),\\
& \notag \textit{choice of cutoff}:z^{-d+1} \left( z^{-d+1} \frac{d}{d z} \right)^2 e^{-\ln f(p,z)}=z^{-d+1} \left(p^2+\frac{m^2}{z^2}\right) e^{-\ln f(p,z)}.
\end{align}
This transforms $\s[\phi]$ to the bulk $AdS$ action $\s [y]$ (in Poincare co-ordinates for $AdS$).
\begin{align}
\s[y]=\int dz \int \frac{d^d p}{(2\pi)^d} \left[ z^{-d+1}\, \frac{\partial y(p,z)}{\partial z}\frac{\partial y(-p,z)}{\partial z} + \left(p^2+ \frac{m^2}{z^2}\right) z^{-d+1}\, y(p,z) y(-p,z)\right].
\end{align}
This completes the brief review of how holographic RG can be recast in terms of boundary ERG. In next two sections we study the symmetry transformations of the bulk actions both in $\phi(p,t)$ and transformed field $y(p,z)$.

\subsection{Composite fields: Scale invariance with finite cutoff}
Symmetry under scale transformations can usually be written as ($\lm = \lo e^{-t}$ is RG scale and ) 
\be  \label{Sigma}
\Sigma \Psi\equiv \int _p \DDx{}{\phi(p)} \Sigma(p)\phi(p) \Psi= \int_p \DDx{}{\phi(p)} (p^\mu\ppx {}{p^\mu} - \Delta^p_\phi)\phi(p) \Psi=0.
\ee
we use 
\[\int_p \equiv \frac{d^d p}{(2\pi)^d}\] throughout the paper.
$\Psi = e^{-S_\lm[\phi]}$ and $\Delta^p_\phi$ is the scaling dimension of $\phi(p)$. For a free field $\Delta_\phi = \frac{D-2}{2}$ and $\Delta_\phi^p = \Delta_\phi-d$. Please note the difference between notations $\Sigma$ and $\Sigma(p)$.
The  Wilson action at finite cutoff $\lm$ is
\be 	
S_\lm[\phi]=\hf \int _p \phi(p) G^{-1}(p,\lm)\phi(-p) + S_{I,\lm}[\phi].
\ee
where $G(p,\lm) = \frac{K(p/\lm)}{p^{2\nu}}$ (see\eqref{eq: G functional form})and $\lm =\lo e^{-t}$. $K=1$ when $\lm=\infty$.  Even in the free case, it is clear that \eqref{Sigma} is satisfied only when $\lm =\infty$. On the other hand the free Wilson action is indeed a fixed point theory and should be scale invariant for any $\lm$. So what is the resolution?

The answer is that $\Sigma$ in \eqref{Sigma} has to also evolve along with the RG flow. Thus, schematically, we define an RG evolution operator $U$ so that
\[
\Psi(\lm)=e^{-S_\lm[\phi]} = U(\Lambda,\Lambda_0) e^{-S_{\lo} [\phi]}.
\]
The bare theory with $\lo=\infty$ is scale invariant. Thus
\[
\Sigma e^{-S_{\lo}}=0~;~\lo\to \infty,
\]
Then we can define, for finite values of $\lm$
(which requires $t\to \infty$ as $\lo\to \infty$)
\[
[\Sigma ]=U\Sigma U^{-1},
\]
So that at $\lm_0=0$
\[
[\Sigma ] U\Psi = U \Sigma \Psi (\lo \to \infty)=0.
\]
The ``composite operator'' $[\Sigma]$ is the symmetry operator (which now will be dependent on RG scale $\lm$), and can be  worked out using the following rules \cite{Igarashi}:
\[
[\phi(p)] = \frac{1}{K(p)}\phi(p)+ \frac{1-K}{K}\DDx{}{\phi(-p)}.
\]
$\Psi$ is understood to be affixed on the right and all operators act on $\Psi$.
Note that for a free theory this gives $[\phi]=\phi$ as expected. Similarly we can compute
\[
\left[\DDx{}{\phi(p)}\right]= K(p)\DDx{}{\phi(p)}.
\]
Thus
\[
[\Sigma]=\int _p K(p)\DDx{}{\phi(p)}( \Sigma (\frac{1}{K(p)}\phi(p)+ \frac{1-K}{K}\DDx{}{\phi(-p)})). 
\]
The condition for scale invariance of $S_\lm$ is thus \footnote{Here we have ignored for simplicity the anomalous dimension that needs to be introduced for an interacting fixed point equation. See \cite{Sonoda:2015} and Appendix \ref{FP} for the more general result.}
\begin{align}\label{eq: scale inv composite}
[\Sigma]e^{-S_\lm[\phi]}=0.
\end{align}
It can be shown (see appendix \ref{FP}) that this is exactly the ERG condition for a fixed point Wilson action \cite{Sonoda:2015} i.e.
\be \label{scaletransf}
({\cal G}_{RG} + \Sigma)\Psi= (\ppx{}t+\Sigma)\Psi= 0.
\ee 
Note that to obtain a fixed point one has to rescale to dimensionless quantities and this is done by 
$\Sigma$. $\Psi$ is a {\em solution} of the ERG equation with special values of the dimensionless parameters (that otherwise run). Here $\ppx{}t$ acting on a  solution of \eqref{eq: Polchinski full ERG} clearly is equivalent to acting by ${\cal G}_{RG}$\footnote{Note that $\ppx{}{t}$ acts on the parameters in the action and not on the field $\phi(p)$ which does not depend on $t$. An example of such a calculation is given in \cite{Dutta:2020}.}.

Equation \eqref{scaletransf} is the resolution to the problem posed in the Introduction. In order to see the scale invariance of fixed point action in the presence of a finite $\lm$  one has to scale $\lm$ along with $p$ so that $p/\lm$ is invariant. It is easy to see for instance that the free Wilson action
\[
S_0=\hf \int_p \phi(p) \frac{p^2}{K(p/\lm)}\phi(-p).
\] is scale invariant and obeys \eqref{scaletransf}: $(\ppx{}t+\Sigma)S_0=0$.  
Thus even though $\lm$ is not a variable or coordinate one has managed to introduce it into the scale transformation. What the mapping to AdS does is to promote $\lm$ to a coordinate so that this can be done without any such subterfuge.\footnote{In the bulk action $\ppx{}{t}$ acts on the bulk fields $\phi(p,t)$ which are functions of $t$.} 

Special conformal transformation in the presence of a finite cutoff are also modified in a similar fashion by the use of composite operators (see \cite{Rosten:2014}-\cite{Sonoda:conf2017}).

\section{Symmetry transformation: $\s[\phi]$}
\label{sec: symmetry phi}

As explained above the presence of a cutoff scale in the boundary Wilson action modifies the form of the generators of scale and conformal transformations. It changes operators of the form $\DDx{}{\phi}G_a \phi $
to composite operators $[\DDx{}{\phi}G_a \phi]$ where $G_a$ is the symmetry generator acting on $\phi$ - for eg dilatations, rotations, special conformal transformations - and form the conformal group $SO(1,d+1)$. These composite  operators introduce a dependence on the cutoff function.
 
In this section we would like to evaluate the symmetry transformations of the evolution action in \eqref{eq: Polchinski evolution int} i.e.
\begin{align}\label{eq: action phi(p,t)}
\mathcal{S}[\phi]= \frac{1}{2} \int_p ~\int d t~ \frac{\dot{\phi}(p,t)\dot{\phi}(-p,t)}{\dot{G}(p,t)}.
\end{align}
 We show that it also has, not unexpectedly, an $SO(1,d+1)$  symmetry  but with modified generators - the modification involves also the cutoff function.\footnote{ For the simplest case of scale transformation they have the same form as \eqref{scaletransf}. But in general they are different. The connection between these modified symmetry generators of the evolution action found in this paper and the modified symmetry generators of the boundary Wilson action is an interesting open question.}

In the next two subsections we explicitly find these modified transformations. In this paper, the total derivative terms in $p$ directions which arises due to variation of an action will be ignored. The boundary terms in $t$, however can modify the boundary states $\psi_{i,f}$ in the functional form of Polchinski's equation \eqref{eq: functional integral}.  We  do not attempt to analyse this in the present work - and the boundary terms are ignored. The RG scale $t$ is a coordinate in this action. Throughout the paper we will refer $t$-independent (or ``scale'' independent)  part of the transformation as \textit{type-I} and $t$-dependent as \textit{type-II}.

\subsection{Scale transformation}
\label{sec: scale functional}

Let us see how the scale transformation changes for a field theory with cut-off. Let's start with the usual scale-independent ansatz.

\begin{align}\label{eq: scale 12}
D^I_{\phi}(p)=p^\mu \frac{\p}{\p p_\mu} +(d-\Delta_\phi).
\end{align}
\textit{where~$\Delta_\phi= \frac{d}{2}-\nu$ is the conformal dimension of the field $\phi(x)$}. This field transformation changes the evolution action as
{\small
\begin{align}
\notag D^I_{\phi} \s[\phi] & =\frac{1}{2} \int_p \int~d t~ \frac{\p }{\p t}\lbrace D_{\phi}\phi(p,t)\rbrace\frac{\dot{\phi}(-p,t)}{\dot{G}(p,t)}+(p \rightarrow -p).
\end{align}
}
It is easy to conclude (ignoring the total derivative terms in momentum)
\begin{align}\label{eq: scale12 on phi action}
& D^{I}_{\phi} \s[\phi]=-\int dt \int_p \left( \frac{\p}{\p t} \frac{1}{\dot{G}(p,t)} \right) \dot{\phi}(p,t) \dot{\phi}(p,t).
\end{align}
We add the following scale dependent transformation to \eqref{eq: scale 12} .
\begin{align}\label{eq: scale 3}
D^{II}_\phi(t)= -\frac{\p}{\p t}.
\end{align}
whose action on $\s[\phi]$  can be written as
{\small
\begin{align}\label{eq: scale 3 on phi action}
D^{II}_{\phi} \s[\phi]= -\int dt \int_p \frac{\p}{\p t}\left[ \frac{1}{\dot{G}(p,t)}\dot{\phi}(p,t) \dot{\phi}(-p,t)\right]+\int dt \int_p\left( \frac{\p}{\p t} \frac{1}{\dot{G}(p,t)} \right) \dot{\phi}(p,t) \dot{\phi}(p,t).
\end{align}
}
The second term in \eqref{eq: scale 3 on phi action} cancels \eqref{eq: scale12 on phi action}. 

\paragraph{Noether Charge:}

The boundary term contributes to the Noether charge: Using standard methods, let
\be \label{eq: action variation scale}
\hat \dd S =\int \p_\mu \chi ^\mu (x,\hat \dd)
\ee
under the symmetry transformation $\hat \dd$.  Also let
\[
\dd S = \int_x \DDx{S}{\phi(x)}\dd \phi(x) + \DDx{S}{\p_\mu \phi(x)}\p_\mu \dd \phi(x)
\]
\be 
 = \int_x[\ppx {{\cal L}}{\phi}-\p_\mu\ppx {{\cal L}}{\p_\mu \phi(x)}]\dd \phi(x) + \int_x \p_\mu (\ppx {{\cal L}}{\p_{\mu} \phi(x)}\dd \phi(x)) =\int_x[\ppx {{\cal L}}{\phi(x)}-\p_\mu\ppx {{\cal L}}{\p_\mu \phi(x)}]\dd \phi(x)  + \int_x \p_\mu K^\mu (x,\dd).
\ee
under an arbitrary variation $\dd$ ($x$ is $d$-dim coordinate in Euclidean signature). So
\be \label{KNoether}
K^\mu (x, \dd) = \DDx{S}{\p_\mu \phi(x)} \dd \phi(x).
\ee
 Then for variation in \eqref{eq: action variation scale}
\be 
j^\mu(x,\hat \dd) =K^\mu(x,\hat \dd)-\chi^\mu(x,\hat \dd). 
\ee
is the Noether current that is conserved on using the EOM.

Note that the above formulae are in position space.
Using fourier transformation in boundary directions\footnote{We require $\DDx{\phi(x)}{\phi(y)}=\dd(x-y)$ and $\DDx{\phi(p)}{\phi(q)}=\dd(p-q)$. }
\[
\phi(x,t) = \int_p e^{ip.x}\phi(p,t),
\]
\[
\DDx{}{\phi(y,t)}= \int_q~ e^{iqy}\DDx{}{\phi(-q,t)}
\]
we obtain
\[
\int_x K^0(x,t)=\int_x \DDx{S}{\dot \phi(x,t)}\dd \phi(x,t)=\int_q~\DDx{S}{\dot\phi(q,t)}\dd \phi(q,t).
\]

We just have a time direction. One can evaluate $\chi^0$ (which is given in \eqref{eq: scale 3 on phi action}) and $K^0$, and we find
that since time here is RG time, the Noether charge is the Hamiltonian when we rotate $t$ to Minkowski space (see \eqref{eq: Hamiltonian along RG time d dim}):

\be 
  \int _x j^0(x,t) = \hf \int _p \frac{\dot \phi(p,t) \dot \phi(-p,t)}{\dot G(p,t)}.
\ee  

From \eqref{eq: scale 12} and \eqref{eq: scale 3} we write the modified scale transformation for evolution action $\s[\phi]$.
\begin{align}\label{eq: scale123 phi action}
\tcbhighmath{
D_{\phi}(p,t)= p^\mu \frac{\p}{\p p_\mu} + (d-\Delta_\phi) - \frac{\p}{\p t}.
}
\end{align} 
Note this modification by a scale dependent transformation is consistent with \eqref{scaletransf}.

\subsection{Conformal transformation}
\label{sec: conformal functional}

The action of the conformal transformation on the action $\s[\phi]$ is given by
{\small
\begin{align}
\notag C_{\mu,\phi} \s[\phi]
& = \frac{1}{2} \int_p ~\int d t~ \frac{\p }{\p t}\lbrace C_{\mu,\phi}\phi(p,t)\rbrace\frac{\dot{\phi}(-p,t)}{\dot{G}(p,t)}-(p \rightarrow -p).
\end{align}
}

The details of the finding the transformation is provided in Appendix \ref{app: conformal transformation phi details}, we state the main results below.

\paragraph*{Scale-independent conformal transformation, $C^I_{\mu,\phi}$} This consists of three transformations 
\begin{align}\label{eq: exp conformal usual}
C^I_{\mu,\phi}(p) = & 2(d-\Delta_\phi) \frac{\partial}{\partial p^\mu}+ 2 p^\rho\frac{\partial^2}{\partial p^\rho \partial p^\mu}-p_\mu \frac{\partial^2}{\partial p^\rho \partial p_\rho}\\
\equiv & \notag C^1_{\mu,\phi}+C^2_{\mu,\phi}+C^3_{\mu,\phi}.
\end{align}

These action of these transformations on $\s[\phi]$ can be evaluated using the property of cutoff function $G(p,t)$ stated in \eqref{eq: property p-sym G}.
\begin{align}\label{eq: conformal123 general}
& (C_{\mu,\phi}^1+ C_{\mu,\phi}^2+ C_{\mu,\phi}^3) \s[\phi]=\int dt \int_p \left[\frac{\partial}{\partial p^\mu}\dot{\phi}(p,t)\right] \dot{\phi}(-p,t) \left( \frac{d-2 \Delta_\phi}{\dot{G}(p,t)}- p^\rho\frac{\partial}{\partial p^\rho} \frac{1}{\dot{G}(p,t)}\right).
\end{align}

\paragraph*{Scale-dependent conformal transformation, $C_{\mu,\phi}^{II}$}

We propose a further transformation
\begin{align}\label{eq: exp conformal scale dep}
C_{\mu,\phi}^4=-2\frac{\partial^2}{\partial t \partial p^\mu}.
\end{align} 
After some cancellations and using the scaling property of $G(p,t)$ we obtain
\begin{align}\label{eq: conformal1234 general simplified}
(C_{\mu,\phi}^I+ C_{\mu,\phi}^4)\s =\int dt\int_p \left[\frac{\partial}{\partial t}\dot{\phi}(p,t)\right]\dot{\phi}(-p,t)\frac{\partial}{\partial p^\mu}\frac{1}{\dot{G}(p,t)}.
\end{align}
We need to add another transformation which is also dependent on cut-off function to cancel \eqref{eq: conformal1234 general simplified}.
\begin{align}\label{eq: exp conformal cutoff dep}
C_{\mu,\phi}^5= -\dot{G}(p,t) \frac{\partial}{\partial p^\mu} \frac{1}{\dot{G}(p,t)} \frac{\partial}{\partial t}.
\end{align}
In this process too we get the following boundary term (see \eqref{eq: boundary conformal phi})
\begin{align} \label{confbdry}
\delta_{bdry} \s[\phi]=-\int_p ~\int_{t_i}^{t_f}d t \frac{\partial}{\partial t}\left[\frac{\frac{\partial}{\partial p^\mu}\dot{\phi}(p,t)\dot{\phi}(-p,t)}{\dot{G}(p,z)}\right].
\end{align} 

\paragraph{Noether Charge:}

We can calculate the Noether charge for conformal transformations.  We focus on new terms involving $t$, i.e. the scale dependent part of $C_\mu$ which we called $C_\mu^{II}$.
Thus we consider
\[
\hat \dd \phi =\bar\eps^\mu [-2 \ppx{}{q^\mu} \dot \phi(q,t) - \dot G(q,t) \ppx{}{q^\mu} (\frac{1}{\dot G(q,t)} )\dot \phi(q,t)].
\]
The contribution to the Noether current $\chi_\mu^\nu$ is read off from \eqref{confbdry}
\be  \label{chi}
\int_x \chi_\mu^0(x,t)=-\int_q (\frac{\dot{\phi}(-q,t)}{\dot{G}(q,t)})\frac{\partial}{\partial q^\mu}\dot{\phi}(q,t).
\ee
Note that the contribution is only to the charge i.e. the time component of the current.

Simlarly we calculate using \eqref{KNoether} the contribution to the charge
$K_\mu^0$ due to $\hat \delta \phi$. It can be calculated as:
\[
\int _x K_\mu^0(x,t)= \int_q (\frac{\dot \phi(-q,t)}{\dot G(q,t)})[-2 \ppx{}{q^\mu} \dot \phi(q,t) - \dot G(q,t) \ppx{}{q^\mu} (\frac{1}{\dot G(q,t)} )\dot \phi(q,t)],
\]
 Combinining with \eqref{chi} gives the Noether charge
\[
\int _x j_\mu^0(x,t)= \int _q \dot \phi (-q,t) \ppx{}{q^\mu} (\frac{\dot \phi(q,t)}{\dot G(q,t)}).
\]
This is the conformal charge in Hamiltonian formalism, which has been also calculated independently in Appendix \ref{app: Hamiltonian details} (the last term in \eqref{eq: conformal typeI typeII}).

Hence from \eqref{eq: exp conformal usual}, \eqref{eq: exp conformal scale dep} and \eqref{eq: exp conformal cutoff dep} we write down the modified conformal transformations of the bulk field $\phi(p,t)$ which keeps the evolution action invariant.
{\small
\begin{equation}\label{eq: phi conformal 12345}
\tcbhighmath{
C_{\mu,\phi}(p,t) = 2(d-\Delta_\phi) \frac{\partial}{\partial p^\mu}+ 2 p^\rho\frac{\partial^2}{\partial p^\rho \partial p^\mu}-p_\mu \frac{\partial^2}{\partial p^\rho \partial p_\rho}-2  \frac{\partial^2}{\partial t \partial p^\mu}-\dot{G}(p,t) \frac{\partial}{\partial p^\mu} \frac{1}{\dot{G}(p,t)}\frac{\partial}{\partial t}.
}
\end{equation}
}

\section{Symmetry transformation: $\s[y]$}
\label{sec: symmetry y}

\paragraph*{}Next we proceed to find how the symmetry transformations changes for the transformed bulk field $y(p,z)= \frac{\phi(p,z)}{f(p,z)}$ (recall $z=e^t$). We can obtain this either directly finding the transformation which keeps the evolution action in $y(p,z)$ invariant or using the field transformation $\delta \phi(p,z)= f(p,z) \delta y(p,z)$. The first one is described in this section. The latter one which serves as a consistency check, is given in Appendix \ref{app: consistency}. We only state the indicative steps of the first method in this section(see appendix~\ref{app: conformal transformation y details} for details).

\paragraph*{}The evolution action in \eqref{eq: action phi(p,t)} after transforming to field $y(p,z)$ becomes
{\footnotesize
\begin{align}\label{eq: bulk y action}
\s[y]= \notag & \int dz \int \frac{d^d p}{(2\pi)^d} \left[ z^{-d+1} \left( \frac{\partial y(p,z)}{\partial z}\right)^2 -\frac{\partial}{\partial z} \lbrace z^{-d+1} \frac{d \ln f(p,z)}{d z}\rbrace y(p,z)^2 + z^{-d+1} \left( \frac{d \ln f(p,z)}{d z}\right)^2 y(p,z)^2\right]\\
+ & \int dz \int \frac{d^d p}{(2\pi)^d} \frac{d}{d z} \left[ z^{-d+1}y(p,z)^2 \frac{d \ln f(p,z)}{d z} \right].
\end{align}
}
We will ignore the boundary term and its effect on the conformal symmetry of the boundary CFT in this paper. \footnote{This boundary term has been discussed in \cite{Sathiapalan:2017,Dharanipragada:2023} for special choices of $f$. For more general $f$ it requires further analysis. }  Before proceeding we define a quantity
\begin{align}\label{eq: def C(p,z)}
C(p,z) z^{-d+1}= -\frac{\partial}{\partial z} \lbrace z^{-d+1} \frac{d \ln f}{d z}\rbrace + z^{-d+1} \left( \frac{d \ln f}{d z}\right)^2.
\end{align}
In terms of $C(p,z)$, the action reduces to the following.
\begin{align}\label{eq: bulk action y in C}
\s[y]=\int dz \int \frac{d^d p}{(2\pi)^d} \left[ z^{-d+1} \left( \frac{\partial y(p,z)}{\partial z}\right)^2 +  C(p,z) z^{-d+1} y(p,z)^2\right].
\end{align}
$C(p,z)$ being a function of $f(p,z)$, satisfy the similar properties as $G(p,z)$.
\begin{align}\label{eq: property p-sym C}
& |p|\,\textit{dependence}:p_\mu \frac{\partial C(p,z)}{\partial p^\rho}= p_\rho \frac{\partial C(p,z)}{\partial p^\mu},\\
& \label{eq: property scale C}\textit{scaling relation}:p^\rho\frac{\partial C(p,z)}{\partial p^\rho}-z\frac{\partial C(p,z)}{\partial z}= 2 C(p,z).
\end{align}
The scaling relation can be obtained from the fact that $C(p,z)$ is the coefficient of $y(p,z)y(-p,z)$ in the action. These properties will be useful in our analysis. We also note that the bulk action \eqref{eq: bulk y action} consists of two kinds of terms, 
\begin{enumerate}

\item
\textit{Type-A}: kinetic term which is proportional to  $\left( \frac{\partial y}{\partial z}\right)^2$ 

\[ \s_A[y]= \int dz \int \frac{d^d p}{(2\pi)^d}  z^{-d+1}\frac{\partial y(p,z)}{\partial z}\frac{\partial y(-p,z)}{\partial z},\]

\item
\textit{Type-B}: mass like term which is proportional to $y^2$,
\[\s_B[y]=\int dz \int \frac{d^d p}{(2\pi)^d}  C(p,z) z^{-d+1} y(p,z)y(-p,z).\]
\end{enumerate}
\subsection{Scale transformation}
We start with the ansatz 
\begin{align}
D^I_{y}+D^{II}_{y} =  p^\mu \frac{\p}{\p p_\mu} - z\frac{\p}{\p z}.
\end{align}
The \textit{type-A} term does not have any cutoff function, hence the effect of the field transformations can be computed easily.
\begin{align}\label{eq: scale SA}
(D^I_{y}+D^{II}_{y}) \s_A= \int dz \int_p \frac{\p}{\p z} \left[ z^{-d+1} z \frac{\p y(p,z)}{\p z}\frac{\p y(-p,z)}{\p z} \right]-2 d~ \s_A.
\end{align}
The corresponding action on the \textit{type-B} term can be found from the following expressions
{\small
\begin{align}\label{eq: scale SB}
p^\mu \frac{\p}{\p p^\mu} \s_B[y]= & -(d) \int dz \int_p ~C(p,z) z^{-d+1} y(p,z)y(-p,z)\\
& \notag -\int dz \int_p~ p^\mu\frac{\p}{\p p^\mu}C(p,z) z^{-d+1} y(p,z)y(-p,z), \\
\notag -z \frac{\p}{\p z}\s_B[y]= & -\int dz \int_p \frac{\p}{\p z} \left[ C(p,z) z^{-d+2} y(p,z)y(-p,z) \right]\\
\notag &+\int dz \int_p~z \frac{\p}{\p z}C(p,z) z^{-d+1} y(p,z)y(-p,z)\\
\notag &+(-d+2) \int dz \int_p~ C(p,z) z^{-d+1}  y(p,z)y(-p,z).
\end{align}
}
The boundary terms contributes to the Noether charge as explained in the previous section. Considering \eqref{eq: scale SA} and  using the scaling property of $C(p,z)$ \eqref{eq: property scale C} in \eqref{eq: scale SB} we can promptly write down the modified scale transformation for $\s[y]$.

\begin{align}\label{eq: y scale123}
\tcbhighmath{
D_{y}(p,z)= p^\mu \frac{\p}{\p p_\mu} + (d-\Delta_y) - z\frac{\p}{\p z}.
}
\end{align}
where the conformal dimension of field $y(x,t)$ is $\Delta_y=0$. This is the scale transformation in $AdS$ space (appendix \ref{app: AdS isometry}). We have checked consistency of $D_y$ with $D_\phi$ in the beginning of the appendix \ref{app: consistency}.

\subsection{Conformal transformation}
We start with the following ansatz
\begin{align}\label{eq: y conformal1234 general}
\nonumber & C_{\mu,y}^1+ C_{\mu,y}^2+ C_{\mu,y}^3+  C_{\mu,y}^4\\
& = 2(d-\Delta_y) \frac{\partial}{\partial p^\mu}+ 2 p^\rho\frac{\partial^2}{\partial p^\rho \partial p^\mu}-p_\mu \frac{\partial^2}{\partial p^\rho \partial p_\rho}-2 z \frac{\partial^2}{\partial z \partial p^\mu}.
\end{align}
The procedure of calculation follows similar line as the evolution action in $\phi$. We provide details in Appendix~\eqref{app: conformal transformation y details}. The \textit{type-A} term can be shown to be invariant under \eqref{eq: y conformal1234 general}. The action on \textit{type-B} term can be summed up as
\begin{align}\label{eq: y conformal1234 type-II main}
  (C_{\mu,y}^1+ C_{\mu,y}^2+C_{\mu,y}^3+C_{\mu,y}^4)\s_B & = 2 \int dz \int_p  \frac{\partial y(p,z)}{\partial z}y(-p,z) \frac{\partial C(p,z)}{\partial p^\mu}z^{-d+1}\\
  & \notag -\int dz \int_p \frac{\p}{\p z} \left[ 2 z^{-d+1} z \frac{\p^2}{\p z \p p^\mu} y(p,z) \frac{\p}{\p z} y(-p,z) \right].
\end{align}
We have used the properties of $C(p,z)$ to arrive at the above expression. We further need to add a cutoff dependent transformation
\begin{align}\label{eq: y conformal5 type-II main} 
    C_{\mu,y}^5=\left[ \int_z z \frac{\partial}{\partial p^\mu} C(p,z)\right].
\end{align}
to cancel the non-boundary term in \eqref{eq: y conformal1234 type-II main}. Hence we write down the modified conformal transformation under which the bulk action $\s[y]$ \eqref{eq: bulk y action} remains invariant.
{\small
\begin{align}\label{eq: y conformal12345}
\tcbhighmath{
 C_{\mu,y}(p,z) = 2(d-\Delta_y) \frac{\partial}{\partial p^\mu}+ 2 p^\rho\frac{\partial^2}{\partial p^\rho \partial p^\mu}-p_\mu \frac{\partial^2}{\partial p^\rho \partial p_\rho}-2 z \frac{\partial^2}{\partial z \partial p^\mu}+\left[ \int_z z \frac{\partial}{\partial p^\mu} C(p,z)\right].
 }
\end{align}
}
$C(p,z)$ is defined in \eqref{eq: def C(p,z)}. As mentioned before we have checked the consistency of $C_{\mu,y}$ with $C_{\mu,\phi}$ in Appendix \ref{app: consistency}.

\section{Hamiltonian formalism}
\label{sec: conformal Hamiltonian}
We consider the Hamiltonian corresponding to the evolution operator of Polchinski's equation and find its symmetry transformation. The symmetry group should comprise of the modified conformal transformations. We consider the evolution operator in $d$-dim momentum space with the RG scale $z=e^t$. The Euclidean action is given by,
\begin{align}
S= -\ha \int \frac{d^d p}{(2\pi)^d}\int dz   \frac{1}{G'(p,z)} \p_z \phi(p,z) \p_z \phi(-p,z),
\end{align}
where \textit{prime} denotes differentiation with $z$. Note the minus sign in front of the action is because $G'<0$. In order to use the Hamiltonian formalism we rotate to the Minkowski space as in the appendix \ref{Full}.

So we let $G=-G_E$ (so that $G_E'>0$) and $G_E=iG_M$. We will also define $z_E=iz_M$ and define
\[
 G'_E \equiv \ppx{G_E}{z_E}=\ppx{G_M}{z_M} \equiv  G'_M.
\]
Then
\begin{align*}
S_E = & \ha \int \frac{d^d p}{(2\pi)^d}\int dG_E  ~ \p_{G_E}\phi(p,G_E) \p_{G_E} \phi(-p,G_E)\\
= &\ha \int \frac{d^d p}{(2\pi)^d}\int dz_E~\frac{1}{ G'_E} \p_{z_E}\phi(p,z_E) \p_{z_E} \phi(-p,z_E),
\end{align*}
with the evolution operator
\[
U=\int \CD \phi~ e^{-S_E}.
\]
After analytic continuation this becomes
\[
U=\int \CD \phi~e^{iS_M}.
\]
with
\[
S_M= \ha \int \frac{d^d p}{(2\pi)^d}\int dG_M~ \p_{G_M}\phi(p,z_M) \p_{G_M} \phi(-p,z_M)
\]
\[
= \int dz_M\underbrace{\hf\int \frac{d^d p}{(2\pi)^d}~\frac{1}{ G'_M} \p_{z_M}\phi(p,z_M) \p_{z_M} \phi(-p,z_M)}_{\cal L}.
\]
The momentum conjugate to the field is 
\begin{align}\label{eq: field momentum along RG scale}
\Pi(p,z_M)=\frac{1}{G_M'(p,z_M)} \p_{z_M} \phi(p,z_M),
\end{align}
 and the cannonical commutation relation with field is
\begin{align}
\left[ \phi(p,z_M),\Pi(q,z_M)\right]= i (2\pi)^d \delta^d(p+q).
\end{align}
The Hamiltonian for $z_M$ evolution can be obtained as
\begin{align}\label{eq: Hamiltonian along RG time d dim}
H(z_M)=\ha \int \frac{d^d p}{(2\pi)^d}  G_M'(p,z_M) \Pi(p,z_M)\Pi(-p,z_M).
\end{align}
For brevity henceforth we use the notation~$\int_p \equiv \int \frac{d^d p}{(2\pi)^d} $ and $z_M \equiv z$. Note that this action is rotation invariant but not Lorentz invariant. Hence the stress tensor components are not symmetric.
\begin{align}
\Theta ^\alpha_{~\beta}= \ppx{\cal L}{\phi_{,\alpha}}\phi_{,\beta}-\dd^\alpha_{~\beta}\cal{L},
\end{align}
where $\alpha= (0,1,2,\cdots, d)=(0,\mu)$. We list the stress energy components which will be required in our computation.
{\small
\begin{align}\label{eq: stress tensor components w cutoff}
& \Theta^0_{~0}(k,z)= \ha \int_q \Pi(k+q,z)\Pi(-q,z)G_M'(q,z)=\ha \int_q \Pi(k+q,z)\Pi(-q,z)G_M'(q,z),\\
\notag & \Theta^0_{~\mu}(l,z)= -i \int_p  \Pi(l+p,z)p_\mu\phi(-p,z).
\end{align}
}
The conformal charge can be computed as follows (see appendix \ref{app: Hamiltonian details}).
{\small
\begin{align}\label{eq: conformal typeI typeII}
\notag \bar{\epsilon}^\mu K^0_\mu\equiv &\bar{\epsilon}^\mu C_\mu\\ 
\notag =& \bar{\epsilon}^\mu \int_k  \delta(k) \bigg[ 2 \Delta_\phi \int_p \frac{\p \phi(p)}{\p p^\mu} \Pi(k-p)+\left( 2 \frac{\p^2}{\p k_\rho \p k^\mu}-\delta_{\rho\mu}\frac{\p^2}{\p k^\sigma \p k_\sigma}  \right)\Theta^{0}_{~\rho}(k,z)+ 2z \frac{\p}{\p k^\mu} \Theta^{0}_{~0}(k,z)\bigg]\\
=&\bar{\epsilon}^\mu( K^0_\mu|_I+K^0_\mu|_{II}) \equiv \bar{\epsilon}^\mu( C_\mu^I+C_\mu^{II}).
\end{align}
}
As in the previous section we have decomposed the conformal charge into a scale independent part $ K^0_\mu|_I \equiv C_\mu^I$ denoted by \textit{type-I} and a scale dependent part $K^0_j|_{II} \equiv C_\mu^{II}$ as \textit{type-II}. We also write down the translation generator
\begin{align}\label{eq: momentum generator}
\epsilon^\nu T_\nu= \epsilon^\nu \int_l \delta(l) \Theta^0_{~\nu}(l,z)= \epsilon^\nu \int_{l} \delta(l) \int_q \Pi(l+q,z)(-i q_\nu) \phi(-q,z)
\end{align}
To check the consistency of our symmetry transformations in $\phi$, we will evaluate the following commutator 
\begin{align}\label{eq: conformal momentum commutator}
& \notag \bar{\epsilon}^\mu \epsilon^\nu [C_\mu,T_\nu]\\
& =\bar{\epsilon}^\mu \epsilon^\nu \left \lbrace [C_\mu^I,T_\nu]+[C_\mu^{II},T_\nu] \right \rbrace,
\end{align}
and also show that the conformal charge in \eqref{eq: conformal typeI typeII} self commutes. The details are given in appendix \ref{app: Hamiltonian details}, we state the results below.

\paragraph*{Commutator of \textit{ type-I} term, $C_\mu^I$ and momentum $T_\nu$}
Using expression of stress tensor components and the conformal charge density and the commutator \eqref{eq: commutator 0i 0j}


\begin{equation}\label{eq: typeI momentum commutator text}
\boxed{
[C_\mu^I,T_\nu]= 2( -M_{\mu \nu}+\delta_{\mu \nu} D )
}
\end{equation}

where $M_{\mu\nu}$ and $D$ are the charges corresponding to rotation operator $M_{\mu \nu}(p)= -i\left( p_\mu\frac{\p}{\p p^\nu}- p_\nu\frac{\p}{\p p^\mu} \right)$ and dilatation operator $D(p)=-i\left(p^\sigma\frac{\p}{\p p^\sigma}+(d-\Delta_\phi)\right)$ along $x$.

\paragraph*{Commutator of\textit{ type-II} term, $C_\mu^{II}$ and momentum $T_\nu$} Using \eqref{eq: commutator 00 0j} we can write
\begin{align}
& \notag \bar{\epsilon}^\mu \epsilon^\nu [C_\mu^{II},T_\nu]=\bar{\epsilon}^\mu \epsilon^\nu z\int_{k} \delta(k)\frac{\p}{\p k^\mu} \int_q q_\nu \left(G_M'(q)+G_M'(k+q) \right)\Pi(k+q)\Pi(-q).
\end{align}
We note an useful identity in \eqref{eq: typeII momentum commutator identity}.
\begin{align}\label{eq: typeII momentum commutator identity}
\int_q \frac{\p \Pi(q,z)}{\p q^\mu} q_\nu G_M'(q) \Pi(-q)= -\ha \int_q \Pi(q) \delta_{\mu \nu} G_M'(q) \Pi(-q)-\ha \int_q \Pi(q) q_\nu \frac{\p G_M'(q)}{\p q^\mu}\Pi(-q).
\end{align}
to obtain
\begin{equation}\label{eq: typeII momentum commutator text}
\boxed{
[C_\mu^{II},T_\nu]=-2 \delta_{\mu \nu} z H(z)
}
\end{equation}
where $H(z)$ is defined in \eqref{eq: Hamiltonian along RG time d dim}.  Combining \eqref{eq: typeI momentum commutator text} and \eqref{eq: typeII momentum commutator text} and replacing back $z \rightarrow z_M $ we conclude
\begin{align}\label{eq: conformal momentum Hamiltonian final}
\tcbhighmath{
[C_\mu,T_\nu]= 2 (-M_{\mu\nu}+\delta_{\mu\nu} D- \delta_{\mu\nu} z_M H(z_M))
}
\end{align}
where $z_M H(z_M)$ generates dilatation along the $z_M$ i.e. $-i z_M \p_{z_M}$. From \eqref{eq: group structure general} we note the transformations follows $SO(1,d+1)$. We can show the full conformal charge $C_\mu$ self commutes (see the end of appendix \ref{app: Hamiltonian details} for details).
\begin{align}
\tcbhighmath{
[C_\mu,C_\nu]=0
}
\end{align}

\subsubsection*{\underline{Reproducing conformal transformation \eqref{eq: phi conformal 12345}}}

From \eqref{eq: Hamiltonian along RG time d dim} we can find the action of conformal transformation on field $\phi(p,z)$ by evaluating the commutator~$[C_\mu,\phi(p,z)]$. 
The part that generates the conformal transformation on $z_M$ is what we are interested in (we have dropped the subscript $M$ on $z$ below):
{\small
\begin{align}
\delta_z \phi(p,z)=\,& \left[\ha \int_q\int_k \delta(k)\,\left(2 \epsilon^\rho z \right)\,\left( \frac{\p }{\p k^\rho}\Pi(q+k,z) \right)G^\prime(q,z) \Pi(-q,z),\phi(p,z)\right]\\
=\,&\notag -i (2\pi)^d\,\left( \epsilon^\rho z \right)  \int_q \int_k \delta(k)\,\,\, \frac{\p}{\p k^\rho} \left\lbrace \delta(p+q+k) G^\prime(q,z) \Pi(-q,z)+ \Pi(q+k,z)G^\prime(q,z)\delta(p-q)\right\rbrace\\
=\,& \notag -i \epsilon^\rho z \left\lbrace \frac{\p}{\p p^\rho}\left(G^\prime(p,z) \Pi(p,z) \right)+ G^\prime(p,z)\frac{\p}{\p p^\rho} \Pi(p,z) \right \rbrace\\
=\notag \,&- 2i \epsilon^\rho z \frac{\p}{\p p^\rho}\frac{\p\phi(p,z)}{\p z}-i \epsilon^\rho z G^\prime(p,z) \frac{\p}{\p p^\rho}\frac{1}{G^\prime(p,z)}\frac{\p\phi(p,z)}{\p z}. 
\end{align}
}

\section{Lie algebra: SO(1,d+1)}
\label{sec: conf algebra}
 In this section, we have checked in detail that the transformations in $\phi$ form the Euclidean conformal group $SO(1,d+1)$. This is somewhat expected as the bulk action in $y$ is mapped to the Euclidean $AdS$ action. The only input we used is properties of cutoff function $f(p,z)$. The symmetry transformations in momentum space  with RG scale $z=e^t$ is writen here for convenience.

{\small
\begin{empheq}[box=\fbox]{align}\label{eq: conformal group elements generalized}
    &\nonumber \textit{Translation}, T_{\mu,\phi}(p): p_\mu,\\
    & \nonumber \textit{Rotation},M_{\mu\nu,\phi}(p) : p_\mu \frac{\partial}{\partial p^\nu}-p_\nu \frac{\partial}{\partial p^\mu},\\
    & \nonumber \textit{Dilatation},D_{\phi}(p): p.\frac{\partial}{\partial p}+ (d-\Delta_\phi)- z \frac{\partial}{\partial z},\\
    & \nonumber \textit{Special conformal: scale independent}: 2(d-\Delta) \frac{\partial}{\partial p^\mu}+ 2 p^\rho\frac{\partial^2}{\partial p^\rho \partial p^\mu}-p_\mu \frac{\partial^2}{\partial p^\rho \partial p_\rho}\equiv C_{\mu,\phi}^I(p),\\
    & \nonumber\textit{Special conformal: scale dependent}: -2 z\frac{\partial^2}{\partial z \partial p^\mu}+2 \left( \frac{\partial}{\partial p^\mu}\ln f\right)z\frac{\partial}{\partial z}\equiv C_{\mu,\phi}^{II}(p,z).
\end{empheq}
}
The computations of the non-trivial commutators have been shown in appendix~\ref{app: conf algebra details}. In the previous section we analytically rotate $z$ to Minskowski signature. We note this does not change the group structure. This is consistent with the fact that the analytic continuation of $z$ is related to the map from $EAdS$ to $dS$ space which also has $SO(1,d+1)$ symmetry.

\section{Map to bulk $AdS$ isometry}
\label{sec: map AdS}

We have computed the modified conformal transformations of the evolution action in $y(p,t)$ for a general cutoff function $f(p,z)$. Now we use the particular form of $f(p,z)$ which maps this action to $AdS$ action. The cutoff parameter or RG scale $z=e^t$ becomes the radial coordinate in bulk $AdS$. We have shown below  that this specific $f(p,z)$ maps the modified conformal transformation to $AdS$ isometry (conformal transformations in $AdS$ isometry is reviewed in appendix \ref{app: AdS isometry}).

Before proceeding we note that the functional form of $f(p,z)$ can be taken as
\begin{align}\label{eq: f func form}
    \frac{1}{f(p,z)}= & p^\nu z^{d/2} h(pz)
\end{align}
which as usual have the following properties
\begin{align}\label{eq: property p-sym f}
& |p|\,\textit{dependence}:p_\mu \frac{\partial}{\partial p^\rho} f(p,z)= p_\rho \frac{\partial}{\partial p^\mu} f(p,z),\\
& \label{eq: property scale f}\textit{scaling relation}:\frac{p^\rho}{f(p,z)} \frac{\partial f(p,z)}{\partial p^\rho} - \frac{z}{f(p,z)} \frac{\partial f(p,z)}{\partial z}= \frac{d}{2}-\nu.
\end{align} 

We note the following points:

\begin{enumerate} 

\item The modified dilatation in \eqref{eq: y scale123} is independent of $f(p,z)$ and same as $AdS$ scale transformation (see \eqref{Dil}). 

\item The conformal transformation of $\phi(p,z)$ action in terms of mapping function $f(p,z)$ is
\begin{align}\label{eq: phi conformal12345 map bulk}
\nonumber C_{\mu,\phi} & = C_\mu^1+ C_\mu^2+ C_\mu^3+C_\mu^4+C_\mu^5\\
\nonumber & = 2(d-\Delta_\phi) \frac{\partial}{\partial p^\mu}+ 2 p^\rho\frac{\partial^2}{\partial p^\rho \partial p^\mu}-p_\mu \frac{\partial^2}{\partial p^\rho \partial p_\rho}\\
&-2 z\frac{\partial^2}{\partial z \partial p^\mu}+2 \left( \frac{\partial}{\partial p^\mu}\ln f(p,z)\right)z\frac{\partial}{\partial z}.
\end{align}

\item 
The bulk action in \eqref{eq: bulk action y in C} maps to the bulk $AdS$ action by substituting $C(p,z)=\left(p^2+ \frac{m^2}{z^2}\right)$. 

\begin{align}
    S=\int dz \int \frac{d^d p}{(2\pi)^d} \left[ z^{-d+1} \left( \frac{\partial y}{\partial z}\right)^2 + \left(p^2+ \frac{m^2}{z^2}\right) z^{-d+1} y^2\right].
\end{align}

\item

Map to $AdS$ ensures
\begin{align}\label{eq: f eq map bulk}
    (pz)h^\prime(pz)+(pz)^2 h^{\prime\prime}(pz)=(p^2z^2+m^2)h(p,z),\, m^2=\nu(\nu-2)
\end{align}
is satisfied.

\item
Hence, the cutoff function dependent part in the modified special conformal transformation of $y$ action, $C_\mu^5$ \eqref{eq: y conformal5 type-II main} becomes
\begin{align}
    \nonumber C_\mu^5=& \int_z z \frac{\partial}{\partial p^\mu}\left(p^2+ \frac{m^2}{z^2}\right)\\
    &= z^2 p_\mu.
\end{align}
which reproduces $AdS$ special conformal transformation \eqref{cftr}.
\begin{align}\label{eq: y conformal12345 map bulk}
\tcbhighmath{
    C_\mu^{AdS}= 2(d-\Delta_y) \frac{\partial}{\partial p^\mu}+ 2 p^\rho\frac{\partial^2}{\partial p^\rho \partial p^\mu}-p_\mu \frac{\partial^2}{\partial p^\rho \partial p_\rho}-2 z \frac{\partial^2}{\partial z \partial p^\mu}+z^2 p_\mu.
    }
\end{align}
In this case also we have explicitly checked the consistency of the conformal transformation for $\phi(p,z)$ \eqref{eq: phi conformal12345 map bulk} and $y(p,z)$ \eqref{eq: y conformal12345 map bulk} at the end of appendix \ref{app: consistency}.

\end{enumerate}

\section{Symmetry transformation: ERG for full action}
\label{sec: full ERG}

In this section we will show that the evolution operator of the ERG equation for the full Wilson action is invariant under a similar conformal transformation as the Polchinski's version. The reason is as follows. The ERG equation \eqref{eq: Polchinski full ERG} for the full action \eqref{eq: full Wilson action} can rewritten for~ $\Psi_t[\phi]=e^{-S_t[\phi]}$ as
\begin{align}
    \partial_t \Psi_t[\phi]= -\ha \dot{G}(p,t) \frac{\p}{\p \phi(p,t)}\left(\frac{\p}{\p \phi(p,t)}+ \frac{2 \phi(p,t)}{G(p,t)}   \right)\Psi_t[\phi].
\end{align}
whose evolution operator can be evaluated as (see appendix~\ref{Full})
\begin{align}\label{eq: evolution full ERG}
    U(t_f,t_i)= \int \mathcal{D}\phi(p,t) e^{\int_{t_i}^{t_f} dt \frac{1}{2 \dot{G}(p,t)}\left( \frac{\p \phi(p,t)}{\p t}- \frac{\dot{G}(p,t)}{G(p,t)} \phi(p,t)\right)^2}.
\end{align}
This evolution operator can be recast in the same form as Polchinski's equation.
\begin{align}\label{source}
    U(t_f,t_i)= \int \mathcal{D}\phi e^{\int_{t_i}^{t_f} dt \frac{\dot{\chi}(p,t)^2}{2\dot{F(p,t)}}}.
\end{align}
The field $\chi(p,t)$ is related to $\phi(p,t)$ by \footnote{In the boundary at leading order $\phi(p) = G(p) J(p)$ where $J$ is a source, so $\chi(p,t) $ is like a bulk field dual to a source in the boundary. The evolution operator for the Wilson ERG for a generating functional is essentially \eqref{source}.}.   
\begin{align}
\chi(p,t)= F(p,t) \phi(p,t),
\end{align}
whose cutoff function is given by
\begin{align}
F(p,t)=\frac{1}{G(p,t)}.
\end{align}
Hence we expect the evolution operator for the full Wilson action will also form $SO(1,d+1)$ group.

The modified scale transformation involves no dependence on cutoff function, hence it is easy to see that this field redefinition does not modify the scale transformation. We can also explicitly check this(see\eqref{eq: consistency scale}). 

Observing \eqref{eq: phi conformal 12345} we can readily write down the modified special conformal transformation for $\chi(p,t)$ ($\Delta_\chi= \frac{d}{2}+\nu$).
{\small
\begin{align}\label{eq: chi conformal full ERG}
\tcboxmath{
 C_{\mu,\chi}(p,t)= 2(d-\Delta_\chi)\frac{\partial}{\partial p^\mu}+ 2 p^\rho\frac{\partial^2}{\partial p^\rho \partial p^\mu}-p_\mu \frac{\partial^2}{\partial p^\rho \partial p_\rho}-2 z\frac{\partial^2}{\partial z \partial p^\mu}-\dot{F}(p,t) \frac{\partial}{\partial p^\mu} \frac{1}{\dot{F}(p,t)}\frac{\partial}{\partial t}.
 }
\end{align} 
}
We can check consistency of this transformation using the relation
\begin{align}
    \delta \phi(p,t)= G(p,t) C_{\mu,\chi} \left( \frac{\chi(p,t)}{G(p,t)} \right).
\end{align}
$F(p,z)$ shares similar property like $G(p,z)$. Using this we reproduce $C_{\mu,\phi}$ in \eqref{eq: phi conformal 12345} from $C_{\mu,\chi}$ \eqref{eq: chi conformal full ERG} (see appendix~\ref{app: full ERG} for details).

\section{Conclusions}\label{concl}

In this paper we have shown that the Exact RG equation for a d-dimensional Euclidean field theory has an $SO(1,d+1)$ invariance. We do this by identifying $SO(1,d+1)$ as the symmetry of the action that occurs in the functional integral representation of the evolution operator of the ERG equation (we denote this action as \textit{evolution action}). This is also studied in a Hamiltonian formalism by analytically continuing to Minkowski signature. The Hamiltonian generating the RG evolution is one of the generators of the $SO(1,d+1)$ algebra.

This study was motivated by the ERG approach to AdS/CFT in which this \textit{evolution action} is mapped to a scalar field action in (Euclidean) AdS space, where it has an $SO(1,d+1)$ symmetry due to the isometry of AdS space \cite{Sathiapalan:2017,Sathiapalan:2019}. The map requires a special choice of cutoff function in the ERG equation. In this paper we have shown that for {\em any} choice of the cutoff function this symmetry exists but the transformation law of the scalar field is non standard and depends on the cutoff function. This also makes it non polynomial in derivatives and hence quasi-local. For the special choice of cutoff function that gives AdS space, the transformations have the standard simpler form. It also explains why AdS space is special in this approach.

This result is, at least in hindsight, not surprising: The scale invariant fixed point Wilson action in $d$ dimensions in general is expected to have an $SO(1,d+1)$ conformal symmetry. The \textit{evolution action}, for {\em any} choice of the cutoff function, preserves a fixed point. So it should have this symmetry as well.

This result lends further credence to the idea that AdS/CFT (and perhaps holography in general) can be understood in terms of ERG evolution. It also suggests that other spaces obtained by this procedure, for example flat space \cite{Sathiapalan:flat}, should also have this symmetry.

It is important to note that in our analysis we did not attempt any analysis on the boundary terms. These terms will in general modify the boundary states or Wilson action in $d$ dimensions. One can try to impose proper boundary conditions to ensure the symmetry of the bulk action. However, this requires careful analysis and leaves room for further work.

The studies relating ERG to Holographic RG and AdS/CFT in \cite{Sathiapalan:2017,Sathiapalan:2019}, \cite{Sathiapalan:2020}-\cite{Dharanipragada:2023} used the Polchinski version of Wilson's ERG equation. This describes the RG evolution of the {\em interacting}~part of the Wilson action and it has the particularly simple form of a diffusion equation. Another result in this paper is that the ERG equation for the {\em full Wilson action} can also be rewritten  as a diffusion equation after a field redefinition and consequently also possesses this $SO(1,D+1)$ symmetry. This should be useful in situations where ERG for the full Wilson action is required.  Furthermore this means that the ERG to Holographic RG maps can be applied in this case also. This makes this approach to holography a little more general.

\begin{appendices}

\section{Fixed point equation and composite fields}\label{FP}

In this section we reformulate the ERG fixed point equation in terms of scale transformation with composite operators as mentioned in section \ref{sec: review}. This was shown in \cite{Sonoda:2015}, we show it here in a different way. We start with the ERG fixed point equation with anomalous dimension $\gamma$
{\small
\[
\Big(-\hf \int_p\dot G(p,t) (\frac{\dd }{\dd \phi(p)}(\frac{\dd}{ \dd \phi(-p)}+ \frac{2}{G(p,t)}\phi(p)) 
+\int_p \DDx{}{\phi(p)} (p^\mu\ppx {}{p^\mu} +\frac{D+2}{2})\phi(p) -\gamma {\cal N}\Big) \Psi[\phi(p),t] =0.
\]
}
Here ${\cal N} = [ \DDx{}{\phi}\phi]$ is the number operator \cite{Igarashi}.
Substituting $G=\frac{K(p/\lm)}{p^2}$ we obtain (we put $\gamma=0$ for the moment)
\be \label{I}
\Big(-\ddx{K}{p^2}\frac{\dd^2}{\dd \phi(p)\dd \phi(-p)}
+(p^\mu \ddx{}{p^\mu}+\frac{D+2}{2})\phi(p)\DDx{}{\phi(p)}- \frac{p^\mu\ddx {K}{p^\mu}}{K}\phi(p)\DDx{}{\phi(p)}\Big)e^{-S}=0.
\ee
We will show this is same as the scale invariance of the Wilson action at finite scale as stated in \eqref{eq: scale inv composite}.
\[
[\Sigma] \Psi=\int_p [\DDx{}{\phi(p)}]\Sigma(p)[\phi(p)]\Psi=0,
\]
or
\be  \label{II}
\int _p K(p)\DDx{}{\phi(p)}( (p^\mu \ddx{}{p^\mu}+\frac{D+2}{2})(\frac{1}{K(p)}\phi(p)+ \frac{1-K}{p^2}\DDx{}{\phi(-p)})) e^{-S}=0.
\ee
The field independent constants are ignored everywhere. To obtain \eqref{II} we have used the definitions of composite operators 
\begin{align}
& [\phi(p)] = \frac{1}{K} \phi(p) + \frac{1-K}{p^2} \DDx{}{\phi(-p)},\\
& \notag [\DDx{}{\phi(p)}]= K(p) \ddp. 
\end{align}
We simplify
\[
[\Sigma(p)][\phi(p)]\Psi=
\]
which is combination of two terms

{\bf I:}
\[
\int _p K(p) \ddp \big[ (\frac{D+2}{2}) [ \underbrace{\frac{1}{K} \phi(p)}_{(a)} -\underbrace{ \frac{1-K}{p^2} \dsmp }_{(b)}] e^{-S}\big],
\]
{\bf II:}
\[
\int _p K(p) \ddp \big[ p^\mu \frac{d}{dp^\mu} [ \underbrace{\frac{1}{K} \phi(p)}_{(a)} -\underbrace{ \frac{1-K}{p^2} \dsmp }_{(b)}]e^{-S}\big].
\]
We define
\[ 
I(a)=\int _p K(p) \ddp \big[ (\frac{D+2}{2})\frac{1}{K} \phi(p)e^{-S}\big] = -(\frac{D+2}{2})\int _p \phi(p) \dsmp e^{-S} +~~~~c-number,~~~
\]
and,
 \begin{align}
 \notag II(a)=\int _p K(p) \ddp \big[ p^\mu \frac{d}{dp^\mu}\frac{1}{K} \phi(p)e^{-S}\big]& = \int _p K(p) \big[-p^\mu \frac{dK}{dp^\mu} \frac{1}{K^2} \phi(p) - \frac{1}{K} p^\mu \frac{d}{dp^\mu} \phi(p)\big]\dsp e^{-S}\\
 \notag &=\int_p (-\frac{1}{K} p^\mu \frac{dK}{dp^\mu}-p^\mu \frac{d}{dp^\mu})\phi(p) \dsp e^{-S}.
\end{align}
Combining we obtain
\be \label{III}
I(a)+II(a)= \int_p [(-p^\mu \frac {dK}{dp_\mu} \frac{1}{K} + \frac{D+2}{2} + p^\mu \frac{d}{dp^\mu} ) \phi(p)] \frac{\dd}{\dd \phi(p)}e^{-S}.
\ee
Next we simplify
\[
I(b)=\frac{D+2}{2} \int_p K(p) \ddmp (\frac{1-K}{p^2} \DDx {}{\phi(-p)} e^{-S} )= \frac{D+2}{2} \int_p \frac{K(1-K)}{p^2} \ddmp ( \ddp e^{-S} ),
\]
and
\begin{align}
II(b) & =\int _p K(p) \ddp \big[ p^\mu\frac{d}{dp^\mu} (\frac{1-K}{p^2} \DDx{}{\phi(-p)} )e^{-S}\big]\\
& \notag = \int _p \frac{K(1-K)}{p^2} \ddp \big[ p^\mu\frac{d}{dp^\mu} ( \ddmp )e^{-S}\big]+ \int_p K(p) p^\mu(\frac{d}{dp^\mu} \frac{1-K}{p^2})\frac{\dd^2}{\dd \phi(p)\dd \phi(-p)} e^{-S}\\
& \notag \equiv (i)+(ii).
\end{align}
We note
\begin{align}
(i) \notag &=\int _p -\hf \frac{d}{dp^\mu} (p^\mu \frac{K(1-K)}{p^2}) \frac{\dd^2}{\dd \phi(p)\dd \phi(-p)} e^{-S},\\
(ii) \notag &=  K(p) p^\mu(\frac{d}{dp^\mu} \frac{1-K}{p^2})\frac{\dd^2}{\dd \phi(p)\dd \phi(-p)} e^{-S}.
\end{align}
Adding we get 
II(b)=(i)+(ii)
\[II(b)= -\frac{D+2}{2}  \frac{K(1-K)}{p^2} - \frac{dK}{dp^2}.
\]
Thus I(b)+II(b) gives
\be \label{IV}
- \frac{dK}{dp^2}\frac{\dd^2}{\dd \phi(p)\dd \phi(-p)} e^{-S}. 
\ee
Adding \eqref{III} and \eqref{IV} we see that \eqref{I} is reproduced.

Including anomalous dimension is now straightforward. Adding $-\gamma {\cal N}$ to \eqref{II} changes $\frac{D+2}{2}$ to $\frac{D+2}{2}-\gamma$ which is just the full scaling dimension of the field in the interacting theory.

\section{Details: ERG equation for the full Wilson action}
\label{Full}
In this section we provide the ERG equation for the full Wilson action and its evolution operator. We have also rewritten the equation in Pochinski's form.
\subsection{Evolution operator}
We denote the finite cutoff as $\lm$ in this section. Polchinski ERG equation for $S_{I,\lm}[\phi_l]$ \eqref{eq: Polchinski diffusion} is
\be 
\p_t \psi[\phi_l(p),t] = -\hf \int_p\dot G(p,t) \frac{\dd^2}{\dd \phi_l(p) \dd \phi_l(-p)} \psi[\phi_l(p),t]
\ee
with $\psi=e^{-S_{I,\lm}[\phi_l]}$, while the full Wilson action 
\[
S[\phi_l]=\hf \int_p \phi_l(p)G^{-1}\phi_l(-p) + S_{I,\lm}[\phi_l]
\]
obeys 
\be 
\p_t \Psi[\phi_l(p),t] = -\hf \int_p\dot G(p,t) (\frac{\dd }{\dd \phi_l(p)}(\frac{\dd}{ \dd \phi_l(-p)}+ \frac{2}{G(p,t)}\phi_l(p)) \Psi[\phi_l(p),t].
\ee
where $\Psi=e^{-S[\phi_l]}$. In simplified notation we write the equation as
\be 
\p_t \Psi (x,t) = -\hf \dot G \ppx{}{x}(\ppx{}{x}+\frac{2}{G}x)\Psi (x,t).
\ee
We proceed to rewrite this equation as
\be 
\ppx{}{G} \Psi (x,G) = -\hf  \ppx{}{x}(\ppx{}{x}+\frac{2}{G}x)\Psi (x,G).
\ee
We will treat $G$ as the Euclidean time. Since $\dot G <0$ we will use the notation $G=-G_E$.
\be 
\ppx{}{G_E} \Psi (x,G_E) = (\hf  \ppx{^2}{x^2}-\frac{1}{G_E}\ppx{}{x}x)\Psi (x,G_E).
\ee
Now Wick rotate: $iG_M=G_E$.
\be 
i\ppx{}{G_M} \Psi (x,G_M) = -(\hf  \ppx{^2}{x^2}+i\frac{1}{G_M}\ppx{}{x}x)\Psi (x,G_M).
\ee
From this we see that the Hamiltonian is: ($-i\ppx{}{x}\to p$)
\begin{align}
H \notag &= \hf p^2 +\frac{1}{G_M}px\\
&= \hf  p^2 +\frac{1}{G_M} \hf(px+xp)-i\frac{1}{2G_M}.
\end{align}
We have written the Hamiltonian as the sum of a manifestly Hermitian term and an imaginary field independent constant that becomes a normalization factor for the wave function.

We now treat the Hamiltonian as a classical object
(we will restore the constant term below)
\[
H= \frac{p^2}{2} + \frac{px}{G_M}.
\]
 and compute the Lagrangian using
\[
\ppx{H}{p}=p+\frac{x}{G_M}=\ppx{x}{G_M} \implies p=\dot x -\frac{x}{G_M}.
\]
Here for compactness we have used $\dot x = \frac{\p x}{\p G_M}$
and 
\[
p\dot x - H =L.
\]
where
\begin{align}
L & \notag = (\dot x - \frac{x}{G_M})x-[\hf(\dot x - \frac{x}{G_M})^2 +(\dot x - \frac{x}{G_M})\frac{x}{G_M}]\\
& =\hf \dot x^2 - \frac{x \dot x}{G_M} + \frac{x^2}{2G_M^2}= \hf(\frac{\p x}{\p G_M} - \frac{x}{G_M})^2.
\end{align}
The evolution operator is given by
\begin{align}
U(G_{Mf},G_{Mi})= & \notag e^{-i\int_{G_{Mi}}^{G_{Mf}} H ~dG_M -i\int_{G_{Mi}}^{G_{Mf}}dG_M~\frac{-i}{2G_M}}\\
=& e^{-i\int_{G_{Mi}}^{G_{Mf}} H ~dG_M - \hf (\ln G_{Mf}-\ln G_{Mi})}.
\end{align}
Using the functional representation of the evolution operator we get
\[
U(G_{Mf},G_{Mi})=\int \CD x~e^{i\int_{G_{Mi}}^{G_{Mf}}
dG_M~\hf(\dot x - \frac{x}{G_M})^2}e^{- \hf (\ln G_{Mf}-\ln G_{Mi})}.
\]
We now analytically continue back to Euclidean time i.e.$G_E=iG_M$.
\begin{align}
U(G_{Ef},G_{Ei})= & \notag \int \CD x~e^{-\int_{G_{Ei}}^{G_{Ef}}
dG_E~\hf[ (\ppx{x}{G_E})^2 -2 \frac{x}{G_E}\ppx{x}{G_E} + (\frac{x}{G_E})^2)}e^{- \hf (\ln (-iG_{Ef})+\ln (-iG_{Ei}))}\\
= & \int \CD x~e^{-\int_{G_{Ei}}^{G_{Ef}}
dG_E~\hf(\dot x - \frac{x}{G_E})^2}\sqrt{\frac{G_{Ei}}{G_{Ef}}}.
\end{align}
Now remembering that $G_E=-G$. We rewrite this as
\be 
U(G_f,G_i)=\int \CD x~e^{\int_{G_{i}}^{G_{f}}
dG~\hf(\ppx{ x}{G} - \frac{x}{G})^2}\sqrt{\frac{G_{i}}{G_{f}}}.
\ee
Finally, we reintroduce our time $t$ and write this as
\be 
U(t_f,t_i)=\int \CD x~e^{\int_{t_i}^{t_f}dt~ \frac{1}{2\dot G} (\ppx{x}{t} - \frac{\dot G}{G}x)^2}\sqrt{\frac{G_{i}}{G_{f}}}
\ee
This expression in the exponential can be rewritten in a different way
by first writing 
\[
\ddt (\frac{x}{G}) =\frac{\dot x}{G} - \frac{x \dot G}{G^2}=\frac{1}{G}(\dot x - \frac{\dot G}{G}x).
\]
Thus
\[
G^2 (\ddt (\frac{x}{G}))^2=(\dot x - \frac{\dot G}{G}x)^2.
\]
We obtain
\[
U(t_f,t_i)=\int \CD x~e^{\hf\int_{t_i}^{t_f}dt~\frac{G^2}{\dot G}(\ddt (\frac{x}{G}))^2}
\sqrt{\frac{G_{i}}{G_{f}}}.
\]
We define $y=\frac{x}{G}$ and $F=\frac{1}{G}$ to conclude
\be \label{fnl}
\tcboxmath{
U(t_f,t_i)=\int \CD y~e^{-\hf\int_{t_i}^{t_f}dt\frac{1}{\dot F} \dot y^2}\sqrt{\frac{G_{i}}{G_{f}}}}
\ee
In path integral we know 
\[
\Psi(y_f,t_f)\approx\int dy_i\frac{1}{\sqrt{F_f-F_i}}e^{-\hf \frac{(y_f-y_i)^2}{F_f-F_i}}\sqrt{\frac{G_{i}}{G_{f}}}\Psi(y_i,t_i).
\]
Reintroducing $y=x/G$ and $F=1/G$ we get
\[
\Psi(x_f,t_f)=\int dx_i\frac{1}{G_i} 
\frac{1}{\sqrt{\frac{1}{G_f}-\frac{1}{G_i}}}e^{-\hf \frac{(\frac{x_f}{G_f}-\frac{x_i}{G_i})^2}{\frac{1}{G_f}-\frac{1}{G_i}}}\sqrt{\frac{G_{i}}{G_{f}}}\Psi(x_i,t_i).
\]
\be 
\tcboxmath{
\Psi(x_f,t_f)=\int dx_i\frac{1}{\sqrt{G_i-G_f}}e^{-\hf \frac{(\frac{x_f}{G_f}-\frac{x_i}{G_i})^2}{\frac{1}{G_f}-\frac{1}{G_i}}}\Psi(x_i,t_i)}
\ee

The action in \eqref{fnl} is the same as usual except for redefinitions.
\begin{align}
    U(t_f,t_i)= \int \mathcal{D}\phi e^{\int_{t_i}^{t_f} dt \frac{\dot{\chi}(p,t)^2}{2\dot{F(p,t)}}}\sqrt{\frac{G(p,t_i)}{G_f(p,t_f)}},
\end{align}
where $\chi(p,t)= \frac{y(p,t)}{G(p,t)}$ and $F(p,t)=\frac{1}{G(p,t)}$. This should have the same conformal invariance. 

\subsection{Rewriting the ERG equation}

We start with
\be \label{eq: full ERG eq simple}
\p_t \Psi (x,t) = -\hf \dot G \ppx{}{x}(\ppx{}{x}+\frac{2}{G}x)\Psi (x,t),
\ee
and employ field refinition: $x(t)=y(t)G(t)$. 
\[
\Psi(x,t)=\Psi(yG,t)=\tilde\Psi(y,t).
\]
Hence the LHS of \eqref{eq: full ERG eq simple} can be simplified as follows.
\begin{align}
\p_t \tilde \Psi(y,t) \notag & = \p_t \Psi(y G,t)|_G+ \ppx{\Psi(yG,t)}{(yG)} y \dot G\\
& \notag = \p_t \Psi(x,t)+ \ppx{\Psi(x,t)}{x} \frac{x}{G}\dot G\\
& =-\hf \dot G \ppx{}{x}(\ppx{}{x}+\frac{2}{G}x)\Psi (x,t)+\ppx{\Psi(x,t)}{x} \frac{x}{G}\dot G,
\end{align}
Thus
\[
\p_t \tilde \Psi(y,t)=-\hf \dot G \frac{\p^2}{\p x^2}\Psi(x,t) -\frac{\dot G}{G}\Psi,
\]
and
\[
\frac{\p^2}{\p x^2}\Psi(x,t)= \frac{\p^2
}{\p (yG)^2}\Psi(yG,t)= \frac{1}{G^2}\frac{\p^2}{\p y^2}\tilde \Psi(y,t).
\]
Thus we get a Polchinski like equation (dropping a field independent term that only affects overall normalization). 
\be \label{FullERG}
\tcboxmath{\p_t \tilde \Psi(y,t)=-\hf \frac{\dot G}{G^2}
\frac{\p^2}{\p y^2}\tilde \Psi(y,t)=\hf\dot F \frac{\p^2}{\p y^2}\tilde \Psi(y,t).}
\ee
We have set $F=\frac{1}{G}$ as before and can compare the results obtained earlier in the functional formalism.
(Note the positive sign on the RHS of \eqref{FullERG}. This is  because while $\dot G<0$, $\dot F>0$.)

\section{Generators in $(E)AdS$}
\label{app: AdS isometry}

The $(E)AdS_3$ spacetime in $3$ dimensions can be written in $4$ dimensional embedding space as
\be \label{ads}
\underbrace{\sum_{i,j=0,1}g_{ij} dX^i dX^j}_{boundary}+(X^{2})^2-(X^4)^2=-1.
\ee

In this Appendix (only) we use $i,j,k..$ for the $d$ boundary directions $(0,1,...,d-1)$ and $\mu,\nu$ for the  $d+2$ coordinate indices  of $R^{d+2}$ in which $AdS_{d+1}$ is embedded. We work with $d=2$ as our example.

Thus as an example for $AdS_3$, the metric on $R^4$  is thus $\left \lbrace{-++-}\right \rbrace$ for Minkowski,  $\left \lbrace{+++-}\right \rbrace$ for Euclidean. We define the generators in covariant form  as
\be\notag
J_{\mu\nu}= X_\mu {\p\over \p X^\nu}- X_\nu {\p\over \p X^\mu}.
\ee
The isometry forms $SO(1,3)$ for Euclidean and $SO(2,2)$ for Minkowski signature. The algebra has 6 generators $(i=0,1)$.
{\small
\begin{align*}
& Translation: T_i= J_{i 2}-J_{i 4},~~Rotation: M_{01} = J_{01},~~Dilatation: D=J_{24},~~SCT: C_i=J_{i 2}+J_{i 4}.
\end{align*}
}
In Euclidean case there are 3 compact and 3 non-compact generators
\[
J_{01},J_{02},J_{12}: compact\ J_{24},J_{14},J_{04}: non-compact
\]
and in Minkowski case only two compact generators
\[
J_{04}, J_{12}: compact
\]

They follows the Lie bracket written below with real structure constant when the generators are chosen real.
\begin{align}
[J_{\mu\nu},J_{\rho \sigma}]= - g_{\mu \rho} J_{\nu \sigma}- g_{\nu \sigma} J_{\mu \rho}+ g_{\mu \sigma} J_{\nu \rho}+ g_{\nu \rho} J_{\mu \sigma}.
\end{align}
We consider an example with $i=1$.
\br\notag
J_{12}&=& X_1{\p\over \p X^2}-X_2{\p\over \p X^1} \implies 
\left\lbrace\begin{array}{c@{~=~}l}
\delta X^2 & \epsilon X_1 = \epsilon X^1, \\ 
\delta X^1 & -\epsilon X_2 = -\epsilon X^2.
\end{array}\right.\\
\notag J_{14}&=&X_1{\p\over \p X^4}-X_4{\p\over \p X^1} \implies 
\left\lbrace\begin{array}{c@{~=~}l}
\delta X^4 & \epsilon X_1 = \epsilon X^1, \\ 
\delta X^1 & -\epsilon X_4 = \ep\, X^4. 
\end{array}\right.
\er
\be
U=X^2+X^4,\,V=X^2-X^4,
\ee
\be	\label{5}
UV =-1 + (X^0)^2 -(X^1)^2 \implies V= {-1-X_\mu X^\mu\over U}.
\ee
One can check that
\begin{subequations}
\begin{eqnarray}
J_{12}-J_{14} &:& \delta X^1= -\epsilon U,\\
J_{12}+J_{14} &:& \delta X^1= -\epsilon V.
\end{eqnarray}
\end{subequations}
We will see that $J_{12}-J_{14}=T_1$ and $J_{12}+J_{14}=C_1$ generates the translation and special conformal transformation respectively. We note
\begin{eqnarray*}
T_1 &:& \delta U=0, \,\delta V= 2\epsilon X^1,\\
C_1 &:& \delta U=2\epsilon X^1; \,\delta V=0.
\end{eqnarray*}
In general, special conformal transformation can be written covariantly as
\be
C_i: \delta U= 2\epsilon X_i, ~~~\delta X^j= -\epsilon V \delta _i^j.
\ee
Poincare patch coordinates can be defined in terms of embedding co-ordinates as
\be 
x^i = {X^i\over U}~,~~~~ z= {1\over U}.
\ee
The action of $C_j$ on $x^i$ can be obtained as
\br
C_j : \delta  x^i &=& {\delta X^i\over U} -{X^i \delta U\over U^2}\\ \nonumber
&=&-{\epsilon V \delta _j ^i\over U} - {X^i 2 \epsilon X_j\over U^2} \\ \nonumber
&=&+\epsilon(z^2+x_k x^k) \delta_j^i - 2\epsilon x_j x^i.
\er
and action on $z$ as
\[
C_j z = C_j {1\over U} : \delta {1\over U} = {-2\epsilon X_j \over U^2} = -2\epsilon z x_j.
\]
Thus combining we get the special conformal transformal transformation as
\be
C_j = (z^2+x_k x^k){\p\over \p x^j} - 2 x_j( x^k {\p\over \p x^k}+z{d\over dz}).
\ee
Finally going into momentum space and including the conformal dimension $\Delta$ admits
\begin{subequations}
\label{cftr}
\begin{align}
C_j & =\,2(d-\Delta) {\p\over \p p^j } + 2 p^k {\p^2\over \p
            p^k \p p^j} -p_j{\p^2\over \p p^k \p p_k}\\ 
\notag & -2z {\p^2\over \p z \p p^j}+ z^2 p_j.
\end{align}
\end{subequations}

\par\noindent\rule{0.4\textwidth}{0.5 pt}

Next we find the scale transformation law from the generator $J_{24}$.
\begin{eqnarray*}
&&J_{24}= X_2{\p\over \p X^4} -X_4{\p\over \p X^2} ~~:\delta X^2 =
   -\epsilon X_4 = + \epsilon X^4;~~\delta X^4 = \epsilon X_2 =
   \epsilon X^2.\\
   \end{eqnarray*}
   
This gives
\begin{eqnarray*} 
&&\delta x^1 = -\epsilon x^1,~~~\delta x^0 = -\epsilon
   x^0,~~~\delta z = -{\delta U\over U^2}=-\epsilon z.\\
\end{eqnarray*}

We can find the scale transformation from the commutator of special conformal transformation and transformation.
\begin{eqnarray*}    
&&[C_1,T_1]=[J_{12}+J_{14},J_{12}-J_{14}]= J_{24} = D = -x^\mu
   {\p\over \p x^\mu} - z{\p\over \p z}. 
\end{eqnarray*}
Again, after including conformal dimension in momentum space this becomes
\be   \label{Dil}
D = p^k {\p\over \p p^k} - z{\p\over \p z}+(d-\Delta).
\ee

\paragraph{ $SO(2,d)$ and $SO(1,d+1)$}  Note that the Minkowskian and Euclidean case differs only by metric $g_{00}$. Hence three commutation relations differs in these two signatures.

\begin{align}
& [C_0,T_{0}]= [J_{02}+J_{04},J_{12}-J_{14}]= 2 g_{00} J_{24}= 2 g_{00} D,\\
& \nonumber [C_{0},M_{01}]=[J_{02}+J_{04},J_{01}]= g_{00} C_1,\\
& \nonumber [C_{1},M_{01}]=[J_{12}+J_{14},J_{01}]= -g_{11} C_0.
\end{align}

This can be generalized to arbitrary dimensions as

\begin{align}\label{eq: group structure general}
& [C_i,T_j]= 2 g_{ij} D - 2 M_{ij},\\
& \nonumber [C_0,M_{0 j}]= g_{00} C_j,\\
& \nonumber [C_i,M_{0 i}]= -g_{ii} C_0.
\end{align}

It is to be noted that in section \ref{sec: conformal Hamiltonian} when we Wick rotated the radial direction $z$ to Minkowski space in the Hamiltonian formalism, we are effectively in $dS$ space, and the isometry group is still $SO(1,d+1)$.

\section{Details: transformation in $\phi(p,t)$}
\label{app: conformal transformation phi details}

The special conformal transformation of $\phi$ in section~\ref{sec: conformal functional} involves careful analysis and is crucial part of this work. In this section we provide important steps of this computation. The cutoff function independent transformation comprises of the following four transformations:

\begin{align}\label{eq: cutoff independent conformal}
\delta \phi &= \notag 2(d-\Delta_\phi) \frac{\partial}{\partial p^\mu}+ 2 p^\rho\frac{\partial^2}{\partial p^\rho \partial p^\mu}-p_\mu \frac{\partial^2}{\partial p^\rho \partial p_\rho}-2 z \frac{\partial^2}{\partial z \partial p^\mu}\\
& \equiv C^1_{\mu}+C^2_{\mu}+C^3_{\mu}+C^4_{\mu}.
\end{align} 
We apply these conformal transformation acts on the evolution action in $\phi(p,t)$ \eqref{eq: action phi(p,t)} 
\begin{align*}
\mathcal{S}[\phi]= \frac{1}{2} \int ~d t \int_p ~ \frac{\dot{\phi}(p,t)\dot{\phi}(-p,t)}{\dot{G}(p,t)}.
\end{align*}
Ignoring the total derivative terms in $p$-directions we obtain

{\small
\paragraph*{1)}

\begin{align}\label{eq: conformal1 general}
C_\mu^1 \s[\phi]& =\frac{1}{2} \int ~dt \int_p \,= 2(d-\Delta_\phi)\int~d t \int_p~\frac{\frac{\partial}{\partial p^\mu}\dot{\phi}(p,t)\dot{\phi}(-p,t)}{\dot{G}(p,t)},
\end{align}
}

\paragraph*{2)}
{\small
\begin{align}\label{eq: conformal2 general exp}
 C_\mu^2 \s[\phi]=&\frac{1}{2} \int ~d t \int_p \frac{2 p^\rho\frac{\partial^2}{\partial p^\rho \partial p^\mu}\dot{\phi}(p,t)\dot{\phi}(-p,t)}{\dot{G}(p,t)}-\frac{1}{2} \int~d t \int_p \frac{2 p^\rho\frac{\partial^2}{\partial p^\rho \partial p^\mu}\dot{\phi}(-p,t)\dot{\phi}(p,t)}{\dot{G}(p,t)}.
\end{align}
}
The first term can be written as

\begin{subequations}
{\small
\begin{align}
\nonumber &= \int~d t \int_p \frac{\partial}{\partial p^\mu}\left[ \frac{p^\rho\frac{\partial}{\partial p^\rho}\dot{\phi}(p,t) \dot{\phi}(-p,t)}{\dot{G}(p,t)} \right]-\int~d t \int_p \left[ \frac{\frac{\partial}{\partial p^\mu}\dot{\phi}(p,t) \dot{\phi}(-p,t)}{\dot{G}(p,t)} \right]\\
\nonumber &- \int~d t \int_p\frac{\partial}{\partial p^\mu}\frac{1}{\dot{G}(p,t)}p^\rho\frac{\partial}{\partial p^\rho}\dot{\phi}(p,t)\dot{\phi}(-p,t) - \int~d t \int_p\frac{ p^\rho\frac{\partial}{\partial p^\rho}\dot{\phi}(p,t) \frac{\partial}{\partial p^\mu}\dot{\phi}(-p,t) }{\dot{G}(p,t)},\\
\end{align}
}
while the second term becomes,
{\small
\begin{align}
\nonumber =&\int~d t \int_p-\frac{\partial}{\partial p^\rho}\left[ \frac{p^\rho\frac{\partial}{\partial p^\mu}\dot{\phi}(-p,t) \dot{\phi}(p,t)}{\dot{G}(p,t)} \right]+ \int~d t \int_p \frac{1}{\dot{G}(p,t)} D \left[\frac{\partial}{\partial p^\mu}\dot{\phi}(-p,t)\right]\dot{\phi}(p,t)\\
&+ \int~d t \int_p\frac{\partial}{\partial p^\rho}\frac{1}{\dot{G}(p,t)} p^\rho \left[\frac{\partial}{\partial p^\mu}\dot{\phi}(-p,t)\right]\dot{\phi}(p,t)+\int~d t \int_p \frac{1}{\dot{G}(p,t)}p^\rho \left[\frac{\partial}{\partial p^\mu}\dot{\phi}(-p,t)\right]\frac{\partial}{\partial p^\rho}\dot{\phi}(p,t).
\end{align}
}
\end{subequations}
Hence  \eqref{eq: conformal2 general exp} becomes,
{\small
\begin{align}\label{eq: conformal2 general}
& \nonumber C_\mu^2 \s[\phi]\\
\nonumber = &- \int~d t \int_p(D+1)\frac{\left[\frac{\partial}{\partial p^\mu}\dot{\phi}(p,t)\right] \dot{\phi}(-p,t)}{\dot{G}(p,t)}-\int~d t \int_p \frac{\partial}{\partial p^\rho} \frac{1}{\dot{G}(p,t)} p^\rho\left[\frac{\partial}{\partial p^\mu}\dot{\phi}(p,t)\right]\dot{\phi}(-p,t)\\
& -\int~d t \int_p\frac{\partial}{\partial p^\mu} \frac{1}{\dot{G}(p,t)} p^\rho\left[\frac{\partial}{\partial p^\rho}\dot{\phi}(p,t)\right]\dot{\phi}(-p,t).
\end{align}
}

\paragraph*{3)}
{\small
\begin{align}\label{eq: conformal3 general exp}
 C_\mu^3 \s[\phi]=&\frac{1}{2} \int~d t \int_p \frac{- p_\mu \frac{\partial^2}{\partial p^\rho \partial p_\rho}\dot{\phi}(p,t)\dot{\phi}(-p,t)}{\dot{G}(p,t)}+\frac{1}{2} \int~d t \int_p \frac{ p_\mu\frac{\partial^2}{\partial p^\rho \partial p_\rho}\dot{\phi}(-p,t)\dot{\phi}(p,t)}{\dot{G}(p,t)}.
\end{align}
}
The first term of the above equation \eqref{eq: conformal3 general exp} can be evaluated as,
{\small
\begin{align}
\nonumber & =-\int~d t \int_p \frac{1}{2}\frac{\partial}{\partial p^\rho}\left[ \frac{p_\mu\frac{\partial}{\partial p^\rho}\dot{\phi}(p,t) \dot{\phi}(-p,t)}{\dot{G}(p,t)} \right]+\frac{1}{2}\int~d t \int_p \frac{\left[\frac{\partial}{\partial p^\mu}\dot{\phi}(p,t)\right] \dot{\phi}(-p,t)}{\dot{G}(p,t)}\\
& \nonumber +\frac{1}{2}\int~d t \int_p \frac{p_\mu \frac{\partial}{\partial p^\rho}\dot{\phi}(p,t) \frac{\partial}{\partial p_\rho}\dot{\phi}(-p,t) }{\dot{G}(p,t)}+ \frac{1}{2}\int~d t \int_p p_\mu \left[\frac{\partial}{\partial p^\rho}\dot{\phi}(p,t) \right]\dot{\phi}(-p,t) \frac{\partial}{\partial p^\rho} \frac{1}{\dot{G}(p,t)}.\\
\end{align}
}
When we compute the $(p \rightarrow -p)$ part the third term cancels and finally we obtain,
{\small
\begin{align}\label{eq: conformal3 general}
\nonumber C_\mu^3 \s[\phi]= \int~d t \int_p\frac{\left[\frac{\partial}{\partial p^\mu}\dot{\phi}(p,t)\right] \dot{\phi}(-p,t)}{\dot{G}(p,t)}+\int~d t \int_p p_\mu \left[\frac{\partial}{\partial p^\rho}\dot{\phi}(p,t) \right]\dot{\phi}(-p,t) \frac{\partial}{\partial p^\rho} \frac{1}{\dot{G}(p,t)}.\\
\end{align}
}
Further, the last term in \eqref{eq: conformal3 general} and \eqref{eq: conformal2 general exp} cancel with each other using the property 
of $G(p,t)$ \eqref{eq: property p-sym G}. Hence, the application of combination of $C_\mu^1$, $C_\mu^2$ and $C_\mu^3$ on the action can be written as,
\begin{align}\label{eq: conformal123 general}
& \nonumber (C_\mu^1+ C_\mu^2+ C_\mu^3) \s[\phi] \\
& = \int~d t \int_p \left[\frac{\partial}{\partial p^\mu}\dot{\phi}(p,t)\right] \dot{\phi}(-p,t) \left( \frac{d-2 \Delta_\phi}{\dot{G}(p,t)}- p^\rho\frac{\partial}{\partial p^\rho} \frac{1}{\dot{G}(p,t)}\right).
\end{align}

\paragraph{4)} Finally we act with $C_\mu^4= -2 z\frac{\partial^2}{\partial z \partial p^\mu}=-2\frac{\partial^2}{\partial t \partial p^\mu} $ on the action
{\small
\begin{align}\label{eq: conformal4 general exp}
 C_\mu^4 \s[\phi]=&-\frac{1}{2} \int~d t \int_p \frac{2 \frac{\partial^2}{\partial t\partial p^\mu}\dot{\phi}(p,t)\dot{\phi}(-p,t)}{\dot{G}(p,z)}+\frac{1}{2} \int~d t \int_p \frac{2 \frac{\partial^2}{\partial t \partial p^\mu}\dot{\phi}(-p,t)\dot{\phi}(p,t)}{\dot{G}(p,t)}.
\end{align}
}
\begin{subequations}
The first term in \eqref{eq: conformal4 general exp} (modulo the total derivative terms in $p$ directions) can be written as
{\small
\begin{align}\label{eq: boundary conformal phi}
\nonumber &= -\int~d t \int_p \frac{\partial}{\partial t}\left[\frac{\frac{\partial}{\partial p^\mu}\dot{\phi}(p,t)\dot{\phi}(-p,t)}{\dot{G}(p,z)}\right]\\
& \nonumber + \int~d t \int_p\frac{\left[\frac{\partial}{\partial p^\mu}\dot{\phi}(p,t)\right] \frac{\partial}{\partial t}\dot{\phi}(-p,t)}{\dot{G}(p,t)}+ \int~d t \int_p \left[\frac{\partial}{\partial p^\mu}\dot{\phi}(p,t)\right]\dot{\phi}(-p,t)\frac{\partial}{\partial t}\frac{1}{\dot{G}(p,t)}.\\
\end{align}
}
Similarly, the second term reduces to
{\small
\begin{align}
\nonumber & =\int~d t \int_p \frac{\partial}{\partial p^\mu}\left[\frac{\frac{\partial}{\partial t}\dot{\phi}(p,t)\dot{\phi}(-p,t)}{\dot{G}(p,z)}\right]\\
& \nonumber - \int~d t \int_p \frac{\left[\frac{\partial}{\partial p^\mu}\dot{\phi}(p,t)\right] \frac{\partial}{\partial t}\dot{\phi}(-p,t)}{\dot{G}(p,t)}-\int~d t \int_p \left[\frac{\partial}{\partial t}\dot{\phi}(-p,t)\right]\dot{\phi}(p,t)\frac{\partial}{\partial p^\mu}\frac{1}{\dot{G}(p,t)}.\\
\end{align}
}
\end{subequations}
Hence the application of $C_\mu^4$ on the action can be summed up as
{\small
\begin{align}\label{eq: conformal4 general}
 \nonumber &C_\mu^4 \s[\phi] \\
\nonumber &= \int~d t \int_p\left[\frac{\partial}{\partial p^\mu}\dot{\phi}(p,t)\right]\dot{\phi}(-p,t)\frac{\partial}{\partial t}\frac{1}{\dot{G}(p,t)}+ \int~d t \int_p\left[\frac{\partial}{\partial t}\dot{\phi}(p,t)\right]\dot{\phi}(-p,t)\frac{\partial}{\partial p^\mu}\frac{1}{\dot{G}(p,t)}.\\
\end{align}
}
The action of four terms in our ansatz of transformations on the action can be written down from \eqref{eq: conformal123 general} and \eqref{eq: conformal4 general}.
{\small
\begin{align}
\nonumber &(C_\mu^1+ C_\mu^2+ C_\mu^3+  C_\mu^4)\s[\phi]\\
\nonumber & = \int~d t \int_p \left[\frac{\partial}{\partial p^\mu}\dot{\phi}(p,t)\right]\dot{\phi}(-p,t)\bigg \lbrace  \frac{\partial}{\partial t}\frac{1}{\dot{G}(p,t)}-p^\rho\frac{\partial}{\partial p^\rho}\frac{1}{\dot{G}(p,t)}+ \frac{d-2\Delta_\phi}{\dot{G}(p,t)} \bigg \rbrace\\
& + \int~d t \int_p \left[\frac{\partial}{\partial t}\dot{\phi}(p,t)\right]\dot{\phi}(-p,t)\frac{\partial}{\partial p^\mu}\frac{1}{\dot{G}(p,t)}.
\end{align}
}

\section{Details: transformation in $y(p,z)$}
\label{app: conformal transformation y details}

In this section we provide the details of the symmetry transformation of bulk action in $y(p,z)$ \eqref{eq: bulk y action}. Following the previous section one can repeat the analysis here fairly easily. Hence in this section we provide details to required extent only. 

\subsubsection*{Action on \textit{type-A} term}  We state the action of four transformations $C_\mu^i, (i=1,4)$ individually below.

\begin{align}
C_{\mu,y}^1 \s_A[y]= & 4(d-\Delta_y) \int dz \int_p z^{-d+1} \frac{\partial}{\partial z}\frac{\partial}{\partial p^\mu} y(p,z) \frac{\partial y(-p,z)}{\partial z},\\
\notag C_{\mu,y}^2 \s_A[y]= & -(2d+2)\int dz \int_p z^{-d+1}\frac{\partial}{\partial z} \frac{\partial}{\partial p^\mu} y(p,z)\frac{\partial y(-p,z)}{\partial z},\\
\notag C_{\mu,y}^3 \s_A[y]= & 2 \int dz \int_p z^{-d+1}   \frac{\partial}{\partial z} \frac{\partial}{\partial p^\mu}y(p,z) \frac{\partial y(-p,z)}{\partial z},\\
\notag C_{\mu,y}^4 \s_A[y]= &-(2d) \int dz \int_p  z^{-d+1}\left[ \frac{\partial^2}{\partial z \partial p^\mu} y(p,z)\right]\frac{\partial y(-p,z)}{\partial z}. 
\end{align}
$\s_A[y]$ resembles the boundary action $S[\phi]$ considered in the previous subsection, multiplied by an extra factor $z^{-d+1}$ but without the cutoff function $G(p,z)$. Hence the action of $\delta y$ on $\s_A$ can be promptly evaluated using the procedure described in previous section.

Summing up all the four contributions above we conclude
\begin{align}
\nonumber & \left(C_{\mu,y}^1+ C_{\mu,y}^2+ C_{\mu,y}^3+  C_{\mu,y}^4\right)\s_A \\
& =- 4 \Delta_y \int~d z \int_p z^{-d+1}\left[ \frac{\partial^2}{\partial z \partial p^\mu} y(p,z)\right]\frac{\partial y(-p,z)}{\partial z}.
\end{align}
We have chosen $y$ in such a way $\Delta_y$ is zero. Hence \textit{type-A} term is invariant under the transformation $C_{\mu,y}^1+ C_{\mu,y}^2+ C_{\mu,y}^3+  C_{\mu,y}^4$.

\subsubsection*{Action on \textit{type-B} term} The \textit{type-B} has simpler structure as it is function of $y(p,z)$. It contains the function $C(p,z)$ \eqref{eq: def C(p,z)}, whose property can be used to simplify the computation. 

{\footnotesize
\begin{align}\label{eq: conformal typeB}
 C_{\mu,y}^1 \s_B[y] = & \int dz \int_p 4(d-\Delta_y)\frac{\partial y(p,z)}{\partial p^\mu}y(-p,z) C(p,z)z^{-d+1},\\
\notag C_{\mu,y}^2 \s_B[y] = & \int dz \int_p \left[ -(2d+2)C(p,z)-2 p^\rho \frac{\p C(p,z)}{\p z} \right]\frac{\partial y(p,z)}{\partial p^\mu}y(-p,z) z^{-d+1},\\
\notag C_{\mu,y}^3 \s_B[y] = &\int dz \int_p 2 p_\mu \frac{\partial y(p,z)}{\partial p^\rho}y(-p,z)\frac{\partial C(p,z)}{\partial p_\rho}z^{-d+1}+2\int \frac{\partial y(p,z)}{\partial p^\mu}y(-p,z) C(p,z)z^{-d+1},\\
\notag C_{\mu,y}^4 \s_B[y] = \notag & -(2d-4)\int dz \int_p \frac{\partial y(p,z)}{\partial p^\mu}y(-p,z) C(p,z)z^{-d+1}+2 \int dz \int_p z\frac{\partial y(p,z)}{\partial z}y(-p,z) \frac{\partial C(p,z)}{\partial p^\mu}z^{-d+1}\\
& \notag +2 \int dz \int_p z\frac{\partial C(p,z)}{\partial z}\frac{\partial y(p,z)}{\partial p^\mu}y(-p,z) z^{-d+1}-\int dz \int_p \frac{\p}{\p z} \left[ 2 z^{-d+1} z \frac{\p^2}{\p z \p p^\mu} y(p,z) \frac{\p}{\p z} y(-p,z) \right].
\end{align}
}
In \eqref{eq: conformal typeB} we observe the following points (we use $\int \equiv \int~d z \int_p$ below for convenience).
\begin{enumerate}

\item
We note the last term in $C_{\mu,y}^2 \s_B[y]$  and first term in $C_{\mu,y}^3 \s_B[y]$.
{\small
\begin{align*}
    -2 \int p^\rho \frac{\partial y(p,z)}{\partial p^\rho}y(-p,z)\frac{\partial C(p,z)}{\partial p^\mu}z^{-d+1}+\int 2 p_\mu \frac{\partial y(p,z)}{\partial p^\rho}y(-p,z)\frac{\partial C(p,z)}{\partial p_\rho}z^{-d+1}
\end{align*}
}
cancel each other due to property of $C(p,z)$ \eqref{eq: property p-sym C}.
\item

We collect the coefficients of $\frac{\partial y(p,z)}{\partial p^\mu}y(-p,z)$
{\small
\begin{align*}
\int &\frac{\partial y(p,z)}{\partial p^\mu}y(-p,z) C(p,z)z^{-d+1} \\
& \left \lbrace 4(d-\Delta_y)-2d-2+2+2+2(-d+1)+2 z \frac{\partial C(p,z)}{\partial z}-2 p^\sigma\frac{\partial C(p,z)}{\partial p^\sigma}\right \rbrace \\
=& \int \frac{\partial y(p,z)}{\partial p^\mu}y(-p,z) C(p,z)z^{-d+1} \left \lbrace (4-4 \Delta_y)+2 z \frac{\partial C(p,z)}{\partial z}-2 p^\sigma\frac{\partial C(p,z)}{\partial p^\sigma}\right \rbrace.
\end{align*}
}
This enables us to conclude that the term proportional to $\frac{\partial y(p,z)}{\partial p^\mu}y(-p,z)$ vanishes for $\Delta_y=0$.

\item

 Hence we can write
\begin{align}\label{eq: y conformal1234 type-II}
  (C_{\mu,y}^1+ C_{\mu,y}^2+C_{\mu,y}^3+C_{\mu,y}^4)S= 2 \int  \frac{\partial y(p,z)}{\partial z}y(-p,z) \frac{\partial C(p,z)}{\partial p^\mu}z^{-d+1}.
\end{align}

\item 
We need to add one more transformation to \eqref{eq: y conformal1234 general} to negate \eqref{eq: y conformal1234 type-II}
\begin{align}
    C_{\mu,y}^5=\left[ \int_z z \frac{\partial}{\partial p^\mu} C(p,z)\right].
\end{align}

\item 

It is easy to see that the transformation $C_\mu^5$ cannot change the type-II term. We compute its action on the type-I term.
{\small
\begin{align}\label{eq: y conformal5 type-II}
    \nonumber C_{\mu,y}^5 \int z^{-d+1} \frac{\partial y(p,z)}{\partial z} \frac{\partial y(-p,z)}{\partial z} &= \int z^{-d+1} \frac{\partial C_\mu^5}{\partial z}\frac{\partial y(-p,z)}{\partial z}y(p,z)+(p \rightarrow -p)\\
    & = -2 \int z^{-d+1} z\frac{\partial C(p,z)}{\partial p^\mu} \frac{\partial y(p,z)}{\partial z}y(-p,z).
\end{align}
}
This cancels the term in \eqref{eq: y conformal1234 type-II}.
 
\end{enumerate}

\section{Consistency check: $C_{\mu,\phi}$ and $C_{\mu,y}$}
\label{app: consistency}

In this section we find the symmetry transformation in $\phi$ using the field redefinition $\delta \phi(p,z)= f(p,z) \delta y(p,z)$. For scale transformation we find
{\small
\begin{align}\label{eq: consistency scale}
\notag \delta \phi(p,z) &= f(p,z) D_{y} \left[\frac{\phi(p,z)}{f(p,z)}\right]\\
& \notag = (p^\mu \frac{\p}{\p p^\mu}-z\frac{\p}{\p z}) \phi(p,z)+ f(p,z) \phi(p,z)(p^\mu \frac{\p}{\p p^\mu}-z\frac{\p}{\p z})\frac{1}{f(p,z)}\\
& + (d-\Delta_y) \phi(p,z).
\end{align}
}
Using scaling property of $f(p,z)$ we note $\delta \phi$ in this case matches with \eqref{eq: scale123 phi action}.

The conformal transformation $C_{\mu,\phi}$ can also be checked along similar line. First we note that in this case \textit{crossed} terms can appear because of double derivatives.
{\small
\begin{align}
    \nonumber \delta_\mu^i\phi(p,z)= C^i_{\mu,y} \phi(p,z)+ f(p,z) \phi(p,z) C^i_{\mu,y}\frac{1}{f(p,z)}+ \text{crossed terms},
\end{align}
}
($i=1,2,\dots,5$). The transformation $\delta \phi(p,z)$ derived in this method match with $C_{\mu,\phi}$ in \eqref{eq: phi conformal 12345} as we show below. 
We write down $\delta \phi(p,z)$ part by part as follows.
\begin{subequations}
{\small
\paragraph{1)}
\begin{align}\label{eq: consistency conformal1}
  \notag \delta^1_\mu \phi(p,z) & = \mathbf{2(d-\Delta_\phi)\frac{\partial \phi(p,z)}{\partial p^\mu}}+ 2(d-\Delta_y)\phi(p,z) f(p,z) \frac{\partial}{\partial p^\mu}\frac{1}{f(p,z)}+ 2 (\Delta_\phi-\Delta_y)\frac{\partial \phi(p,z)}{\partial p^\mu},\\
\end{align}

\paragraph{2)}
\begin{align}\label{eq: consistency conformal2}
    \nonumber \delta^2_\mu \phi(p,z) = & \mathbf{2 p^\rho\frac{\partial^2\phi(p,z)}{\partial p^\rho \partial p^\mu}}+2p^\rho f(p,z) \frac{\partial}{\partial p^\mu}\frac{1}{f(p,z)}\frac{\partial \phi(p,z)}{\partial p^\rho}+2p^\rho f(p,z) \frac{\partial}{\partial p^\rho}\frac{1}{f(p,z)}\frac{\partial \phi(p,z)}{\partial p^\mu}\\
& + f(p,z)\phi(p,z) 2 p^\rho\frac{\partial^2}{\partial p^\rho \partial p^\mu}\left[\frac{1}{f(p,z)}\right],
\end{align}

\paragraph{3)}

\begin{align}\label{eq: consistency conformal3}
 \delta^3_\mu \phi(p,z)=\mathbf{-p^\mu\frac{\partial^2 \phi(p,z)}{\partial p^\rho \partial p_\rho}}- f(p,z) \phi(p,z) p_\mu \frac{\partial^2}{\partial p^\rho \partial p_\rho}\frac{1}{f(p,z)}-2p_\mu f(p,z) \frac{\partial}{\partial p^\rho}\frac{1}{f(p,z)}\frac{\partial \phi(p,z)}{\partial p_\rho},
\end{align}

\paragraph{4)}

\begin{align}\label{eq: consistency conformal4}
    \nonumber \delta^4_\mu \phi(p,z)= & \mathbf{- 2 z\frac{\partial^2\phi(p,z)}{\partial z\partial p^\mu}}-2 z f(p,z) \frac{\partial}{\partial p^\mu}\frac{1}{f(p,z)}\frac{\partial \phi(p,z)}{\partial z}-2 z f(p,z) \frac{\partial}{\partial z}\frac{1}{f(p,z)}\frac{\partial \phi(p,z)}{\partial p^\mu}\\
&- 2 z f(p,z)\phi(p,z) \frac{\partial^2}{\partial z \partial p^\mu}\left[\frac{1}{f(p,z)}\right],
\end{align}

\paragraph{5)}

\begin{align}\label{eq: consistency conformal5}
    \nonumber \delta^5_\mu \phi(p,z)=\left[ \int_z z \frac{\partial}{\partial p^\mu} C(p,z)\right]\phi(p,z).\\    
\end{align}
}
\end{subequations}

From \eqref{eq: phi conformal 12345} we observe the bold-typed terms above match with scale independent term $C^I_{\mu,\phi}$ and scale dependent but cutoff function independent term on the identification of $z=e^t$. Hence, we should be able to show that the rest of terms in \eqref{eq: consistency conformal1}-\eqref{eq: consistency conformal5} should equal to cutoff function dependent part $C_{\mu,\phi}^5 \phi(p,z)= 2 \left( \frac{\partial}{\partial p^\mu}\ln f\right)z\frac{\partial \phi(p,z)}{\partial z}$ in \eqref{eq: phi conformal 12345} i.e.
{\small
\begin{align}\label{eq: consistency conformal full}
    & \nonumber 2(d-\Delta_y)\phi(p,z) f(p,z) \frac{\partial}{\partial p^\mu}\frac{1}{f(p,z)}+ 2 (\Delta_\phi-\Delta_y)\frac{\partial \phi(p,z)}{\partial p^\mu}\\
    & \nonumber +2p^\rho f(p,z) \frac{\partial}{\partial p^\mu}\frac{1}{f(p,z)}\frac{\partial \phi(p,z)}{\partial p^\rho} +2p^\rho f(p,z) \frac{\partial}{\partial p^\rho}\frac{1}{f(p,z)}\frac{\partial \phi(p,z)}{\partial p^\mu}\\
    & \nonumber + f(p,z)\phi(p,z) 2 p^\rho\frac{\partial^2}{\partial p^\rho \partial p^\mu}\left[\frac{1}{f(p,z)}\right]- f(p,z) \phi(p,z) p_\mu \frac{\partial^2}{\partial p^\rho \partial p_\rho}\frac{1}{f(p,z)}\\
    & \nonumber -2p_\mu f(p,z) \frac{\partial}{\partial p^\rho}\frac{1}{f(p,z)}\frac{\partial \phi(p,z)}{\partial p_\rho}-2 z f(p,z) \frac{\partial}{\partial p^\mu}\frac{1}{f(p,z)}\frac{\partial \phi(p,z)}{\partial z}\\
    & \nonumber -2 z f(p,z) \frac{\partial}{\partial z}\frac{1}{f(p,z)}\frac{\partial \phi(p,z)}{\partial p^\mu}- 2 z f(p,z)\phi(p,z) \frac{\partial^2}{\partial z \partial p^\mu}\left[\frac{1}{f(p,z)}\right]\\
    & +\left[ \int_z z \frac{\partial}{\partial p^\mu} C(p,z)\right]\phi(p,z)= 2 \left( \frac{\partial}{\partial p^\mu}\ln f\right)z\frac{\partial \phi(p,z)}{\partial z}.
\end{align}
}
We prove this step by step below.

\begin{enumerate}

\item

First note that the 3rd and 7th term in LHS in \eqref{eq: consistency conformal full} cancels each other using the property of $f(p,z)$ in \eqref{eq: property p-sym f}.
\item

Using the scaling relation of $f(p,z)$ in \eqref{eq: property scale f} we note that the terms proportional to $\frac{\partial \phi(p,z)}{\partial p^\mu}$ vanishes i.e.
{\small
\begin{align}
   \nonumber  & \frac{\partial \phi(p,z)}{\partial p^\mu} \left[2 (\Delta_\phi-\Delta_y)+2p^\rho f(p,z) \frac{\partial}{\partial p^\rho}\frac{1}{f(p,z)} -2 z f(p,z) \frac{\partial}{\partial z}\frac{1}{f(p,z)}\right]\\
    & \notag= \frac{\partial \phi(p,z)}{\partial p^\mu}\left[2\Delta_\phi- 2\left(\frac{d}{2}-\nu\right)\right]\\
    & =0.
\end{align}
}

\item
Next we consider the terms proportional to $\phi(p,z)$.
\paragraph{a)}
The 5th and 10th term in LHS of \eqref{eq: consistency conformal full} can be simplified as follows:
{\small
\begin{align}\label{eq: consistency check phi1}
 & \nonumber f(p,z)\phi(p,z) 2 p^\rho\frac{\partial^2}{\partial p^\rho \partial p^\mu}\left[\frac{1}{f(p,z)}\right]- 2 z f(p,z)\phi(p,z) \frac{\partial^2}{\partial z \partial p^\mu}\left[\frac{1}{f(p,z)}\right]\\
 & \nonumber = 2 f(p,z) \phi(p,z)\left \lbrace \frac{\partial }{\partial p^\mu} \left[p^\rho \frac{\partial}{\partial p^\rho} \frac{1}{f(p,z)}\right]- \frac{\partial}{\partial p^\mu} \frac{1}{f(p,z)}- z\frac{\partial^2}{\partial z \partial p^\mu}\left[\frac{1}{f(p,z)}\right]\right\rbrace\\
 &= 2 f(p,z) \phi(p,z) \frac{\partial}{\partial p^\mu}\frac{1}{f(p,z)} \left[ -\frac{d}{2}+\nu-1\right].
\end{align}
}
We have used the scaling property of $f(p,z)$ to deduce the last line.

\paragraph{b)} The 6th term in LHS of \eqref{eq: consistency conformal full} can be manipulated using $|p|$ dependence of $f(p,z)$ \eqref{eq: property p-sym f}.
{\small
\begin{align}\label{eq: consistency check phi2}
   & \nonumber - f(p,z) \phi(p,z) p_\mu \frac{\partial^2}{\partial p^\rho \partial p_\rho}\frac{1}{f(p,z)}\\
   & = \nonumber f(p,z) \phi(p,z) \left \lbrace -\frac{\partial}{\partial p^\rho}\left[ p^\mu \frac{\partial}{\partial p^\rho} \frac{1}{f(p,z)}\right]+ \frac{\partial}{\partial p^\mu}\frac{1}{f(p,z)}\right \rbrace\\
   & \nonumber = f(p,z) \phi(p,z) \left \lbrace -\frac{\partial}{\partial p^\rho}\left[ p^\rho \frac{\partial}{\partial p^\mu} \frac{1}{f(p,z)}\right]+ \frac{\partial}{\partial p^\mu}\frac{1}{f(p,z)}\right \rbrace\\
   & \nonumber =-(d-1) \phi(p,z)f(p,z)\frac{\partial}{\partial p^\mu}\frac{1}{f(p,z)}\\
   & +\phi(p,z) \left \lbrace - p^\rho\frac{\partial}{\partial p^\rho} \left(f(p,z)\frac{\partial}{\partial p^\mu}\frac{1}{f(p,z)}\right) + p^\rho\frac{\partial f(p,z)}{\partial p^\rho}\frac{\partial}{\partial p^\mu}\frac{1}{f(p,z)} \right \rbrace. 
\end{align}
}
We collect the terms proportional to $\phi(p,z)$ i.e. \eqref{eq: consistency check phi1}
,\eqref{eq: consistency check phi2}, the 1st term  and last term in LHS of \eqref{eq: consistency conformal full}:
{\small
\begin{align}\label{eq: consistency check conformal phi coefficient}
    & \nonumber \left[ \int_z z \frac{\partial}{\partial p^\mu} C(p,z)\right]\phi(p,z)+(2 \nu-1)\phi(p,z)f(p,z)\frac{\partial}{\partial p^\mu}\frac{1}{f(p,z)}\\
   & +\phi(p,z) \left \lbrace - p^\rho\frac{\partial}{\partial p^\rho} \left(f(p,z)\frac{\partial}{\partial p^\mu}\frac{1}{f(p,z)}\right) + p^\rho\frac{\partial f(p,z)}{\partial p^\rho}\frac{\partial}{\partial p^\mu}\frac{1}{f(p,z)} \right \rbrace. 
\end{align}
}

We provide the expressions of some functions of $f(p,z)$ below.
{\footnotesize
\begin{align}\label{eq: C(p,z) explicit}
 & \nonumber \bullet \frac{\partial}{\partial p^\mu} \frac{1}{f(p,z)}= p^\nu z^{d/2} h^\prime(pz)\frac{z p_\mu}{p}+ p^{\nu-1}\frac{\nu p_\mu}{p} z^{d/2} h(pz),\\
& \nonumber \bullet p^\rho \frac{\partial}{\partial p^\rho} f(p,z)= - \nu p^{-\nu} \frac{z^{-d/2}}{h(pz)}-z p^{1-\nu}z^{-d/2}\frac{h^\prime(pz)}{h^2(pz)},\\
 & \nonumber \bullet p^\rho \frac{\partial}{\partial p^\rho} \left[f(p,z)\frac{\partial}{\partial p^\mu}\frac{1}{f(p,z)}\right]= \left[p^\rho \frac{\partial}{\partial p^\rho}\frac{h^\prime(pz)}{h(pz)}\right]\frac{z p_\mu}{p}-\frac{\nu p_\mu}{p^2},\\
 & \nonumber \bullet C(p,z)=\frac{1}{z^2 h(pz)}\left[ -\frac{d^2}{4} h(pz)+ z^2 \frac{\partial^2 h(pz)}{\partial z^2}+ z \frac{\partial h(pz)}{\partial z} \right],\\
 & \nonumber \bullet z \frac{\partial}{\partial p^\mu}C(p,z)=  \frac{p z^2}{h(pz)}p_\mu h^{\prime \prime \prime}(pz)- \frac{z^2 p p_\mu}{h^2(pz)}h^\prime(pz)h^{\prime \prime}(pz)+3 \frac{z p_\mu}{h(p,z)}h^{\prime\prime}(pz)\\
 & -\frac{1}{h^2(pz)} z p_\mu (h^\prime(pz))^2+ \frac{p_\mu}{p}\frac{h^\prime(pz)}{h(pz)}.
\end{align}
}
The prime denotes the differentiation w.r.t $(pz)$. We denote the sum of the 2nd, 3rd and 4th term in \eqref{eq: consistency check conformal phi coefficient} as $A(p,z)\phi(p,z)$. $A(p,z)$ can be readily computed as,
{\small
\begin{align}\label{eq: A(p,z) consistency check conformal}
  \nonumber & A(p,z)\\
   \nonumber & =(2 \nu-1)f(p,z)\frac{\partial}{\partial p^\mu}\frac{1}{f(p,z)}\\
   &\nonumber +\left \lbrace - p^\rho\frac{\partial}{\partial p^\rho} \left(f(p,z)\frac{\partial}{\partial p^\mu}\frac{1}{f(p,z)}\right) + p^\rho\frac{\partial f(p,z)}{\partial p^\rho}\frac{\partial}{\partial p^\mu}\frac{1}{f(p,z)} \right \rbrace\\
    & =\left[(\nu-2)\frac{\nu p_\mu}{p^2}-\frac{h^\prime(pz)}{h(pz)}\frac{z p_\mu}{p}-z^2 p_\mu\frac{h^{\prime\prime}(pz)}{h(pz)}\right].
\end{align}
}
In order to see that the \eqref{eq: consistency check conformal phi coefficient} vanishes we note the following identity 
\begin{align}\label{eq: identity A(p,z)}
    \frac{\partial A(p,z)}{\partial z}=- z \frac{\partial}{\partial p^\mu}C(p,z).
\end{align}
One can prove this by substituting the functional form of various functions listed in \eqref{eq: C(p,z) explicit} or directly from the properties of $f(p,z)$ (see Appendix \eqref{app: Identities}). Finally the only term left in the LHS of \eqref{eq: consistency conformal full} is the 8th term  i.e.

\begin{align}
    & \nonumber - 2 z f \frac{\partial \phi(p,z)}{\partial z} \frac{\partial}{\partial p^\mu} \frac{1}{f(p,z)}\\
    & = 2 z \frac{\partial}{\partial p^\mu} \ln f(p,z)\frac{\partial \phi(p,z)}{\partial z}.
\end{align}
In fact, this matches the RHS of \eqref{eq: consistency conformal full}. Hence the conformal transformations in $y$ \eqref{eq: y conformal12345} are consistent with  $C_{\mu,\phi}$.
\end{enumerate}

\paragraph{Details of map to $AdS$}~Along similar lines we can explicitly check the consistency of the conformal transformation in $\phi(p,z)$ \eqref{eq: phi conformal12345 map bulk} and $y(p,z)$ \eqref{eq: y conformal12345 map bulk}, when we map the bulk action to $AdS$. We need to consider only  \eqref{eq: consistency check conformal phi coefficient} as only that depends on $f(p,z)$. After substituting $C(p,z)$ this becomes
\begin{align}\label{eq: consistency check conformal map bulk phi coefficient}
    & \nonumber z^2 p_\mu \phi(p,z) +(2 \nu-1)\phi(p,z)f(p,z)\frac{\partial}{\partial p^\mu}\frac{1}{f(p,z)}\\
   & +\phi(p,z) \left \lbrace - p^\rho\frac{\partial}{\partial p^\rho} \left(f(p,z)\frac{\partial}{\partial p^\mu}\frac{1}{f(p,z)}\right) + p^\rho\frac{\partial f(p,z)}{\partial p^\rho}\frac{\partial}{\partial p^\mu}\frac{1}{f(p,z)} \right \rbrace. 
\end{align}
Using \eqref{eq: C(p,z) explicit} one can explicitly compute $A(p,z)\phi(p,z)$ in \eqref{eq: A(p,z) consistency check conformal} as
{\small
\begin{align}
& \nonumber A(p,z)\phi(p,z)\\
&\nonumber =(2 \nu-1)\phi(p,z)f(p,z)\frac{\partial}{\partial p^\mu}\frac{1}{f(p,z)}\\
   &\nonumber  +\phi(p,z) \left \lbrace - p^\rho\frac{\partial}{\partial p^\rho} \left(f(p,z)\frac{\partial}{\partial p^\mu}\frac{1}{f(p,z)}\right) + p^\rho\frac{\partial f(p,z)}{\partial p^\rho}\frac{\partial}{\partial p^\mu}\frac{1}{f(p,z)} \right \rbrace\\
   &\nonumber = \phi(p,z)\left \lbrace \frac{p_\mu \nu}{p^2}(\nu-2)-\frac{z p_\mu}{p}\frac{h^\prime(pz)}{h(pz)}-z^2 p_\mu\frac{h^{\prime\prime}(pz)}{h(pz)}\right \rbrace\\
   & \nonumber = \phi(p,z)\frac{p_\mu}{p^2 h(p,z)}\left[\nu(\nu-2)h(p,z)-(p^2z^2+m^2) h(p,z)\right]\\
   &=-z^2 p_\mu \phi(p,z).  
\end{align}
}
In last two lines we have used \eqref{eq: f eq map bulk}. Hence we see \eqref{eq: consistency check conformal map bulk phi coefficient} vanishes.

\section{Details: Hamiltonian formalism}
\label{app: Hamiltonian details}
In this section we provide the details of the computations done in Section~\ref{sec: conformal Hamiltonian}. First we list the various \textit{equal time} commutators between stress tensor components which will be used here, where we refer the RG scale $t$ or $z$ as time. For brevity of notation we omit the time $z$ from argument of the fields.

\subsubsection*{\underline{(1+1)-dim position space without cutoff}}
\begin{subequations}
\begin{align}
& [\Theta_{01}(x),\Theta_{01}(y)]=i \p_x \delta(x-y) \left \lbrace\p_y \phi(y) \Pi(x)+\p_x \phi(x) \Pi(y)\right \rbrace,\\
& [\Theta_{00}(x),\Theta_{01}(y)]=i \p_x \delta(x-y) \Pi(x) \Pi(y).
\end{align}
\end{subequations}

\subsubsection*{\underline{$d$-dim momentum space with cutoff}}
{\small
\begin{subequations}
\begin{align}
& \notag [\Theta_{0 \mu}(k),\Theta_{0 \nu}(l)]\\
& \notag =-[\int_p \Pi(k+p)p_\mu \phi(-p,z),\int_q \Pi(l+q)q_\nu \phi(-q,z) ]\\
& \notag = -i\int_{p,q} [- \delta^d(p+k-q) p_\mu \phi(-p,z) \Pi(l+q) q_\nu + \delta^d(l+q-p) \Pi(k+p)p_\mu  q_\nu \phi(-q)]\\
& \notag \text{We perform integral on $p$ in the first term and integral on $q$ in the second to obtain}\\
& \label{eq: commutator 0i 0j}=-i \int \left \lbrace q_\mu(q-l)_\nu \Pi(k+q)\phi(l-q)- q_\nu (q-k)_\mu \Pi(l+q)\phi(k-q)\right \rbrace,\\
& \notag [\Theta_{0 0}(k),\Theta_{0 \mu}(l)]\\
& \notag =[\ha \int_q \Pi(k+q)\Pi(-q) G'(q), \int_p \Pi(l+p)(-i p_\mu)\phi(-p)]\\
& \notag = \ha \int_{p,q}\left[(-i p_\mu) \Pi(k+q)\Pi(l+p) G'(q) (-2 \pi i \delta(p+q))+ (-i p_\mu) \Pi(-q)\Pi(l+p) G'(q) (-2 \pi i \delta(k+q-p)) \right]\\
& \notag \text{Performing integral as mentioned above we obtain}\\
& \label{eq: commutator 00 0j}= \ha \int_q q_\mu [G'(q)+G'(k+q)]\Pi(k+q)\Pi(l-q).
\end{align}
\end{subequations}
}
We consider the Hamiltonian corresponding to the evolution operator of Polchinski's equation and find its symmetry group. The symmetry group should comprise of modified scale and conformal transformation. As a warm up we consider a simple action without cutoff.
\begin{align}\label{eq: action euclidean wo cutoff}
S= \ha \int dt d^d x ~  \p_t \phi(x,t) \p_t \phi(x,t).
\end{align}
$t$ is the RG scale. The action is rotation invariant in $d$ spatial direction, but not Lorentz invariant.  The Hamiltonian corresponding to this action is given as
\begin{align}
H=\ha \int dt\, d^d x  ~ \Pi(x,t) \Pi(x,t),
\end{align}
 where 
\begin{align}
\Pi(x,t)= \p_t \phi(x,t)
\end{align}
is the field momentum. The canonical commutation relation is set to be 
\begin{align}
 [\phi(x,t),\Pi(x,t)]=i\delta(x-y).
 \end{align}
The stress-energy tensor for this theory is 
\begin{align}
&\Theta^\alpha_\beta= \frac{\p \mathcal{L}}{\p (\p_\alpha\phi(x,t))}\p_\beta\phi(x,t)- \delta^\alpha_\beta \mathcal{L},
\end{align}
where 
\[\Theta^0_0=\ha (\p_t \phi(x,t))^2= \frac{\Pi(x,t)^2}{2}=\mathcal{H}\]
generates time translation and 
\[\Theta^0_\mu=  \frac{\p \mathcal{L}}{\p (\p_t\phi(x,t))} \p_\mu \phi(x,t)= \Pi(x,t) \p_\mu\phi(x,t)\]
generates translation in $\mu$th $(\mu=1 \dots d)$ spatial direction. Note that $\Theta^\mu_0=0$, hence the energy momentum tensor is not symmetric. 
The variation of the Lagrangian density in \eqref{eq: action euclidean wo cutoff} under any arbitrary variation $\delta x^\mu=\ep^\mu(x)$  can be obtained as (neglecting the total derivative terms)
\begin{align}
\delta \mathcal{S}= - \frac{\partial \ep^\mu(x)}{\p x^\nu}\Theta^\nu_\mu,
\end{align} 
Two cases of variation $\delta x_\mu$ are given below:
\begin{align*}
\ep^\mu(x)& =\ep^\mu: \text{\textit{translation}}\\
&= \epsilon x^\mu: \text{\textit{Dilatation}}
\end{align*}
Hence $\epsilon^\mu(x) \Theta^0_{\mu}$ generates the variation $\delta x^\mu=\ep^\mu(x)$ and the corresponding current density is $\ep^\mu \Theta^\nu_\mu$. $\epsilon x^\nu \Theta^0_\nu(x,t) \equiv \mathcal{D}(x,t)$ is the \textit{dilatation charge density}. We consider the following variant of the conformal transformation
\begin{align}\label{eq: conformal transformation x space}
& \ep^\mu(x)= \bar{\epsilon}_\nu (2 x^\nu x^\mu- \delta^{\nu \mu} x^2),\,\,\ep^0(x)= 2 \bar{\epsilon}_\mu x^\mu t.
\end{align}
This is usual conformal transformation supplemented with a transformation $\delta t$ along the RG scale $t$. The action \eqref{eq: action euclidean wo cutoff} is invariant under this transformation. The conformal charge density is defined as
\begin{align}\label{eq: conformal charge density x space}
\epsilon^\nu \Theta^0_\nu+ \ep^0 \Theta^0_0 = \bar{\epsilon}^\mu K^0_\mu(x,t)\equiv \bar{\epsilon}^\mu C_\mu(x,t).
\end{align} 
The conformal current density is $K^\mu_\alpha(x,t)$. We would like to see how this modified symmetry \eqref{eq: conformal transformation x space} becomes manifested in Hamiltonian description.

\subsubsection{Conformal algebra in $1$-dim position space}

As a warm up, we would like to study the conformal algebra of the above theory in one spatial dimension $(d=1)$. The conformal transformation in this case can be determined from \eqref{eq: conformal transformation x space}.
\begin{align}
\delta x= \bar{\ep} x^2 \equiv \ep^1(x,t), \ \delta t=2 \bar{\ep}x t \equiv \ep^0(x,t).
\end{align}
It can be shown that the action \eqref{eq: action euclidean wo cutoff} is invariant under the above conformal transformation. Subsequently one can find the modified conformal algebra. Suppose we consider the commutator between the generators of  spacial conformal transformation and translation.  The only non-zero components of the respective charge densities are $K^0_1(x,t) \equiv C_1(x,t)$ and $T_1(x,t)$. The corresponding charges are
\begin{align*}
C \equiv \int dx~ C_1(x,t)=& \int dx (\epsilon^1(x,t) \Theta^0_1(x,t)+\epsilon^0(x,t) \Theta^0_0(x,t))\\
=&\int dx~  \left( \bar{\epsilon} \, x^2 \Pi(x,t) \p_x \phi(x,t)+2 \bar{\ep} x t \frac{\Pi(x,t)^2}{2}\right),\\
T \equiv \int dy ~T_1(y,t)  = & \int dy~ \epsilon^1(y,t) \Theta^0_1(y,t)\\
= & \int dy~ \bar{\epsilon} \,\p_y \phi(y,t) \Pi(y,t).\\
\end{align*}
Using the expressions of the commutators provided above, one can obtain the expression of $[C,T]$ as
{\small
\begin{align}
& [C,T]\\
& \notag =  \int dx \int dy \left[ i x^2 \p_x \delta(x-y) \lbrace \p_x \phi(x,t) \Pi(y,t) + \p_y \phi(x,t)\Pi(x)\rbrace+ 2 i x t \p_x \delta(x-y) \Pi(x,t) \Pi(y,t) \right].
\end{align}
}
We integrate by parts on $y$, integrating out $y$ and finally integrating by parts on $x$ to obtain
\begin{align}
[C,T]= -\int dx\, (2 i t H+ 2 i x P)= -2i \int dx\,\mathcal{D}(x,t)  \equiv -2 i D(t). 
\end{align}
We obtain the dilatation charge $D(t)$ which generates RG time evolution. Hence we observed in this case that the action and the Hamiltonian have same conformal symmetry which is possible because the scalar field is dimensionless and we have not imposed any cutoff in the theory.

\subsubsection{Conformal algebra in $d$-dim momentum space}

We consider the evolution operator in $d$-dim momentum space. The conformal charge  can be constructed following \eqref{eq: conformal transformation x space} and \eqref{eq: conformal charge density x space}~(replacing $t$ with $z$ and adding the suitable term for conformal dimension).
{\small
\begin{align}\label{eq: conformal typeI typeII}
\notag \bar{\epsilon}^\mu K^0_\mu=& \bar{\epsilon}^\mu \int_k  \delta(k)  \left[ 2 \Delta_\phi \int_p\frac{\p \phi(p)}{\p p^\mu}\Pi(-p+k)+\left( 2 \frac{\p^2}{\p k_\sigma \p k^\mu}-\delta_{\sigma\mu}\frac{\p^2}{\p k^\rho \p k_\rho}  \right)\Theta^{0}_{\sigma}(k,z)+ 2z \frac{\p}{\p k^\mu} \Theta^{0}_{0}(k,z)\right]\\
& =\bar{\epsilon}^\mu( K^0_\mu|_I+K^0_\mu|_{II}) \equiv \bar{\epsilon}^\mu( C_\mu^I+C_\mu^{II}).
\end{align}
}
It is to be noted $z$ refers to $z_M$ in Minkowski signature. As per the the main text we have decomposed the conformal transformation into a cutoff independent part $ C_\mu^I$ denoted by \textit{type-I} and a cutoff dependent part $ C_\mu^{II}$ known as \textit{type-II}.
We need to compute 
\begin{align}\label{eq: conformal momentum commutator}
& \notag \bar{\epsilon}^j \epsilon^m [C_\mu,T_\nu]\\
& = \bar{\epsilon}^\mu \epsilon^\nu \left \lbrace [C_\mu^I,T_\nu]+[C_\mu^{II},T_\nu] \right \rbrace.
\end{align}
Momentum generator $T_\nu$ is given in \eqref{eq: momentum generator} i.e.
\begin{align}\label{eq: momentum generator}
\epsilon^\nu T_\nu= \epsilon^\nu \int_l \delta(l) \Theta^0_{~\nu}(l,z)= \epsilon^\nu \int_{l} \delta(l) \int_q \Pi(l+q,z)(-i q_\nu) \phi(-q,z).
\end{align}
We evaluate the commuatator \eqref{eq: conformal momentum commutator} step by step below. Note that \textit{prime} denotes derivative w.r.t $z$. We have removed the $z$ from the arguments of fields and cutoff function for brevity of notations.

\subsubsection*{Commutator of \textit{ type-I} term, $C_\mu^I$ and momentum $T_\nu$}
{\small
\begin{align}\label{eq: typeI momentum commutator step1}
& \notag \bar{\epsilon}^\mu \epsilon^\nu[C_\mu^I,T_\nu]\\
&= \notag [\epsilon^\mu \int_k  \delta(k) 2 \Delta_\phi \int_p\frac{\p \phi(p)}{\p p^\mu}\Pi(-p+k), \epsilon^\nu \int_{l} \delta(l) \Theta^0_\nu(l,z) ]\\
&\notag +[\bar{\epsilon}^\mu \int_k  \delta(k) \left( 2 \frac{\p^2}{\p k^\sigma \p k^\mu}-\delta_{\sigma\mu}\frac{\p^2}{\p k^\rho \p k_\rho}  \right)\Theta^{0 \sigma}(k,z), \epsilon^\nu \int_{l} \delta(l) \Theta^0_\nu(l,z) ]\\
& \notag\text{From \eqref{eq: commutator 0i 0j} we can write}\\
& \notag =-i \bar{\epsilon}^\mu \epsilon^\nu \int_k \delta(k)\left( 2 \frac{\p^2}{\p k^\sigma \p k^\mu}-\delta_{\sigma\mu}\frac{\p^2}{\p k^\rho \p k_\rho}  \right) \int_q \left \lbrace q^\sigma q_\nu \Pi(k+q) \phi(-q) - q_\nu (q-k)^\sigma \Pi(q) \phi(k-q)\right \rbrace\\
& + 2\Delta_\phi \bar{\epsilon}^\mu \epsilon^\nu \delta_{\mu\nu}\int_q \Pi(q) \phi(-q).
\end{align}
}
The first term in \eqref{eq: typeI momentum commutator step1}   has $k$ dependence only in $\Pi(k+q,z)$, hence this can be readily written as
\begin{align}\label{eq: typeI momentum commuatator term1}
=-i \bar{\epsilon}^\mu \epsilon^\nu \int_q \phi(-q,z) q^\sigma q_\nu\left(2 \frac{\p^2}{\p q^\sigma \p q^\mu}-\delta_{\sigma \mu}\frac{\p^2}{\p q^\rho \p q_\rho}\right) \Pi(q).
\end{align}
The second term in \eqref{eq: typeI momentum commutator step1} has $k$ dependence in both $(q-k)^\sigma$ and $\phi(k-q,z)$. After integrating over k we integrate by parts on $q$ to transfer the action of conformal transformation on $\Pi(q,z)$. After these manipulations the second term becomes
{\small
\begin{align}\label{eq: typeI momentum commutator term2}
= i \bar{\epsilon}^\mu \epsilon^\nu &\int_q q^\sigma\phi(-q,z) \bigg\lbrace q_\nu \left( 2 \frac{\p^2}{\p q^\sigma \p q^\mu}-\delta_{\sigma \mu}\frac{\p^2}{\p q^\rho \p q_\rho} \right)+ 2 \left( \delta_{\sigma \nu } \frac{\p}{\p q^\mu}+ \delta_{\mu \nu}\frac{\p}{\p q^\sigma}-\delta_{\sigma \mu} \frac{\p}{\p q^\nu} \right)\bigg\rbrace \Pi(q)
\end{align}
}
Combining \eqref{eq: typeI momentum commuatator term1} and \eqref{eq: typeI momentum commutator term2} we finally obtain the commutator \eqref{eq: typeI momentum commutator step1} as
\begin{align}
& \notag \bar{\epsilon}^\mu \epsilon^\nu [C_\mu^I,T_\nu]\\
&= \notag i \bar{\epsilon}^\mu \epsilon^\nu \int_q \phi(-q,z)q^\sigma\left \lbrace 2\left( \delta_{\sigma \nu} \frac{\p}{\p q^\mu}+ \delta_{\mu \nu}\frac{\p}{\p q^\sigma}-\delta_{\sigma \mu} \frac{\p}{\p q^\nu} \right)+ 2\Delta_\phi \delta_{\mu\nu}\right \rbrace\Pi(q).
\end{align}
We further integrate by parts and simplify.
{\small
\begin{align}\label{eq: typeI momentum commutator step2}
& =\notag i \bar{\epsilon}^\mu \epsilon^\nu \int_q\left\lbrace-2   \left( q_\nu\frac{\p}{\p q^\mu}- q_\mu\frac{\p}{\p q^\nu} \right) \phi(-q,z) -2\delta_{\mu \nu} q^\sigma\frac{\p \phi(-q,z)}{\p q^\sigma}-2 (d-\Delta_\phi)\,\delta_{\mu \nu}\phi(-q,z) \right\rbrace\Pi(q).\\
\end{align}
}
The action of the RHS of \eqref{eq: typeI momentum commutator step2} on field $\phi(p,z)$ is given by,
\begin{align}
\delta \phi(p,z)= \left(2 p_\mu\frac{\p}{\p p^\nu}- 2 p_\nu\frac{\p}{\p p^\mu}-2\delta_{\mu \nu} q^\sigma\frac{\p}{\p q^\sigma} -2 (d-\Delta_\phi) \,\delta_{\mu \nu}\right) \phi(p,z).
\end{align}
Hence we conclude
\begin{align}\label{eq: typeI momentum commutator}
[C_\mu^I,T_\nu]= 2(-M_{\mu\nu}+\delta_{\mu\nu} D).
\end{align}
where $M_{\mu\nu}$ and $D$ are the charges corresponding to rotation operator $M_{\mu \nu}(p)= -i\left( p_\mu\frac{\p}{\p p^\nu}- p_\nu\frac{\p}{\p p^\mu} \right)$ and dilatation operator $D(p)=-i\left(p^\sigma\frac{\p}{\p p^\sigma}+(d-\Delta_\phi)\right)$ along $x$.

\subsubsection*{Commutator of\textit{ type-II} term, $C_\mu^{II}$ and momentum $T_\nu$}

Using \eqref{eq: commutator 00 0j} we can write
\begin{align}\label{eq: typeII momentum commutator step1}
& \notag \bar{\epsilon}^\mu \epsilon^\nu [C_\mu^{II},T_\nu]\\
& \notag =2 \bar{\epsilon}^\mu \epsilon^\nu z \int_{k,l} \delta(k)\delta(l)[\frac{\p}{\p k^\mu} \Theta^0_0(k,z),\Theta^0_\nu(l,z)]\\
& =\bar{\epsilon}^\mu \epsilon^\nu z\int_{k} \delta(k)\frac{\p}{\p k^\mu} \int_q q_\nu \left(G'(q)+G'(k+q) \right)\Pi(k+q)\Pi(-q).
\end{align}

\par\noindent\rule{0.25\textwidth}{1 pt}

Before proceeding we note an useful identity. To obtain this we first note that due to $(q\rightarrow -q)$ symmetry we can write
\begin{align}
\int_q \frac{\p \Pi(q)}{\p q^\mu}q_\nu G^\prime(q) \Pi(-q)= \int_q \frac{\p \Pi(-q)}{\p q^\mu}q_\nu G^\prime(q) \Pi(q). 
\end{align}
After integrating by parts we obtain
\begin{align}
\notag &\int_q \frac{\p \Pi(q)}{\p q^\mu}q_\nu G^\prime(q) \Pi(-q)\\
 \notag =&- \int_q \Pi(q) \delta_{\mu \nu} G^\prime(q) \Pi(-q)-\int_q \Pi(q) q_\nu \frac{\p G^\prime(q)}{\p q_\mu} \Pi(-q)\\
& - \int_q \frac{\p \Pi(-q)}{\p q^\mu} q_\nu G^\prime(q)\Pi(q).
\end{align}
which immediately gives
\begin{align}\label{eq: typeII momentum commutator identity}
\int \frac{\p \Pi(q)}{\p q^\mu} q_\nu G'(q) \Pi(-q)= -\ha \int \Pi(q) \delta_{\mu \nu} G'(q) \Pi(-q)-\ha \int_q \Pi(q) q_\nu \frac{\p G'(q)}{\p q_\mu}\Pi(-q).
\end{align}

\par\noindent\rule{0.38\textwidth}{1 pt}

Using this identity, we simplify the first term in \eqref{eq: typeII momentum commutator step1}.
\begin{align}\label{eq: typeII momentum commutator term1}
& \notag \bar{\epsilon}^\mu \epsilon^\nu z\int_k  \delta(k)\frac{\p}{\p k^\mu} \int_q q_\nu G'(q) \Pi(k+q)\Pi(-q)\\
\notag = & \,\bar{\epsilon}^\mu \epsilon^\nu z \int_q q_\nu G'(q)\Pi(-q)\frac{\p}{\p q^j}\Pi(q)\\
 = & -\ha \bar{\epsilon}^\mu \epsilon^\nu z \int_q \left[ \delta_{\mu \nu} G'(q) \Pi(q)\Pi(-q)+ q_\nu \frac{\p G'(q)}{\p q_\mu} \Pi(q)\Pi(-q) \right].
\end{align}
Similarly we simplify the second term in \eqref{eq: typeII momentum commutator step1} using \eqref{eq: typeII momentum commutator identity} to obtain
\begin{align}\label{eq: typeII momentum commutator term2}
& \notag 2 \bar{\epsilon}^\mu \epsilon^\nu z\int_k  \delta(k)\frac{\p}{\p k^\mu} \ha \int_q q_\nu G'(k+q) \Pi(k+q)\Pi(-q)\\
& = -\ha \bar{\epsilon}^\mu \epsilon^\nu z \int_q \left[ \delta_{\mu \nu} G'(q) \Pi(q)\Pi(-q)- q_\nu \frac{\p G'(q)}{\p q_\mu} \Pi(q) \Pi(-q) \right].
\end{align}
Combining \eqref{eq: typeII momentum commutator term1} and \eqref{eq: typeII momentum commutator term2} we finally obtain
\begin{align}\label{eq: typeII momentum commutator}
& [C_\mu^{II},T_\nu]=- \delta_{\mu\nu} \, z\int_q  G'(q) \Pi(q)\Pi(-q)\equiv -2 \delta_{\mu\nu} z H(z).
\end{align}
where the Hamiltonian $H(z)$ is defined in \eqref{eq: Hamiltonian along RG time d dim}. Combining \eqref{eq: typeI momentum commutator} and \eqref{eq: typeII momentum commutator} we write down the full commutator \eqref{eq: conformal momentum commutator} in terms of action on $\phi$ (replacing $z \rightarrow z_M)$.
\begin{center}
\boxed{
[C_\mu,T_\nu]= 2 \left (-M_{\mu\nu}+\delta_{\mu\nu} D-\delta_{\mu\nu} z_M H(z_M)\right)
}
\end{center}
where $z_M H(z_M)$ generates dilatation $-i z_M \frac{\p}{\p z_M}$ along RG scale.

\subsubsection*{\underline{Self commutation of $C_\mu$}}
Next we show the self-commutator of the conformal charge vanishes. Self commutator of $C_\mu^{II}$ being dependent on $\Theta_0^0$ does not contribute. It is straightforward to show $[C_\mu^I,C_\nu^I]=0$. Hence we need to show
\begin{align}
\bar{\epsilon}^\mu \bar{\epsilon}^\nu[C_\mu,C_\nu]= \bar{\epsilon}^\mu \bar{\epsilon}^\nu[C_\mu^I, C_\nu^{II}]-(\mu\rightarrow \nu)=0.
\end{align}
Using the commutator in \eqref{eq: commutator 00 0j} we can evaluate the above expression as
\begin{align}\label{eq: commutator K K Hamiltonian}
\bar{\epsilon}^\mu \bar{\epsilon}^\nu[C_\mu^I,C_\nu^{II}] &=  -\bar{\epsilon}^\mu \bar{\epsilon}^\nu \int_{k,q} \left( 2 \frac{\p^2}{\p k_\sigma \p k^\mu}-\delta_{\sigma \mu} \frac{\p^2}{\p k^\rho \p k_\rho} \right) 2z \frac{\p}{\p q^\nu} \delta(k) \delta(q)\\
\notag & \times \int_l l_\sigma \left[ G^\prime(l)+G^\prime(q+l) \right]\Pi(l+q)\Pi(k-l) - (\mu \rightarrow \nu)\\
\notag & = (-2z)\bar{\epsilon}^\mu \bar{\epsilon}^\nu\int_l \bigg \lbrace  2 l_\sigma\, \frac{\p G^\prime(l)}{\p l^\nu} \,\Pi(l)\, \frac{\p^2 \Pi(-l)}{\p l_\sigma \p l^\mu} + 2 l_\sigma \,(2 G^\prime(l))\,\frac{\p \Pi(l)}{\p l^\nu}\frac{\p^2 \Pi(-l)}{\p l_\sigma \p l^\mu}\\
& \notag -\delta_{\sigma \mu} l_\sigma\, \frac{\p G^\prime(l)}{\p  l^\nu} \Pi(l)\frac{\p^2 \Pi(-l)}{\p l^\rho \p l_\rho}-\delta_{\sigma \mu} l_\sigma\,(2 G^\prime(l))\, \frac{\p \Pi(l)}{\p l^\nu}\frac{\p^2 \Pi(-l)}{\p l^\rho \p l_\rho}\bigg \rbrace -(\mu \rightarrow \nu).
\end{align}
From \eqref{eq: commutator K K Hamiltonian} we note the following points (we omit the multipilicative $(-2z)$ factor for convenience).
\begin{enumerate}
\item
The third term vanishes when we use the property of $G(l)$ i.e. 
\begin{align}
l_\mu \frac{\p}{\p l^\nu} G^\prime(l)=l_\nu \frac{\p}{\p l^\mu} G^\prime(l).
\end{align}

\item

The second term on integrating by parts in $l^\nu$ and subsequently combining with first term admits 
\begin{align}\label{eq: commutator KK Hamiltonian 2nd term}
=2 \int_l G^\prime(l)\, l_\sigma \,\frac{\p \Pi(l)}{\p l^\nu} \frac{\p^2 \Pi(-l)}{\p l_\sigma l^\mu}-(\mu \rightarrow \nu).
\end{align}
We can further simplify \eqref{eq: commutator KK Hamiltonian 2nd term} by integrating by parts in $l_\sigma$. After using the antisymmetry in $\mu$ and $\nu$ and ignoring the total derivative terms the only non-vanishing terms are
\begin{align} \label{eq: commutator KK Hamiltonian 2nd term step2}
& = -2\int_l l_\sigma \frac{\p G^\prime(l)}{\p l^\mu}\,\frac{\p \Pi(l)}{\p l^\nu}\frac{\p \Pi(-l)}{\p l_\sigma}-(\mu \rightarrow \nu)\\
& \notag = -2\int_l l_\mu \frac{\p G^\prime(l)}{\p l^\sigma}\,\frac{\p \Pi(l)}{\p l^\nu}\frac{\p \Pi(-l)}{\p l_\sigma}-(\mu \rightarrow \nu)\\
& \notag =  2 \int_l G^\prime(l) \bigg \lbrace \delta_{\sigma \mu} \frac{\p \Pi(l)}{\p l^\nu}\frac{\p \Pi(-l)}{\p l_\sigma}+ l_\mu\,\frac{\p^2 \Pi(l)}{\p l_\sigma \p l^\nu}\, \frac{\p \Pi(-l)}{\p l^\sigma}+ l_\mu\,\frac{\p \Pi(l)}{\p l^\nu}\frac{\p^2 \Pi(-l)}{\p l^\rho \p l_\rho} \bigg \rbrace -(\mu \rightarrow \nu).
\end{align}
We have used the property of $G(l)$ in second line and integrated by parts on $l_\sigma$ in third line. In the final expression the first term vanishes due to antisymmetry of $\mu$ and $\nu$, the third term cancels with the fourth term in \eqref{eq: commutator K K Hamiltonian}. Finally we integrate by parts in $l^\nu$ on the second term in \eqref{eq: commutator KK Hamiltonian 2nd term step2} to show that term too vanishes i.e.

\begin{align}
& \int_l l_\mu\, G^\prime(l) \,\frac{\p^2 \Pi(l)}{\p l_\sigma \p l^\nu} \frac{\p \Pi(-l)}{\p l^\sigma}-(\mu\rightarrow \nu)\\
\notag & =-\ha \int_l \bigg \lbrace l_\mu \, \frac{\p G^\prime(l)}{\p l^\nu} \frac{\p \Pi(l)}{\p l^\sigma}\frac{\p \Pi(-l)}{\p l_\sigma}+ \delta_{\mu\nu} G^\prime(l) \frac{\p \Pi(l)}{\p l^\sigma}\frac{\p \Pi(-l)}{\p l_\sigma}\bigg \rbrace-(\mu \rightarrow \nu)\\
& \notag =0,
\end{align}
where we have used the property of $G(l)$ in the first term. Hence we can write
\begin{align}
\tcboxmath{
[C_\mu,C_\nu]=0}
\end{align}
\end{enumerate}

\section{Details: Lie algebra}
\label{app: conf algebra details}

In this section we evaluate the non-trivial commutators of modified conformal transformations in $\phi$ \eqref{eq: phi conformal 12345}. The full conformal group transformation in momentum space is given in section \eqref{sec: conf algebra}. We drop the subscript $\phi$ in this section for convenience. We note functional form of $\ln f(p,z)$.

\begin{align}\label{eq: ln f expression}
    & \frac{1}{f(p,z)}= \nonumber p^\nu z^{d/2} h(pz)\\
    & \implies \frac{1}{f(p,z)} =\nonumber z^{d/2-\nu} \Tilde{h}(pz)\\
    & \implies \lnf= (\nu-d/2)\ln z+ \Tilde{f}(pz).
\end{align}
 Using the fact the $\ln f(p,z)$ is function $|p|$ we can conclude the following identity
\begin{align}
    p_\mu\frac{\partial^3 \lnf}{\partial p^\nu \partial p_\rho \partial p^\rho}=p_\nu\frac{\partial^3 \lnf}{\partial p^\mu \partial p_\rho \partial p^\rho}.
\end{align}
and the scaling relation
\begin{align}
\frac{p}{f}\frac{\partial f}{\partial p} - \frac{z}{f}\frac{\partial f}{\partial z}= \frac{d}{2}-\nu.
\end{align}
We provide the details of the non-trivial commutators below.

\subsubsection*{Commutator of special conformal transformation and translation}

It is known that $[C_\mu^I,T_\mu]= -2 M_{\mu\nu}+ 2 \delta_{\mu\nu} \left( p.\frac{\partial}{\partial p}+ (d-\Delta_\phi)\right)$, so we need to evaluate $ \left[ C_\mu^{II},T_\nu\right]$. We note,

\begin{align}
    & \left[ -2 z \frac{\partial^2}{\partial z \partial p^\mu},p_\nu\right]\phi(p,z)=-2 z \frac{\partial \phi(p,z)}{\partial z}\delta_{\mu\nu},\\
    & \nonumber \left[2 \left( \frac{\partial}{\partial p^\mu}\ln f\right)z\frac{\partial}{\partial z}, p_\nu\right]\ppz=0.
\end{align}
Hence we conclude
\begin{align}
\tcboxmath{
    \left[ C_\mu,T_\nu\right]=-2 R_{\mu\nu}+ 2 \delta_{\mu\nu} D.}
\end{align}

\subsubsection*{Self commutator of special conformal transformation}

We require to show
\begin{align}
     [C_\mu,C_\nu]=[C_\mu^I,C_\nu^{II}]+[C_\mu^{II},C_\nu^I]+[C_\mu^{II},C_\nu^{II}]=0
\end{align}
We will use the anti-symmetry in $\mu$ and $\nu$.
\begin{align}\label{eq: symmetry mu nu}
   [C_\mu^{II},C_\nu^I]= -[C_\mu^I,C_\nu^{II}]|_{\mu \leftrightarrow \nu}. 
\end{align}
We start with computing the crossed commutator $[C_\mu^0,C_\nu^2]$
\begin{align}
 [C_\mu^I,C_\nu^{II}]= \left[ 2(d-\Delta_\phi) \frac{\partial}{\partial p^\mu}+ 2 p^\rho\frac{\partial^2}{\partial p^\rho \partial p^\mu}-p_\mu \frac{\partial^2}{\partial p^\rho \partial p_\rho},-2 z\frac{\partial^2}{\partial z \partial p^\nu}+2 \left( \frac{\partial}{\partial p^\nu}\ln f\right)z\frac{\partial}{\partial z}\right].   
\end{align}
This comprises of
\footnotesize
\begin{align}\label{eq: commutator C0 Ct}
   \bullet & \left[  2(d-\Delta_\phi) \frac{\partial}{\partial p_\mu}, 2 \frac{\partial \ln f(p,z)}{\partial p_\nu}z\frac{\partial}{\partial z} \right]\phi(p,z)=4(d-\Delta) \frac{\partial}{\partial p_\mu}\frac{\partial}{\partial p_\nu}\ln f(p,z)z\frac{\partial \phi(p,z) }{\partial z},\\
     \bullet \nonumber & \left[2 p^\rho\frac{\partial^2}{\partial p^\rho \partial p^\mu}, 2 \left( \frac{\partial}{\partial p^\mu}\ln f\right)z\frac{\partial}{\partial z}\right]\phi(p,z)\\
     & \nonumber = 4 p^\rho z\bigg[\frac{\partial^3 }{\partial p^\rho p^\mu p^\nu} \ln f(p,z) \frac{\partial \phi(p,z)}{\partial z} + \frac{\partial^2}{\partial p^\mu \partial p^\nu} \ln f(p,z) \frac{\partial^2 \phi(p,z)}{\partial p^\rho\partial z}\\
     & \nonumber+ \frac{\partial^2}{\partial p^\rho \partial p^\nu} \ln f(p,z) \frac{\partial^2 \phi(p,z)}{\partial p^\mu\partial z}\bigg],\\
     \bullet & \nonumber \left[ -p_\mu \frac{\partial^2}{\partial p^\rho \partial p_\rho}, 2 \left( \frac{\partial}{\partial p^\mu}\ln f\right)z\frac{\partial}{\partial z} \right]\phi(p,z)\\
     & \nonumber= -2 p_\mu z \bigg[\frac{\partial^3 }{\partial p^\rho p_\rho p^\nu} \ln f(p,z) \frac{\partial \phi(p,z)}{\partial z}+ 2\frac{\partial^2}{\partial p^\rho \partial p^\nu} \ln f(p,z) \frac{\partial^2 \phi(p,z)}{\partial p^\rho\partial z}\bigg],\\
\bullet \nonumber & \left[ 2(d-\Delta) \frac{\partial}{\partial p^\mu},-2 z \frac{\partial^2}{\partial z \partial p^\nu} \right]\ppz=0,\\    
    \bullet & \nonumber \left[ 2 p^\rho \frac{\partial^2}{\partial p^\rho \partial p^\mu}, -2 z \frac{\partial^2}{\partial z \partial p^\nu}\right]\ppz= 4 z \frac{\partial^3 \ppz}{\partial z \partial p^\nu \partial p^\mu},\\
    \bullet & \nonumber \left[-p_\mu \frac{\partial^2}{\partial p^\rho \partial p_\rho}, -2 z \frac{\partial^2}{\partial z \partial p^\nu} \right]\ppz =-2z \frac{\partial^3}{\partial z \partial p_\rho \partial p^\rho}\phi(p,z) \delta_{\mu\nu}.
\end{align}
\normalsize
Using the relation \eqref{eq: symmetry mu nu} we can readily write down  the full expression of the crossed commutator.
\begin{align}\label{eq: commutator C0 Ct+Ct C0 step1}
    & [C_\mu^I,C_\nu^{II}]+[C_\mu^{II},C_\nu^I]\\
    & \nonumber = 4 p^\rho z \left[\frac{\partial^2}{\partial p^\rho \partial p^\nu} \ln f(p,z) \frac{\partial^2 \phi(p,z)}{\partial p^\mu\partial z}\right]- 4 p_\mu z \frac{\partial^2}{\partial p^\rho \partial p^\nu}\ln f(p,z) \frac{\partial^2 \phi(p,z)}{\partial p^\rho \partial z}\\
    & \nonumber - 2p_\mu z \frac{\partial^3 \lnf}{\partial p^\nu \partial p_\rho \partial p^\rho}\frac{\partial \phi(p,z)}{\partial z}-(\mu \leftrightarrow \nu).
\end{align}
We use properties of $\ln f(p,z)$ to further simplify \eqref{eq: commutator C0 Ct+Ct C0 step1}
\begin{align}\label{eq: commutator C0 Ct+Ct C0}
    & [C_\mu^I,C_\nu^{II}]+[C_\mu^{II},C_\nu^I]\\
    & \nonumber = 4 p^\rho z \left[\frac{\partial^2}{\partial p^\rho \partial p^\nu} \ln f(p,z) \frac{\partial^2 \phi(p,z)}{\partial p^\mu\partial z}\right]- 4 p_\mu z \frac{\partial^2}{\partial p^\rho \partial p^\nu}\ln f(p,z) \frac{\partial^2 \phi(p,z)}{\partial p^\rho \partial z}-(\mu \leftrightarrow \nu).
\end{align}

Finally we proceed to find the expression of self commutator of scale dependent conformal transformation $C_\mu^{II}$.
\begin{align}
    \nonumber & [C_\mu^{II},C_\nu^{II}]=\left[ -2 z\frac{\partial^2}{\partial z \partial p^\mu}+2 \left( \frac{\partial}{\partial p^\mu}\ln f\right)z\frac{\partial}{\partial z} ,-2 z\frac{\partial^2}{\partial z \partial p^\nu}+2 \left( \frac{\partial}{\partial p^\nu}\ln f\right)z\frac{\partial}{\partial z}\right].\\
\end{align}
This can be evaluated term by term.
\footnotesize
\begin{align}\label{eq: commutator Ct Ct part 1}
    \bullet &\left[-2 z\frac{\partial^2}{\partial z \partial p^\mu},-2 z\frac{\partial^2}{\partial z \partial p^\nu}\right]\ppz=0,\\
    \bullet \nonumber & \left[-2 z\frac{\partial^2}{\partial z \partial p^\mu}, 2 \left( \frac{\partial}{\partial p^\nu}\ln f\right)z\frac{\partial}{\partial z}\right]\ppz= -4 z \frac{\partial^2}{\partial p^\mu \partial p^\nu}\ln f(p,z) \frac{\partial \phi(p,z)}{\partial z}-4 z^2 \frac{\partial^3 \ln f}{\partial p^\mu \partial p^\nu \partial z} \frac{\partial \phi(p,z)}{\partial z}\\
    & \nonumber -4 z^2 \frac{\partial^2}{\partial p^\mu \partial p^\nu} \ln f \frac{\partial^2 \ppz}{\partial z^2} -4 z^2 \frac{\partial^2 \ln f}{\partial p^\nu \partial z} \frac{\partial^2 \ppz}{\partial p^\mu \partial z}.
 \end{align}
 \normalsize
 and interchanging $\mu$ and $\nu$ indices results in
 \begin{align}\label{eq: commutator Ct Ct part 2}
& \nonumber \left[2 \left( \frac{\partial}{\partial p^\mu}\ln f\right)z\frac{\partial}{\partial z},-2 z\frac{\partial^2}{\partial z \partial p^\nu}\right]\ppz =-\left[-2 z\frac{\partial^2}{\partial z \partial p^\mu}, 2 \left( \frac{\partial}{\partial p^\nu}\ln f\right)z\frac{\partial}{\partial z}\right]\ppz|_{\mu \leftrightarrow \nu}.\\
\end{align}
Further, the self commutator of $-2 z\frac{\partial^2}{\partial z \partial p^\mu}$ and $2 \left( \frac{\partial}{\partial p^\mu}\ln f\right) z\frac{\partial}{\partial z}$ vanishes as can be seen from the functional form of $f(p,z)$. Hence, we obtain the self commutator of scale dependent conformal transformation as
\begin{align} \label{eq: commutator Ct Ct}
   & [C_\mu^{II},C_\nu^{II}]= -4 z^2 \frac{\partial^2}{\partial p^\nu \partial z}\lnf \frac{\partial^2 \ppz}{\partial p^\mu \partial z} -(\mu \leftrightarrow \nu).
\end{align}
Finally using \eqref{eq: commutator C0 Ct+Ct C0} and \eqref{eq: commutator Ct Ct}  we obtain the expression of self commutator of modified conformal transformation.
\begin{align}\label{eq: commutator Cmu Cnu step 1}
    & [C_\mu,C_\nu]\\
    & \nonumber = 4 p^\rho z \left[\frac{\partial^2}{\partial p^\rho \partial p^\nu} \ln f(p,z) \frac{\partial^2 \phi(p,z)}{\partial p^\mu\partial z}\right]- 4 p_\mu z \frac{\partial^2}{\partial p^\rho \partial p^\nu}\ln f(p,z) \frac{\partial^2 \phi(p,z)}{\partial p^\rho \partial z}\\
    & \nonumber -4 z^2 \frac{\partial^2}{\partial p^\nu \partial z}\lnf \frac{\partial^2 \ppz}{\partial p^\mu \partial z} -(\mu \leftrightarrow \nu).
\end{align}
Next, we use the scaling property of $f(p,z)$ to simplify the last term on RHS of \eqref{eq: commutator Cmu Cnu step 1}.
\begin{align}\label{eq: commutator Cmu Cnu term 3}
  & -4 z \frac{\partial^2}{\partial p^\nu \partial z}\lnf z \frac{\partial^2 \ppz}{\partial p^\mu \partial z} -(\mu \leftrightarrow \nu)\\
  & = \nonumber \left[ -4z \frac{\partial \lnf}{\partial p^\nu} -4 z p^\rho \frac{\partial^2}{\partial p^\nu \partial p^\rho}\lnf\right] \frac{\partial^2 \ppz}{\partial p^\mu \partial z}-(\mu \leftrightarrow \nu).
\end{align}
Substituting \eqref{eq: commutator Cmu Cnu term 3} in \eqref{eq: commutator Cmu Cnu step 1} we obtain
\begin{align}
    & \nonumber [C_\mu,C_\nu]\\
    & \nonumber = -4z \frac{\partial \lnf}{\partial p^\nu}\frac{\partial^2 \ppz}{\partial p^\mu \partial z} - 4 p_\mu z \frac{\partial^2}{\partial p^\rho \partial p^\nu}\ln f(p,z) \frac{\partial^2 \phi(p,z)}{\partial p^\rho \partial z}+(\mu \leftrightarrow \nu). 
\end{align}
To simplify it further we use the expression of $\ln f(p,z)$ in \eqref{eq: ln f expression} and obtain
\begin{align}\label{eq: identity double p derivative lnf}
    & p_\mu \frac{\partial^2 \lnf}{\partial p^\rho \partial p^\nu}-(\mu \leftrightarrow \nu)\\
    & \nonumber =z p_\mu\left[ \left( \frac{\delta_{\rho \nu}}{p}+ \frac{p_\nu}{-p^2}\frac{p_\rho}{p} \right) \frac{\partial \Tilde{f}(pz)}{\partial (pz)} + z \frac{p_\rho p_\nu}{p^2} \frac{\partial^2 \Tilde{f}(pz)}{\partial (pz)^2}  \right] -(\mu \leftrightarrow \nu)\\
    & =\nonumber  z p_\mu \frac{\delta_{\rho \nu}}{p} \frac{\partial \Tilde{f}(pz)}{\partial (pz)}-(\mu \leftrightarrow \nu)\\
    & \nonumber = \delta_{\rho \nu}\frac{\partial \lnf}{\partial p^\mu}-(\mu \leftrightarrow \nu).
\end{align}
Using \eqref{eq: identity double p derivative lnf} we conclude

\begin{align}
\tcboxmath{
[C_\mu,C_\nu]=0.}
\end{align}

\subsubsection*{Commutator of special  conformal transformation and rotation} 
We obtain the commutation between the scale dependent part of $C_\mu$ and rotation operator                   
\begin{align}                                                                
& \nonumber [C_\mu,M_{\alpha \beta}] \\                                    \nonumber = &\delta_{\mu \alpha} C_\beta^I-\delta_{\mu \beta}C_\alpha^I + [C_\mu^{II}, M_{\alpha \beta}]. 
\end{align}                                                               
We need to check whether
 \begin{align} \label{eq: commutator C2 R}                                    [C_\mu^{II},M_{\alpha \beta}]= [ -2 z\frac{\partial^2}{\partial z \partial p^\mu}+2 \left( \frac{\partial}{\partial p^\mu}\ln f\right)z\frac{\partial}{\partial z},p_\alpha \frac{\partial}{\partial p^\beta}-p_\beta \frac{\partial}{\partial p^\alpha}]=\delta_{\mu \alpha} C_\beta^{II}-\delta_{\mu \beta}C_\alpha^{II}.                                               \end{align}                                                                   
is true. It is sufficient to compute the commutation of $C_\mu^{II}$ with  $p_\alpha \frac{\partial}{\partial p^\beta}$ because of antisymmetry in $\alpha$ and $\beta$.
 
\begin{subequations}                                                            \begin{enumerate}                                                             
\item                                                                   \begin{align}\label{eq: commutator C2 R part 1}
    & [-2 z\frac{\partial^2}{\partial z \partial p^\mu},p_\alpha \frac{\partial}{\partial p^\beta}-p_\beta \frac{\partial}{\partial p^\alpha}]\phi(p,z)=- 2z \frac{\partial^2 \ppz}{\partial z \partial p^\beta} \delta_{\alpha \mu} +(\alpha \leftrightarrow \beta ),
\end{align}                                                                 
\item                                                                     \begin{align}\label{eq: commutator C2 R part 2}
   & [2 \left( \frac{\partial}{\partial p^\mu}\ln f\right)z\frac{\partial}{\partial z}, p_\alpha \frac{\partial}{\partial p^\beta}-p_\beta \frac{\partial}{\partial p^\alpha}]\ppz\\                  & =\nonumber 2 \frac{\partial \ln f }{\partial p^\mu} p_\alpha \frac{\p}{\p p^\beta} \left( z \frac{\p \ppz}{\p z}\right)-p_\alpha \frac{\p}{\p p^\beta}\left[ 2\frac{\partial \ln f }{\partial p^\mu}z\frac{\partial}{\partial z}\right]\ppz -(\alpha \leftrightarrow \beta )\\     & =\nonumber -2 p_\alpha \frac{\p^2 \ln f}{\p p^\beta \p p^\mu} \frac{\p \ppz}{\p z}+ (\alpha \leftrightarrow \beta )\\                                      
& =\nonumber -2 z \frac{\p \ln f}{\p p^\alpha}\delta_{\beta \mu} \frac{\p \ppz}{\p z} + (\alpha \leftrightarrow \beta ).   
 \end{align}                                                           
 We have used \eqref{eq: identity double p derivative lnf} to derive the last line in RHS of \eqref{eq: commutator C2 R part 2} i.e.                                                                           \[ -p_\alpha\frac{\p^2 \ln f}{\p p^\beta \p p^\mu} + (\alpha \leftrightarrow \beta )= -\delta_{\beta \mu} \frac{\p \ln f}{\p p^\alpha}+(\alpha \leftrightarrow \beta).\]                         \end{enumerate}                                                       \end{subequations}                                                     Combining \eqref{eq: commutator C2 R part 1} and \eqref{eq: commutator C2 R part 2} we conclude \eqref{eq: commutator C2 R} is true. Hence we finally obtain
 \begin{align}
\tcboxmath{
 [C_\mu,M_{\alpha \beta}]= \delta_{\mu \alpha} C_\beta-\delta_{\mu \beta}C_\alpha}
\end{align}

\subsubsection*{Commutator of special conformal transformation and dilatation} 
We proceed to find the commutator between the conformal transformation and dilatation i.e.
\begin{align}
[C_\mu,D]=&[C_\mu^I+C_\mu^{II},D]= C_\mu^I+[C_\mu^{II},D].
\end{align}                                                             
Hence we need to verify
 \begin{align}\label{eq: commutator C D}
        [C_\mu^{II},D]=[-2 z\frac{\partial^2}{\partial z \partial p^\mu}+2 \left( \frac{\partial}{\partial p^\mu}\ln f\right)z\frac{\partial}{\partial z}, p^\rho \frac{\p}{\p p^\rho}-z \frac{\p}{\p z}]=C_\mu^{II}.
    \end{align}                                                           
It is easy to see
\begin{align}\label{eq: commutator C D part 1}
[-2 z\frac{\partial^2}{\partial z \partial p^\mu},p^\rho \frac{\p}{\p p^\rho}-z \frac{\p}{\p z}]\ppz=-2 z \frac{\partial^2 \ppz}{\partial z \partial p^\mu} .  
\end{align}
and using scaling relation of $f(p,z)$\eqref{eq: property scale f} we can find
\begin{align}\label{eq: commutator C D part 2}
 & [2 \left( \frac{\partial}{\partial p^\mu}\ln f\right)z\frac{\partial}{\partial z}, p^\rho \frac{\p}{\p p^\rho}-z \frac{\p}{\p z}]\ppz\\                     & =\nonumber -2 p^\rho\frac{\p}{\p p^\rho}\left( \frac{\partial}{\partial p^\mu}\ln f\right)z\frac{\partial \ppz }{\partial z}+2 z\frac{\p}{\p z}\left( \frac{\partial}{\partial p^\mu}\ln f\right)z\frac{\partial \ppz }{\partial z} \\                     & = \nonumber 2 \frac{\p \ln f}{\p p^\mu}\frac{\partial \ppz }{\partial z}.   
\end{align}                                                                      
 Combining \eqref{eq: commutator C D part 1} and \eqref{eq: commutator C D part 2} one can prove \eqref{eq: commutator C D}. Hence we obtain
 
 \begin{align}
 \tcboxmath{
 [C_\mu,D]=C_\mu.}
 \end{align}

Similarly one can check other commutators. Combining all the these above commutation relations we conclude from \eqref{eq: group structure general} that the Lie algebra for $\phi$ transformations form $SO(1,d+1)$.

\section{Details: Symmtery transformation in full ERG equation}
\label{app: full ERG}

In this section we provide the consistency check of the special conformal transformation of $\chi(p,t)$ in \eqref{eq: chi conformal full ERG}. We have noted in Section~\ref{sec: full ERG} that the transformation of $\phi$ can be obtained from the following relation.
\begin{align}
    \delta_\mu^i \phi(p,t)= G(p,t) C^i_{\mu,\chi} \left( \frac{\chi(p,t)}{G(p,t)} \right),
\end{align}
$(i=1,2,\cdots 5)$. Hence we need to simplify the following expressions (substituting $\Delta_\phi= \frac{d}{2}-\nu$).
\begin{subequations}
{\small
\begin{enumerate}
    \item 
    \begin{align}\label{eq: consistency chi conformal1}
    & G(p,t) C_{\mu,\chi}^1 \left[ \frac{\phi(p,t)}{G(p,t)} \right] = 2 \left(\frac{d}{2}-\nu\right)\frac{\p \phi(p,t)}{\p p^\mu}+ 2 \left(\frac{d}{2}-\nu\right) G(p,t) \phi(p,t) \frac{\p}{\p p_\mu} \frac{1}{G(p,t)},
\end{align}
\item 
\begin{align}\label{eq: consistency chi conformal2}
   G(p,t) C_{\mu,\chi}^2 \left( \frac{\phi(p,t)}{G(p,t)} \right) &= \mathbf{2 p^\rho\frac{\partial^2\phi(p,t)}{\partial p^\rho \partial p^\mu}}+2p^\rho G(p,t) \frac{\partial}{\partial p^\mu}\frac{1}{G(p,t)}\frac{\partial \phi(p,t)}{\partial p^\rho}\\
    & \nonumber +2p^\rho G(p,t) \frac{\partial}{\partial p^\rho}\frac{1}{G(p,t)}\frac{\partial \phi(p,t)}{\partial p^\mu}+ G(p,t)\phi(p,t) 2 p^\rho\frac{\partial^2}{\partial p^\rho \partial p^\mu}\frac{1}{G(p,t)},
\end{align}
\item
 \begin{align}\label{eq: consistency chi conformal3}
  G(p,t) C_{\mu,\chi}^3 \left[\frac{\phi(p,t)}{G(p,t)}\right] & =\mathbf{-p^\mu\frac{\partial^2 \phi(p,t)}{\partial p^\rho \partial p_\rho}}- G(p,t) \phi(p,t) p_\mu \frac{\partial^2}{\partial p^\rho \partial p_\rho}\frac{1}{G(p,t)}\\
    & \nonumber -2p_\mu G(p,t) \frac{\partial}{\partial p^\rho}\frac{1}{G(p,t)}\frac{\partial \phi(p,t)}{\partial p_\rho},
\end{align}   

\item 

\begin{align}\label{eq: consistency chi conformal4}
    G(p,t) C_{\mu,\chi}^4 \left[\frac{\phi(p,t)}{G(p,t)}\right] &= \mathbf{- 2 \frac{\partial^2\phi(p,t)}{\partial t\partial p^\mu}}\color{black}-2 f(p,z) \frac{\partial}{\partial p^\mu}\frac{1}{G(p,t)}\frac{\partial \phi(p,t)}{\partial t}\\
    & \nonumber -2 G(p,t) \frac{\partial}{\partial t}\frac{1}{G(p,t)}\frac{\partial \phi(p,t)}{\partial p^\mu}- 2  G(p,t)\phi(p,t) \frac{\partial^2}{\partial t\partial p^\mu}\left[\frac{1}{G(p,t)}\right],
    \end{align}

\item 

\begin{align}\label{eq: consistency chi conformal5}
    & G(p,t) C_{\mu,\chi}^5 \left[\frac{\phi(p,z)}{G(p,t)}\right]\\
    & \nonumber  = -\dot{F}(p,t)\frac{\partial}{\partial p^\mu} \frac{1}{\dot{F}(p,t)}\frac{\partial \phi(p,t)}{\partial t}- G(p,t) \dot{F}(p,t)\frac{\partial}{\partial p^\mu} \frac{1}{\dot{F}(p,t)} \phi(p,t) \frac{\p}{\p t} \frac{1}{G(p,t)}.
    \end{align}

\end{enumerate}
}
\end{subequations}

From eq.\eqref{eq: consistency chi conformal1}-\eqref{eq: consistency chi conformal5} we observe the following :

\begin{subequations}
\begin{enumerate}
\item
\underline{\textit{Coefficient of $\phi(p,t)$}}
{\small
\begin{align}\label{eq: consistency chi phi part1}
    & \notag 2 \left(\frac{d}{2}-\nu\right) G(p,t) \frac{\p}{\p p_\mu} \frac{1}{G(p,t)}+ G(p,t)2 p^\rho\frac{\partial^2}{\partial p^\rho \partial p^\mu} \frac{1}{G(p,t)}-G(p,t) p_\mu \frac{\p^2}{\p p^\rho p_\rho} \frac{1}{G(p,t)}\\
    & \nonumber - 2  G(p,t)\frac{\partial^2}{\partial t\partial p^\mu}\left[\frac{1}{G(p,t)}\right]- G(p,t) \dot{F}(p,t)\frac{\partial}{\partial p^\mu} \frac{1}{\dot{F}(p,t)} \frac{\p}{\p t} \frac{1}{G(p,t)}\\
    & \notag = 2 \left(\frac{d}{2}-\nu\right) G(p,t) \frac{\p}{\p p_\mu} \frac{1}{G(p,t)}-G(p,t) p_\mu \frac{\p^2}{\p p^\rho p_\rho} \frac{1}{G(p,t)}\\
    & +(4 \nu-2)G(p,t)\frac{\p}{\p p_\mu} \frac{1}{G(p,t)}- G(p,t) \dot{F}(p,t)\frac{\partial}{\partial p^\mu} \frac{1}{\dot{F}(p,t)} \frac{\p}{\p t} \frac{1}{G(p,t)},
\end{align}
}
where we have used the property of $\frac{\p F(p,t)}{\p p^\mu}$.
\begin{align}\label{eq: property F}
    p^\rho \frac{\p^2}{\p p^\rho \p p^\mu} F(p,t)- \frac{\p^2}{\p t \p p^\mu} F(p,t)= (2 \nu-1) F(p,t)
\end{align}
Following \eqref{eq: consistency check phi2} we can further simplify \eqref{eq: consistency chi phi part1} as follows:
\begin{align}
    & \nonumber (2 \nu-1)G(p,t)\frac{\p}{\p p_\mu} \frac{1}{G(p,t)} - G(p,t) \dot{F}(p,t)\frac{\partial}{\partial p^\mu} \frac{1}{\dot{F}(p,t)} \frac{\p}{\p t} \frac{1}{G(p,t)}\\
    & - G(p,t) p^\rho \frac{\p^2}{\p p^\rho \p p^\mu}\frac{1}{G(p,t)}.
\end{align}
We substitute the following expression in second term 
\begin{align}
   \dot{F}(p,t)\frac{\partial}{\partial p^\mu} \frac{1}{\dot{F}(p,t)}= - \frac{1}{\dot{G}(p,t)} \frac{\p}{\p p^\mu} \dot{G}(p,t)+ \frac{2}{G(p,t)}\frac{\p G(p,t)}{\p p^\mu},
\end{align}
and the property of $F(p,t)$ \eqref{eq: property F} in the third term  to note the coefficient of $\phi(p,t)$ vanishes. 

\item

\underline{\textit{Coefficient of $\frac{\p \phi(p,t)}{\p t}$}}

Similarly we can show
{\small
\begin{align}
    & \notag -2 G(p,t) \frac{\p}{\p p^\mu} \frac{1}{G(p,t)}-\dot{F}(p,t)\frac{\partial}{\partial p^\mu} \frac{1}{\dot{F}(p,t)}= \dot{G}(p,t)\frac{\partial}{\partial p^\mu} \frac{1}{\dot{G}(p,t)}.
\end{align}
}
\item
\underline{\textit{Coefficient of $\frac{\p \phi(p,t)}{\p p^\mu}$}}

Using the scaling relation of $F(p,t)$
\begin{align}
   \left( p^\rho \frac{\p}{\p p^\rho}- z\frac{\p}{\p z} \right) F(p,t)=2 \nu F(p,t),
\end{align}
we can show that the coefficient of $\frac{\p \phi(p,t)}{\p p^\mu}$ becomes
\begin{align}
    & 2 \left(\frac{d}{2}-\nu\right) + 2 G(p,t)  p^\rho \frac{\p}{\p p^\rho} \frac{1}{G(p,t)}-2 z\frac{\p}{\p z} \frac{1}{G(p,t)}= 2\left(\frac{d}{2}+\nu \right).
\end{align}
\item 
The 2nd term of \eqref{eq: consistency chi conformal2} and 3rd term of \eqref{eq: consistency chi conformal3} cancel each other as $G(p,t)$ is a function of $|p|$.
\end{enumerate}
\end{subequations}
Hence we can promptly write down the special conformal transformation of $\phi(p,t)$
\begin{align}
    \delta_\mu \phi  \notag &= 2\left(\frac{d}{2}+\nu \right)  \frac{\partial}{\partial p^\mu}+ 2 p^\rho\frac{\partial^2}{\partial p^\rho \partial p^\mu}-p_\mu \frac{\partial^2}{\partial p^\rho \partial p_\rho}\\
&  -2 \frac{\partial^2}{\partial t \partial p^\mu}-\dot{G}(p,t) \frac{\partial}{\partial p^\mu} \frac{1}{\dot{G}(p,t)}\frac{\partial}{\partial t}.
\end{align}
This matches with \eqref{eq: phi conformal 12345} on substitution of $\Delta_\phi= \frac{d}{2}-\nu$.

\section{Useful Identity}
    \label{app: Identities}

We would like to prove the following identity in \eqref{eq: identity A(p,z)} (for brevity of notation we have denoted $f(p,z)$ as $f$ here)
\begin{align}
   \frac{\partial A(p,z)}{\partial z}=- z \frac{\partial}{\partial p^\mu}C(p,z), 
\end{align}
where
   \[ A(p,z)=(2 \nu-1)f\frac{\partial}{\partial p^\mu}\frac{1}{f}+ \left \lbrace - p^\rho\frac{\partial}{\partial p^\rho} \left(f\frac{\partial}{\partial p^\mu}\frac{1}{f}\right) + p^\rho\frac{\partial f}{\partial p^\rho}\frac{\partial}{\partial p^\mu}\frac{1}{f} \right \rbrace,\]
and 
\[C(p,z) z^{-d+1}= -\frac{\partial}{\partial z} \lbrace z^{-d+1} \frac{d \ln f}{d z}\rbrace + z^{-d+1} \left( \frac{d \ln f}{d z}\right)^2.\]
First we note that $C(p,z)$ consists of terms involving $z$ derivatives. Hence one can easily replace $p_\mu$ derivatives in terms of $p$ derivatives.
\begin{align}
    -z\frac{\partial}{\partial p^\mu}C(p,z)=-z \frac{p_\mu}{p}\frac{\partial}{\partial p}C(p,z).
\end{align}
We will see below that an overall multiplicative factor of $\frac{p_\mu}{p}$ can also be extracted out from the LHS. In that anticipation we only compute $\frac{\partial}{\partial p}C(p,z)$.
{\small
\begin{align}
    &\nonumber \frac{\partial C(p,z)}{\partial p}=(d-1) \frac{1}{z}\frac{\partial}{\partial z}\frac{\partial \ln f}{\partial p}-\frac{\partial^2}{\partial z^2} \frac{\partial \ln f}{\partial p}+ 2 \frac{\partial \ln f}{\partial z} \frac{\partial}{\partial z}\frac{\partial \ln f}{\partial p}.
\end{align}
}
We use the scaling relation of $f$ to write $p$ derivatives as $z$ derivatives.
{\small
\begin{align}\label{eq: identity A(p,z) proof RHS}
    =\nonumber \frac{1}{p}\left[\frac{d-1}{z} \left\lbrace\frac{\partial \ln f}{\partial z}+z \frac{\partial^2 \ln f}{\partial z^2} \right\rbrace-\frac{\partial^2}{\partial z^2}\left(z \frac{\partial \ln f}{\partial z}\right)+ 2 \left(\frac{\partial \ln f}{\partial z}\right)^2+2z \frac{\partial \ln f}{\partial z}\frac{\partial^2 \ln f}{\partial z^2}\right].
\end{align}
}
Similarly LHS  can be simplifies as
{\small
\begin{align}
    \frac{\partial A(p,z)}{\partial z}=& \frac{\partial}{\partial z} \bigg[(2 \nu-1)f(p,z)\frac{\partial}{\partial p^\mu}\frac{1}{f(p,z)}\\
   &\nonumber  + \left \lbrace - p^\rho\frac{\partial}{\partial p^\rho} \left(f(p,z)\frac{\partial}{\partial p^\mu}\frac{1}{f(p,z)}\right) + p^\rho\frac{\partial f(p,z)}{\partial p^\rho}\frac{\partial}{\partial p^\mu}\frac{1}{f(p,z)} \right \rbrace\bigg].
\end{align}
}
Further, $p_\mu$ derivatives can be written in terms of $p$ derivatives.
{\small
\begin{align}
\nonumber \frac{\partial A(p,z)}{\partial z} &=\frac{p_\mu}{p}\bigg[\left(\frac{d}{2}+\nu-1\right)\frac{\partial}{\partial z}\left(f\frac{\partial}{\partial p} \frac{1}{f}\right)+ \frac{z}{f}\frac{\partial f}{\partial z} \frac{\partial}{\partial z}\left( f\frac{\partial}{\partial p} \frac{1}{f}\right)\\
   &\nonumber+\frac{\partial}{\partial z} \left(\frac{z}{f}\frac{\partial f}{\partial z}\right)f\frac{\partial}{\partial p}\frac{1}{f}-p^\rho\frac{\partial}{\partial p^\rho} \frac{\partial}{\partial z}\left(f\frac{\partial}{\partial p}\frac{1}{f}\right)\bigg].
   \end{align}
   }
We simplify the coefficient of $\frac{p_\mu}{p}$ in $\frac{\partial A(p,z)}{\partial z}$ further by replacing $p$ derivatives with $z$ derivatives.
{\small
\begin{align}
    \nonumber & -\left( \frac{d}{2}+\nu-1\right)\frac{\partial}{\partial z}\left[\frac{1}{p}\left \lbrace \frac{d}{2}-\nu+\frac{z}{f}\frac{\partial f}{\partial z}\right \rbrace\right]-\frac{z}{f}\frac{\partial f}{\partial z}\frac{\partial}{\partial z}\left[\frac{1}{p}\left \lbrace \frac{d}{2}-\nu+\frac{z}{f}\frac{\partial f}{\partial z}\right \rbrace\right]\\
    \nonumber & -\frac{1}{p}\frac{\partial}{\partial z} \left(\frac{z}{f}\frac{\partial f}{\partial z}\right)\left \lbrace \frac{d}{2}-\nu+\frac{z}{f}\frac{\partial f}{\partial z} \right \rbrace+p^\rho\frac{\partial}{\partial p^\rho}\left[\frac{1}{p}\frac{\partial}{\partial z} \left \lbrace \frac{d}{2}-\nu+\frac{z}{f}\frac{\partial f}{\partial z} \right \rbrace \right]\\
    \nonumber & =\frac{1}{p}\frac{\partial}{\partial z} \left(\frac{z}{f}\frac{\partial f}{\partial z}\right)\left \lbrace -\frac{d}{2}-\nu+1-\frac{d}{2}+\nu -2\frac{z}{f}\frac{\partial f}{\partial z}\right \rbrace\\
    & \nonumber + \frac{p^\rho}{p}\frac{\partial}{\partial p^\rho}\frac{\partial}{\partial z}\left( \frac{z}{f}\frac{\partial f}{\partial z}\right)-\frac{1}{p}\frac{\partial}{\partial z} \left(\frac{z}{f}\frac{\partial f}{\partial z}\right).\\
\end{align}
}
Next we write the above expression in terms of $\ln f$ and use the identity
\[\left[p^\rho \frac{\partial}{\partial p^\rho}-z\frac{\partial}{\partial z}\right]\left(\frac{z}{f}\frac{\partial f}{\partial z}\right)=0.\]
to finally obtain
\begin{align}\label{eq: identity A(p,z) proof LHS}
    & \nonumber =\frac{1}{p} \bigg[ (-d+1) \frac{\partial \ln f}{\partial z}-2z\left(\frac{\partial \ln f}{\partial z}\right)^2\\
    & +\left(-d+1-2z\frac{\partial \ln f}{\partial z} \right) z \frac{d^2 \ln f}{d z^2}+z \frac{\partial^2}{\partial z^2}\left( z \frac{\partial \ln f}{\partial z}\right)\bigg].
\end{align}
Comparing \eqref{eq: identity A(p,z) proof LHS} and \eqref{eq: identity A(p,z) proof RHS} we can prove the identity.

\end{appendices}


\newpage
\begin{thebibliography}{99}



\bibitem{Maldacena} 
  J.~M.~Maldacena,
  ``The Large N limit of superconformal field theories and supergravity,''
  Int.  J. Theor. Phys.  {\bf 38}, 1113 (1999)
  [Adv. Theor. Math. Phys.  {\bf 2}, 231 (1998)]
  doi:10.1023/A:1026654312961
  \texttt{arXiv:hep-th/9711200}.
  
\bibitem{Polyakov}  
S.~S.~Gubser, I.~R.~Klebanov, and A.~M.~Polyakov,
``Gauge theory correlators from non-critical string
theory,'' Phys. Lett. \textbf{B428} (1998) 105-114,
\texttt{arXiv:hep-th/9802109}.

\bibitem{Witten1} 
E.~Witten, ``Anti-de Sitter space and holography,'' 
Adv. Theor. Math. Phys. \textbf{2} (1998) 253-291,
\texttt{arXiv:hep-th/9802150}.
  
\bibitem{Witten2} 
  E.~Witten,
  ``Anti-de Sitter space, thermal phase transition, and confinement in
  gauge theories,'' 
  Adv. Theor. Math. Phys.  {\bf 2}, 505 (1998)
  \texttt{arXiv:hep-th/9803131}.


\bibitem{tHooft:1993} 
  G.~'t Hooft,
  ``Dimensional reduction in quantum gravity,''
  Conf.\ Proc.\ C {\bf 930308}, 284 (1993)
  [gr-qc/9310026].
  
  \bibitem{Susskind:1994} 
  L.~Susskind,
  ``The World as a hologram,''
  J.\ Math.\ Phys.\  {\bf 36}, 6377 (1995)
  doi:10.1063/1.531249
  [hep-th/9409089].

  
  


\bibitem{Akhmedov}
E.~T.~Akhmedov, ``A Remark on the AdS / CFT
correspondence and the renormalization group flow,''
Phys. Lett.  \textbf{B442} (1998) 152-158,
\texttt{arXiv:hep-th/9806217 [hep-th]}.

\bibitem{Akhmedov1}
E.~T.~Akhmedov, ``Notes on multitrace operators and holographic renormalization
group".
Talk given at 30 Years of Supersymmetry, Minneapolis, Minnesota, 13-27 Oct
2000, and at Workshop on Integrable Models, Strings and Quantum Gravity,
Chennai, India, 15-19 Jan 2002.
\texttt{arXiv: hep-th/0202055}

\bibitem{Akhmedov2}
E.~T.~ Akhmedov, I.B. Gahramanov, E.T. Musaev,`` Hints on integrability in the Wilsonian/holographic renormalization group"
\texttt{ arXiv:1006.1970 [hep-th]}



\bibitem{Alvarez} 
E.~Alvarez and C.~Gomez, ``Geometric holography, the
renormalization group and the c theorem,'' Nucl.Phys.
\textbf{B541} (1999) 441-460, \texttt{arXiv:hep-th/9807226
[hep-th]}.





V.~Balasubramanian and P.~Kraus, ``Space-time and the
holographic renormalization group,'' Phys. Rev. Lett. \textbf{83}
(1999) 3605-3608, \texttt{arXiv:hep-th/9903190 [hep-th]}.

\bibitem{Warner} 
D.~Freedman, S.~Gubser, K.~Pilch, and N.~Warner,
``Renormalization group flows from holography
supersymmetry and a c theorem,''
Adv. Theor. Math. Phys. \textbf{3} (1999) 363-417,
\texttt{arXiv:hep-th/9904017 [hep-th]}.

\bibitem{Verlinde} 
J.~de~Boer, E.~P.~Verlinde, and H.~L.~Verlinde, ``On the
holographic renormalization group,'' JHEP \textbf{08} (2000)
003, \texttt{arXiv:hep-th/9912012}.

\bibitem{Boer} J.~de~Boer, 
``The Holographic renormalization group,''
Fortsch.~Phys. \textbf{49} (2001) 339-358,
\texttt{arXiv:hep-th/0101026 [hep-th]}.

\bibitem{Faulkner} 
  T.~Faulkner, H.~Liu, and M.~Rangamani,
  ``Integrating out geometry: Holographic Wilsonian RG and the membrane paradigm,''
  JHEP {\bf 1108}, 051 (2011)
  doi:10.1007/JHEP08(2011)051
\texttt{arXiv:1010.4036 [hep-th]}.
\bibitem{Igarashi} 
  Y.~Igarashi, K.~Itoh, and H.~Sonoda,
  ``Realization of Symmetry in the ERG Approach to Quantum Field Theory,''
  Prog.\ Theor.\ Phys.\ Suppl.\  {\bf 181}, 1 (2010)
  doi:10.1143/PTPS.181.1
  \texttt{arXiv:0909.0327 [hep-th]}
\bibitem{Sonoda:2015}
H.~Sonoda,
``Construction of the Energy-Momentum Tensor for Wilson Actions,''
Phys. Rev. D \textbf{92}, no.6, 065016 (2015)
doi:10.1103/PhysRevD.92.065016
[arXiv:1504.02831 [hep-th]].

\bibitem{Klebanov:1999tb} 
  I.~R.~Klebanov and E.~Witten,
  ``AdS / CFT correspondence and symmetry breaking,''
  Nucl. Phys. {\bf B556}, 89 (1999)
  doi:10.1016/S0550-3213(99)00387-9
  \texttt{arXiv:hep-th/9905104}.
 
 \bibitem{Heemskerk} 
  I.~Heemskerk and J.~Polchinski,
  ``Holographic and Wilsonian Renormalization Groups,''
  JHEP {\bf 1106}, 031 (2011)
  doi:10.1007/JHEP06(2011)031
\texttt{arXiv:1010.1264 [hep-th]}.   

\bibitem{Morris} 
  J.~M.~Lizana, T.~R.~Morris, and M.~Perez-Victoria,
  ``Holographic renormalisation group flows and renormalisation from a Wilsonian perspective,''
  JHEP {\bf 1603}, 198 (2016)
  doi:10.1007/JHEP03(2016)198
\texttt{arXiv:1511.04432 [hep-th]}.

\bibitem{Bzowski:2015pba} 
  A.~Bzowski, P.~McFadden, and K.~Skenderis,
  ``Scalar 3-point functions in CFT: renormalisation, beta functions and anomalies,''
  JHEP {\bf 1603}, 066 (2016)
  doi:10.1007/JHEP03(2016)066
\texttt{arXiv:1510.08442 [hep-th]}.
  
\bibitem{deHaro:2000} 
  S.~de Haro, S.~N.~Solodukhin, and K.~Skenderis,
  ``Holographic reconstruction of space-time and renormalization in the AdS / CFT correspondence,''
  Comm.\ Math.\ Phys.\  {\bf 217}, 595 (2001)
  doi:10.1007/s002200100381
\texttt{arXiv:hep-th/0002230}.

\bibitem{Skendris:2002}
``Lecture Notes on Holographic Renormalization,"
Class.Quant.Grav.19:5849-5876,2002,
\texttt{arXiv:hep-th/0209067}
  
\bibitem{SSLee:2010} S.-S. Lee, ``Holographic description of quantum field theory”, Nuclear Physics B 832 (Jun, 2010) 567585,
\texttt{arXiv:0912.5223}.

\bibitem{SSLee:2012}
 S.-S. Lee, ``Background independent holographic description: from matrix field theory to quantum gravity”,
Journal of High Energy Physics 2012 (Oct, 2012) 160, 
\texttt{arXiv:1204.1780}.
\bibitem{Wilson} 
  K.~G.~Wilson and J.~B.~Kogut,
  ``The Renormalization group and the epsilon expansion,''
  Phys. Rept. {\bf 12}, 75 (1974).
  doi:10.1016/0370-1573(74)90023-4

\bibitem{Wegner}
F.~J.~Wegner and A.~Houghton, ``Renormalization group
equation for critical phenomena,'' Phys.\ Rev.\ \textbf{A8} (1973)
401-412.

\bibitem{Wilson2} 
K.~G.~Wilson, ``The renormalization group and critical
phenomena,''  Rev. Mod. Phys. \textbf{55} (1983) 583-600.
\bibitem{Sathiapalan:2017} 
  B.~Sathiapalan and H.~Sonoda,
  ``A Holographic form for Wilson's RG,''
  Nucl.\ Phys.\ B {\bf 924}, 603 (2017)
  doi:10.1016/j.nuclphysb.2017.09.018
  [arXiv:1706.03371 [hep-th]].



  
  \bibitem{Sathiapalan:2019} 
  B.~Sathiapalan and H.~Sonoda,
  ``Holographic Wilson's RG,''
  Nucl.\ Phys.\ B {\bf 948}, 114767 (2019)
  doi:10.1016/j.nuclphysb.2019.114767
  [arXiv:1902.02486 [hep-th]].

\bibitem{Polchinski} 
  J.~Polchinski,
  ``Renormalization and Effective Lagrangians,''
  Nucl. Phys. {\bf B231}, 269 (1984).
  doi:10.1016/0550-3213(84)90287-6

  

\bibitem{Sathiapalan:2020}
B.~Sathiapalan,
``Holographic RG and Exact RG in O(N) Model,''
Nucl. Phys. B \textbf{959}, 115142 (2020)
doi:10.1016/j.nuclphysb.2020.115142
[arXiv:2005.10412 [hep-th]].

\bibitem{Dharanipragada:2020}
P.~Dharanipragada, S.~Dutta and B.~Sathiapalan,
``Bulk gauge fields and holographic RG from exact RG,''
JHEP \textbf{23}, 174 (2020)
doi:10.1007/JHEP02(2023)174
[arXiv:2201.06240 [hep-th]].


\bibitem{Dharanipragada:2022}
P.~Dharanipragada, S.~Dutta and B.~Sathiapalan,
``Aspects of the map from exact RG to holographic RG in AdS and dS,''
Mod. Phys. Lett. A \textbf{37}, no.37n38, 2250235 (2022)
doi:10.1142/S0217732322502352
[arXiv:2301.13605 [hep-th]].
\bibitem{Dharanipragada:2023}
P.~Dharanipragada and B.~Sathiapalan,
``Holographic RG from an exact RG: Locality and general coordinate invariance in the bulk,''
Phys. Rev. D \textbf{109}, no.10, 106017 (2024)
doi:10.1103/PhysRevD.109.106017
[arXiv:2306.07442 [hep-th]].

\bibitem{Sathiapalan:flat}
B.~Sathiapalan
``Mapping from Exact RG to Holographic RG in Flat Space,''
JHEP \textbf{09} (2025) 022
doi:10.1007/JHEP09(2025)022
[arXiv:2408.00628]



%
%
%
%
%
%
%
%
%
%
%
%
%
%
%
%
%
%
%
%
%
%
%
%
%

\bibitem{Wang:2015}
Z-L Wang, Y.Yan
``Bulk Local Operators, Conformal Descendants and Radial Quantization,"
Adv.High Energy Phys. 2017 (2017) 8185690,
doi:10.1155/2017/8185690,
[arXiv:1507.05550].

\bibitem{Rosten:2014}
O.J. Rosten
``On Functional Representations of the Conformal Algebra,"
Phys. J. C (2017) 77: 477,
[arXiv:1411.2603].

\bibitem{Rosten:2016}
O.J. Rosten
``A Wilsonian Energy-Momentum Tensor,"Eur. Phys. J. C (2018) 78: 312,
[arXiv:1605.01055].

\bibitem{Sonoda:conf2017}
H.Sonoda
``Conformal invariance for Wilson actions," PTEP2017.083B05,
[arXiv:1705.01239].


\end{thebibliography}
\end{document}